\documentclass[namedreferences]{SolarPhysics}
\usepackage{spr-sola-addons} 
\usepackage{color}           
\usepackage{url}             

\usepackage[pdftex]{graphicx} 

\newcommand{\kms}{\,km\,s$^{-1}$}
\newcommand{\dcm}{\,dyn\,cm$^{-2}$}
\newcommand{\intun}{\,$\times10^{4}$\,erg\,cm$^{-2}$\,s$^{-1}$\,sr$^{-1}$}

\newcommand{\degr}{{\hbox{$^\circ$}}}

\newcommand{\arcsec}{{\hbox{$^{\prime\prime}$}}}
\renewcommand{\sun}{_\odot}
\newcommand{\ha}{H$_\alpha$}
\newcommand{\hb}{H$_\beta$}
\newcommand{\Mg}{Mg{\sc\,ii}}
\newcommand{\Cb}{C{\sc\,ii}}
\newcommand{\Si}{Si{\sc\,iv}}
\newcommand{\Cl}{Cl\,{\sc i}}
\newcommand{\OI}{O\,{\sc i}}
\newcommand{\OIV}{O\,{\sc iv}}

\newcommand{\be}{\begin{equation}}
\newcommand{\ee}{\end{equation}}
\newcommand{\bea}{\begin{eqnarray}}
\newcommand{\eea}{\end{eqnarray}}
\newcommand{\beas}{\begin{eqnarray*}}
\newcommand{\eeas}{\end{eqnarray*}}

\newcommand{\apjl}{   {\it Astrophys. J. Lett.}}

\chardef\us=`\_ 

\begin{document}

\begin{article}

\begin{opening}

\title{IRIS Observations of Spicules and Structures Near the Solar Limb}

\author[addressref={aff1},corref,email={calissan@cc.uoi.gr}]{\inits{C.E.}\fnm{C. E.}~\lnm{Alissandrakis}} 
\author[addressref={aff2}]{\inits{J.-C.}\fnm{J.-C.}~\lnm{Vial}}
\author[addressref={aff1}]{\inits{A.}\fnm{A.}~\lnm{Koukras}}
\author[addressref={aff2}]{\inits{E.}\fnm{E.}~\lnm{Buchlin}}
\author[addressref={aff2}]{\inits{M.}\fnm{M.}~\lnm{Chane-Yook}}

\address[id=aff1]{Section of Astro-Geophysics, Department of Physics, University of Ioannina, GR-45110 Ioannina, Greece}
\address[id=aff2]{Institut d' Astrophysique Spatiale, CNRS (UMR 8617) Universit\' e Paris-Sud 11, Orsay,  France} 

\runningauthor{C. E. Alissandrakis et al.}
\runningtitle{IRIS Observations of Spicules}

\begin{abstract}
We have analyzed {\it Interface Region Imaging Spectrograph} (IRIS) spectral and slit-jaw observations of a quiet region near the South Pole. In this article we present an overview of the observations, the corrections, and the absolute calibration of the intensity. We focus on the average profiles of strong (Mg\,{\sc ii} h and k, \Cb\ and \Si), as well as of weak spectral lines in the near ultraviolet (NUV) and the far ultraviolet (FUV), including the Mg\,{\sc ii} triplet, thus probing the solar atmosphere from the low chromosphere to the transition region. We give the radial variation of bulk spectral parameters as well as line ratios and turbulent velocities. We present measurements of the formation height in lines and in the NUV continuum, from which we find a linear relationship between the position of the limb and the intensity scale height. We also find that low forming lines, such as the \Mg\ triplet, show no temporal variations above the limb associated with spicules, suggesting that such lines are formed in a homogeneous atmospheric layer and, possibly, that spicules are formed above the height of 2\arcsec. We discuss the spatio-temporal structure near the limb from images of intensity as a function of position and time. In these images, we identify p-mode oscillations in the cores of lines formed at low heights above the photosphere, slow moving bright features  in \OI\ and fast moving bright features in \Cb. Finally, we compare the \Mg\ k and h line profiles, together with intensity values of the Balmer lines from the literature, with computations from the PROM57Mg non-LTE model developed at the Institut d' Astrophysique Spatiale and estimated values of the physical parameters. We obtain electron temperatures in the range of $\sim8000$\,K at small heights to $\sim20000$\,K at large heights, electron densities from $1.1\times10^{11}$ to $4\times10^{10}$\,cm$^{-3}$ and a turbulent velocity of $\sim24$\kms.
\end{abstract}
\end{opening}

\section{Introduction} \label{intro} 
Spicules, first described by \cite{1875leso.book.....S} almost one and a half century ago, still elude us as far as their physical parameters, dynamics and origin are concerned, in spite of significant recent progress (see \citealp{2012SSRv..169..181T} for a recent review and the classic reviews of \citealp{1968SoPh....3..367B, 1972ARA&A..10...73B}). Recognized as the basic component of the upper chromosphere, these narrow, nearly vertical, spike-like structures protrude into the low corona, reaching heights in excess of 10\,000\,km, well above the base of the chromosphere-corona transition region (TR), which one-dimensional  models put around 2\,000\,km. They are traditionally observed in \ha, but they are seen in all chromospheric lines and in the continuum outside the optical spectral range. On the disk, they appear as dark structures (dark mottles) in \ha, where they are not randomly located but are grouped in rosettes or bushes \citep{1968SoPh....3..367B, 1972ARA&A..10...73B}, in close connection with the chromospheric network the elements of which appear as bright mottles at their roots.

Spicules are highly dynamic, as shown in imaging and spectral observations, ascending and then descending. Many spicules appear to diffuse into the corona, depositing energy and mass; estimates \citep{1968SoPh....3..367B, 1972ARA&A..10...73B, 2004A&A...424..279T} show that they can provide more than enough mass and less than enough heating to replenish coronal losses. Our understanding of spi\-cu\-le dynamics has greatly improved thanks to observations from the {\it Hinode} spacecraft. \cite{2007PASJ...59S.655D}, using Ca\,{\sc ii} H-line observations with the {\it Solar Optical Telescope} (SOT) onboard {\it Hinode}, recognized two kinds of spicules: Type-I, similar to the traditional spicules and the fast, short-lived Type-II spicules. This distinction has been contested by \cite{2012ApJ...750...16Z}, but confirmed by \cite{2012ApJ...759...18P}. In addition to radial motions, spicules exhibit transverse oscillations (see \citealp{2009SSRv..149..355Z} for a review), potentially useful as magnetic field diagnostics. 

The mechanism of formation of spicules is not clear and several models have been proposed (see the review of \citealp{2000SoPh..196...79S}); moreover, different mechanisms may apply in type I and type II spicules. Magnetic reconnection has been suggested, as well as shocks produced by pressure pulses from reconnection events or p-mode oscillations.

The determination of physical conditions (temperature and density) in spicu\-les and their variation with height is complicated by  the fact that the radiation in all spectral lines is formed under non-LTE (NLTE) conditions. Due to this difficulty, not much work has been done on this subject in recent years (see \citealp{2012SSRv..169..181T} for references). An important issue is whether spicules are hotter or cooler than their surroundings, which will have implications on the center-to-limb variation of the intensity in the microwave range (see the review by \citealp{2011SoPh..273..309S}).

{\it Interface Region Imaging Spectrograph} (IRIS) observations of chromospheric and transition region (TR) lines, with their unique spatial, spectral and temporal capabilities, allow for a complete diagnostic of the highly dynamical spicules from their roots to their expansion and disappearance into the corona. Moreover, IRIS provides information on the Mg\,{\sc ii} h and k lines, valuable for the computation of physical parameters and important in terms of radiative losses.
 
So far, there have been few works on spicules near the limb using IRIS observations (\citealp{2014ApJ...792L..15P, 2015ApJ...806..170S} ), mostly on the dynamics of spicules, with only one example of average \Mg\ spectrum given in \cite{2014ApJ...792L..15P}. In this article we analyze in depth IRIS spectral observations centered at the South Pole. In Section 2 we describe the observations, the required corrections, and the absolute calibration. In Section 3 we present our measurements of bulk parameters for time-averaged spectra, as a function of distance from the limb. In Section 4 we discuss the spatial structure and temporal variations of the intensity near the limb, while in Section 5 we compare our measured parameters of the Mg\,{\sc ii} h and k lines with NLTE model computations. Finally, in the last section, we discuss the results and present our conclusions. Our results on the dynamics will be presented in a subsequent publication.

\section{Observations and Data Reduction} \label{obs}
IRIS is a small explorer spacecraft providing simultaneous spectra and images \citep{2014SoPh..289.2733D}, specifically designed for the study of the upper chromosphere and the low transition region. It consists of a 19\,cm telescope that feeds a dual-bandpass imaging spectrograph, which obtains spectra in two wavelength ranges in the far ultraviolet (FUV), from 1332 to 1358\,\AA\ (FUV1) and from 1389 to 1407\,\AA\ (FUV2), and one in the near ultraviolet (NUV), from 2783 to 2834\,\AA. Slit-jaw images in four different passbands can also be obtained simultaneously with the spectra.

\begin{table}[h]
\begin{center}
\caption{IRIS spectral windows}
\label{Table01}
\begin{tabular}{rrlcl}
\hline 
Window & Band &Designation& Wavelength range & Principal lines \\
            &         &                &  (\AA)                 &                     \\ 
\hline 
      1 & FUV1    &C II 1336  &  1332.1 - 1340.4 & C\,{\sc ii} doublet \\
      2 & FUV1    &O I 1356   &  1347.3 - 1357.2 &  Cl\,{\sc i}, O\,{\sc i} \\
      3 & FUV2    &Si IV 1394 &  1391.6 - 1396.0 & Si\,{\sc iv} 1393.8\,\AA\ \\
      4 & FUV2    &Si IV 1403 &  1398.6 - 1406.6 & Si\,{\sc iv} 1402.8\,\AA, O\,{\sc iv} 1401.2\\
      5 & NUV     &Mg II k 2796&2790.8 - 2809.5 & Mg\,{\sc ii} h and k, Mg\,{\sc ii} triplet \\
      6 & NUV     &2814        &   2812.9 - 2816.0 & Far red wing of Mg\,{\sc ii} h \\
      7 & NUV     &2832        &   2831.5 - 2833.8 & Far red wing of Mg\,{\sc ii} h \\
\hline 
\end{tabular}
\end{center}
\end{table}

We used a set of observations near the South Pole obtained on 24 February 2014, from 11:05 to 12:45 UT (OBSID 3800259459), in a region where no coronal hole was present. The spectrograph slit was oriented in the North-South (NS) direction, about 8\arcsec\ west of the South Pole. Every 9.5\,s the position of the slit alternated between two locations, 2\arcsec\ apart; we will refer to the position nearest to the pole as position 1 and to the other slit position as position 2. With a cadence of 19 s, 318 spectra were taken at each slit position during the 100 min long observing sequence. The slit extended from $\sim50$\arcsec\ inside the disk ($\mu=0.32$) to $\sim65$\arcsec\ beyond the limb, providing a good coverage of structures both near and above the limb.

\begin{figure}[h]
\begin{center}
\includegraphics[width=\textwidth]{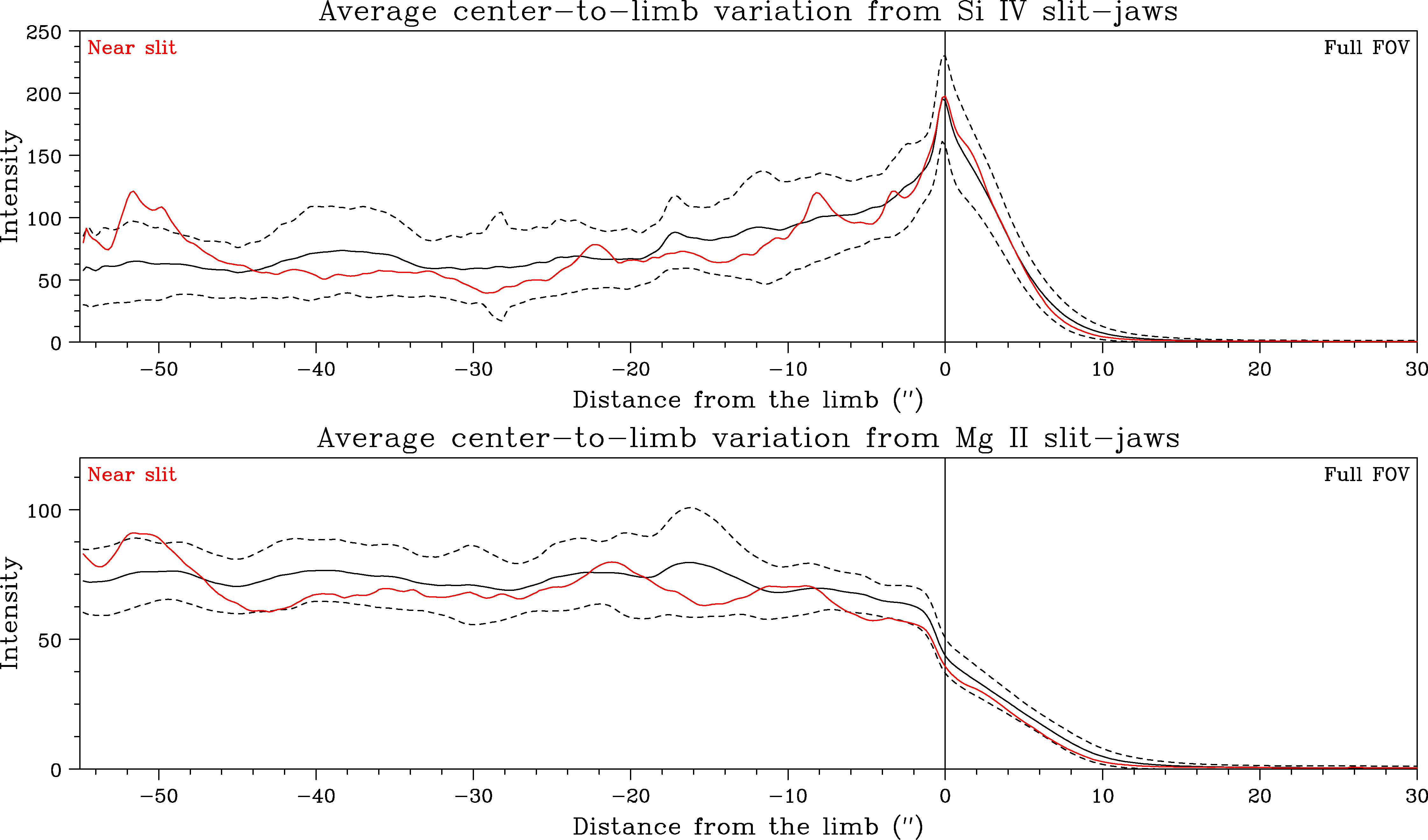}
\end{center}
\caption{Center to limb variation of the intensity from slit-jaw images, averaged over position and time, for the entire field of view (black curves) and near the slits (red curves). The dashed lines are at $\pm 1\sigma$ with respect to the average curve over the entire field of view. Top is for \Si, bottom for \Mg.}
\label{CLV_SJ}
\end{figure}

We note that, in spite of the averaging in time, some residual structure remained due to persistent network features ({\it c.f.} Figure \ref{AvSpec}, last row). In order to quantify the differences between the mean quiet Sun and our data, we computed the average and the root mean square (rms) of the intensity from all slit-jaw images for the entire field of view and for a narrow region (4\arcsec\ wide) around the position of the slit. The results are shown in Figure \ref{CLV_SJ}, where dashed lines are at $\pm 1\sigma$ with respect to the average curve over the entire field of view. The curves near the slit (in red) are mostly within the dashed lines, in particular at the region of interest near the limb (for \Si\ they are practically identical). We conclude that our data set is fairly representative of the average quiet Sun, although not coincident. Moreover, this data set is suitable for the study of the temporal evolution of the spectra with relatively high cadence at each slit position.

Spectra were recorded in seven windows, four in the FUV range and three in the NUV range, covering the C\,{\sc ii} doublet at 1335\,\AA, the O\,{\sc i} and C\,{\sc i} lines near 1353\,\AA, the Si\,{\sc iv} doublet near 1400\,\AA\ and the Mg\,{\sc ii} h and k lines near 2800\,\AA, as well as two narrow spectral regions in the red wing of  Mg\,{\sc ii} h line (Table \ref{Table01}). In our data set the sampling step in the direction of dispersion was 0.025\,\AA, which corresponds to 5.4\,km\,s$^{-1}$ in the FUV band and 2.7\,km\,s$^{-1}$ in the NUV band of IRIS; we note that the resolution of the spectrograph is 0.026\,\AA\ and  0.053\,\AA\ respectively \citep{2014SoPh..289.2733D}. Along the slit the sampling step was 0.17\arcsec, which is about two times smaller than the effective spatial resolution of IRIS ($\sim0.4$\arcsec). We thus have about two pixels {\it per} resolution element in the spatial direction, while in the spectral direction we have two samples {\it per} resolution element in the NUV and one on the FUV. The width of the slit was 0.33\arcsec\ and the exposure time 8 s.

Slit-jaw (SJ) images were obtained in the 1400\,\AA\ wavelength band, which includes the Si\,{\sc iv} lines, with a spectral width of 55\,\AA\ and in the 2796\,\AA\ band (Mg\,{\sc ii} k line) with a width of 4\,\AA. The effective field of view was 114\arcsec$\times$122\arcsec, the sampling step 0.17\arcsec, the cadence 19 s and the exposure time 8 s. The 1400\,\AA\ images were taken when the slit was at position 1 and the 2796\,\AA\ images when the slit was at position 2; this makes a total of 318 images at each wavelength.

We used level 2 data from the IRIS site, which are properly oriented and corrected for dark current and flat field. Still, further corrections were necessary for pointing and jitter, stray light and absolute calibration which will be described below. Hot pixels were rather scarce;  when it was deemed necessary, they were corrected by comparing each image with the previous and the next one.

\subsection{Pointing and Jitter Corrections}
We used the sharp drop of intensity at the limb to correct for jitter effects in the NUV spectra. We computed a time-averaged spectrum and measured the relative displacement of individual spectra with respect to the average spectrum by cross-correlation; subsequently a new average spectrum was computed from the corrected individual spectra and the procedure was repeated until \begin{figure}[h]
\begin{center}
\includegraphics[width=0.9\textwidth]{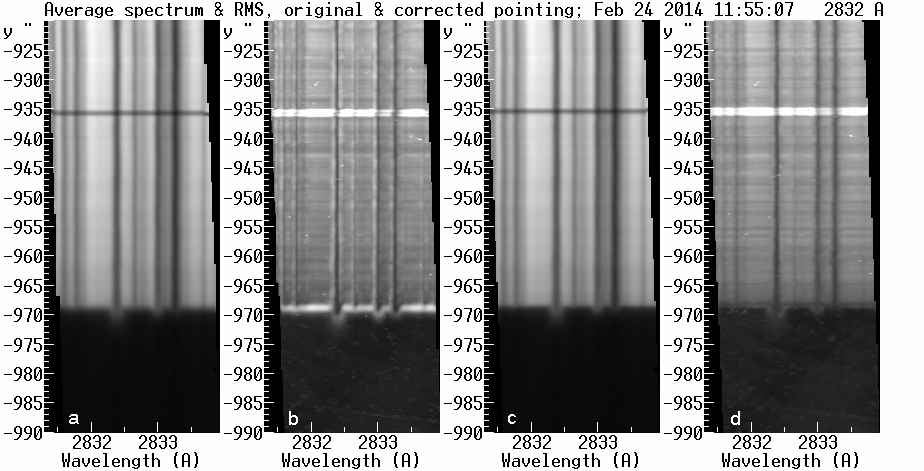}
\end{center}
\caption{Average (a, c) and rms (b, d) of the intensity in the 2832\,\AA\ spectral band, with the original pointing (a, b) and after jitter correction (c, d). The limb is at $y=-969.7$\arcsec\ and the dark horizontal lane near $y=-935$\arcsec\ is the fiducial mark.}
\label{jitter}
\end{figure}
convergence was achieved. The accuracy of this correction was about 0.1 pixel. The maximum displacement along the slit was 6 pixels (1\arcsec) and the rms value was 1.3 pixels (0.2\arcsec); with respect to the values recorded in the header of the IRIS FITS files, the rms of the jitter was slightly higher, 2.1 pixels (0.35\arcsec). In the direction of dispersion a small jitter, of the order of $\pm0.5$ pixels ($\pm0.012$\,\AA) in the NUV band and $\pm0.2$ pixels ($\pm0.005$\,\AA) in the FUV band, was also detected and corrected.
 
Figure \ref{jitter} shows the effect of jitter in the average spectra and the rms of the intensity variation. Although there is no appreciable difference in the average spectra (Figures \ref{jitter}a and \ref{jitter}c), the effect of jitter is well visible as a bright band near the limb in the uncorrected rms image (Figure \ref{jitter}b) that disappeared after the correction (Figure \ref{jitter}d). Note that the minute jitter along the direction of dispersion has a pronounced effect in the rms images. 

In order to establish the zero height we used the position of the limb, determined from the inflection point of the center-to-limb intensity variation. Since there is no true continuum in the IRIS NUV spectral range, we computed the position of the inflection point at the brightest point of the photospheric spectrum, at 2832.0\,\AA. This position was displaced by 1.7 pixels (0.3\arcsec) with respect to the value recorded in the IRIS FITS file. We note that according to Figure 1 of \cite{1981ApJS...45..635V} the emission at the center of the disk in the wings of the h and k lines is formed in the height range of 430\,--\,630\,km above the $\tau_{5000}=1$ level. 

The pointing and jitter corrections were passed to the FUV spectra by means of the position of the fiducial mark. The position of the fiducial mark was also used in applying the jitter and pointing corrections along the slit to the SJ images. The pointing perpendicular to the slit (in the East-West (EW) direction) was determined by cross-correlation of the average SJ images with the corresponding average {\it Atmospheric Imaging Assembly} (AIA) images in the 1600 and 1700\,\AA\ bands (see images in the last row of Figure \ref{AvSpec}). Subsequently, the correction for individual images was computed by cross-correlation with the average; in the EW direction the jitter was less than one pixel. 

\begin{figure}[h]
\begin{center}
\includegraphics[width=0.9\textwidth]{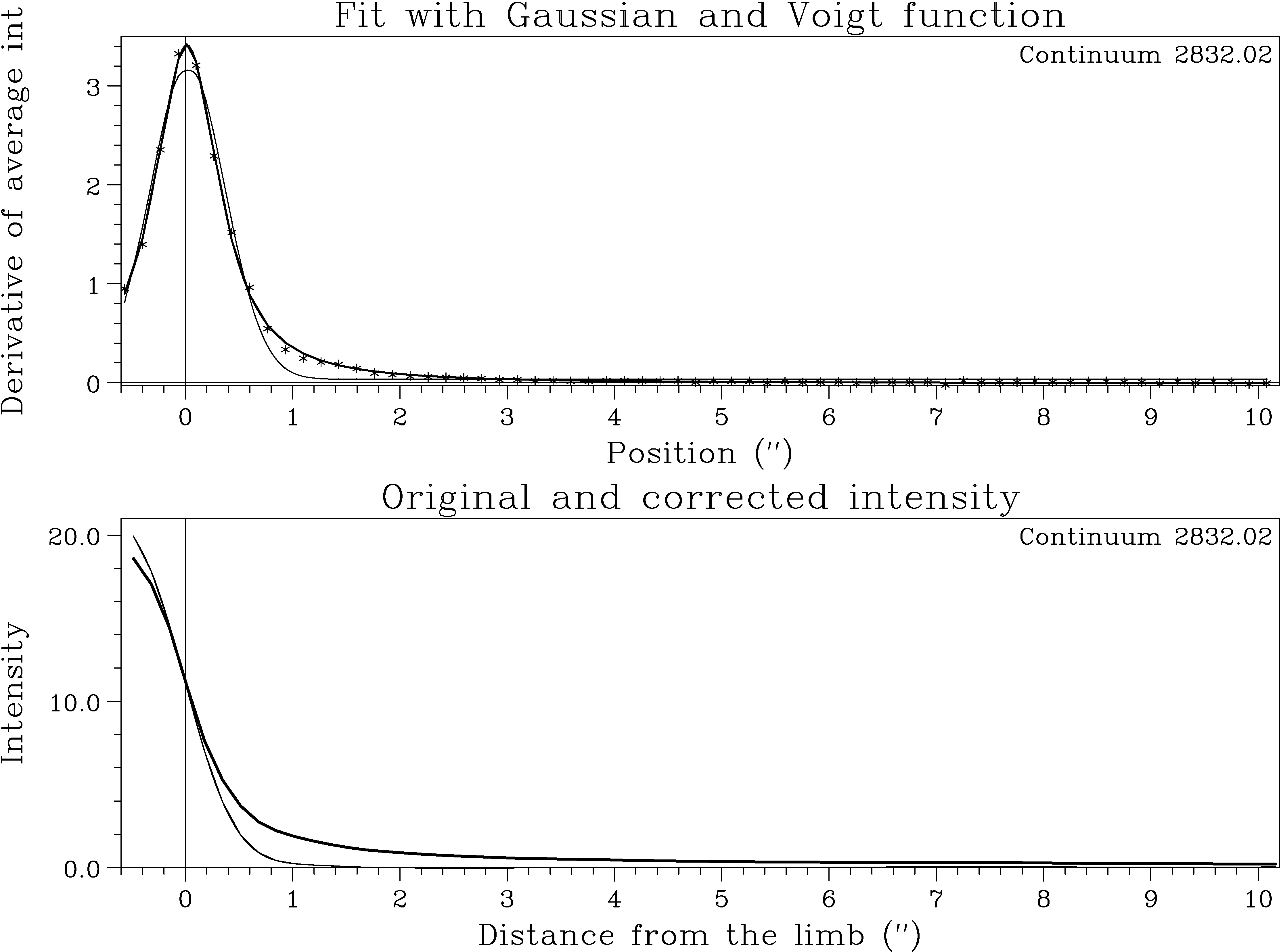}
\end{center}
\caption{Top: the derivative of the average radial variation of the intensity at 2832.0\,\AA. The thin line is the Gaussian fit, the thick one is the fit with a Voigt function. Bottom: Original (thick line) and corrected (thin line) limb profile.}
\label{strayr}
\end{figure}

\subsection{Stray-Light Correction}
Traces of absorption lines were detectable in the NUV spectra as far as $\sim20$\arcsec\ beyond the photospheric limb, a clear indication of stray light in the direction along the slit, which had to be corrected for precise photometric work. To this end we computed the derivative of the average radial variation of the intensity. Were the photospheric limb a sharp edge, this would have given us the point spread function (PSF) of the instrument; although the limb is not a straight edge, it is sharp enough to give an upper limit of the spatial resolution along the slit and a fair estimate of the stray light, which appeared as a long tail in the derivative.

We computed the derivative of the average center-to-limb variation at the brightest point of the photospheric spectrum (2832.0\,\AA), as well as at 2793.33 \AA\ (blue wing of k) and 2808.48 \AA\ (red wing of h). Apart from the position of the limb, we got similar results at all three wavelengths. The full width at half maximum (FWHM) was  0.73\arcsec\ which, as expected, is larger than the effective resolution of 0.4\arcsec\ quoted above. 

The wings of the PSF (Figure \ref{strayr}, top) were too extended to be fitted to a Gaussian function, we therefore chose to fit it with a Voigt function, $H(a,Ax)$, which can have extended wings; here $a$ is the damping parameter and $x$ the distance from the limb in arcseconds. The parameters derived from the fit were $a=2.58$ and $A=7.5$\arcsec\,$^{-1}$.

In order to correct for stray light only, while preserving the original spatial resolution, we deconvolved the spectra with the Voigt function and reconvolved them with the corresponding Gaussian. In order to avoid over-correction near the limb, we had to adjust the parameters to $a=2.1$ and $A=9.25$\arcsec\,$^{-1}$. Judging from the plot in the lower panel of Figure \ref{strayr}, where the intensity above $\sim1$\arcsec\ is practically zero and the fact that no absorption lines are seen beyond the limb after the correction, our method provides a satisfactory correction of the stray light.  A broader correction function would produce negative intensity values.

We note that there is probably stray light in the direction of dispersion as well, but we had no way to correct for that; also note that this method could not be applied to the FUV spectra where the limb is neither sharp nor well visible in the continuum. We did not attempt to apply the NUV correction to the FUV spectra, due to possible instrumental differences. 

\subsection{Absolute Intensity Calibration}
Following \cite{2016SoPh..291...67V}, we used the results of Kohl and Parkinson (1976, hereafter KP) on the Mg\,{\sc ii} lines, to perform the absolute calibration of the NUV spectra. These authors provided absolute spectral intensity for two positions on the disk, at $\mu=1.0$ and $\mu=0.23$. The latter is within our field of view, but we could also use their results at $\mu=1.0$ because a short run of six spectra for focusing purposes had been performed at disk center at 20:20 UT on the day of our observations  (OBSID 4202400003).

\begin{figure}[h]
\begin{center}
\includegraphics[width=\textwidth]{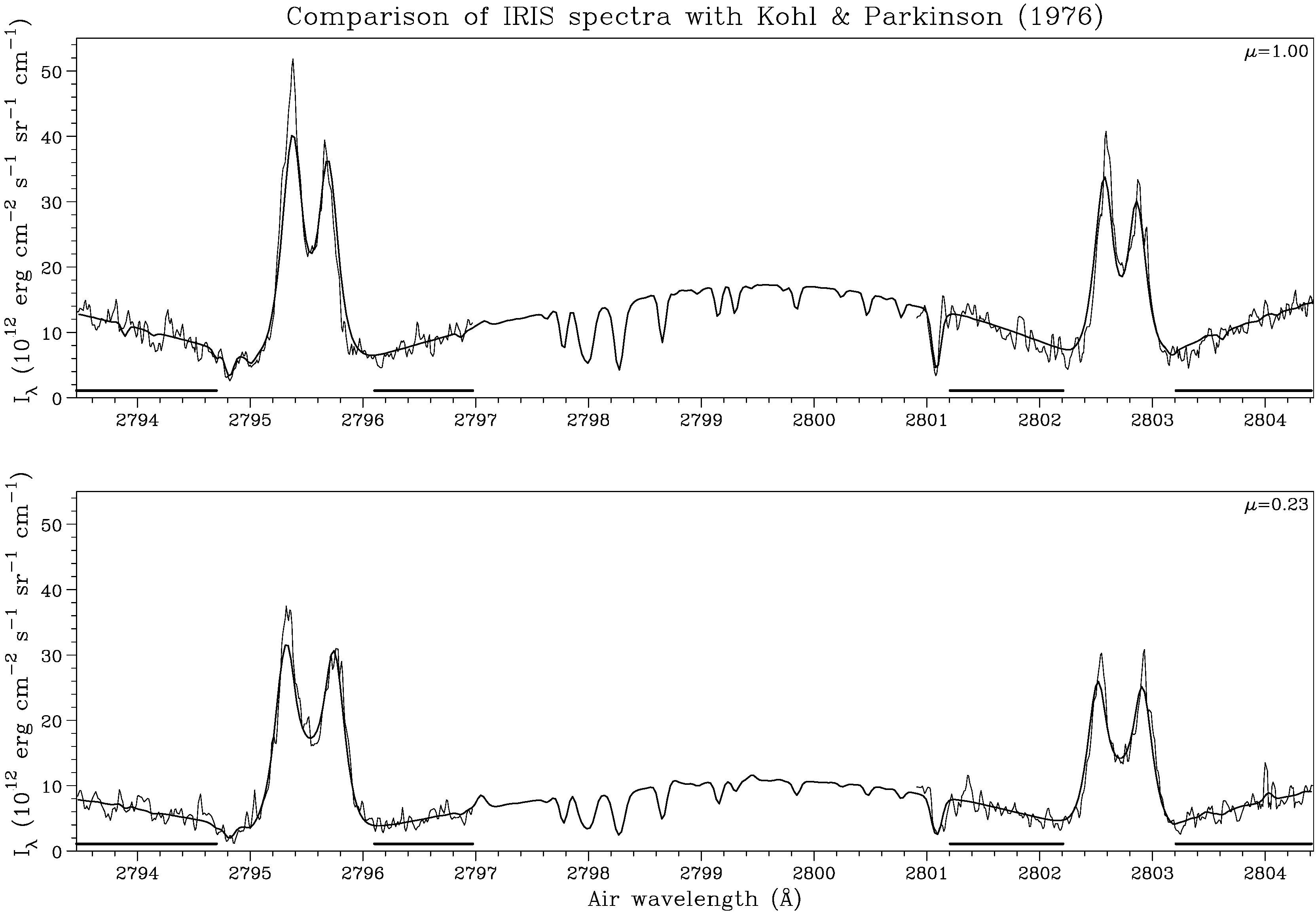}
\end{center}
\caption{Comparison of the spectra of Kohl and Parkinson (1976, thin lines) with calibrated IRIS spectra (thick lines) at $\mu=1.00$ and $\mu=0.23$. The horizontal bars show the wavelength ranges used for the comparison of the two data sets.}
\label{CalNUV}
\end{figure}

The calibration factor was computed by comparison of our values, after background correction, with those of Table 1 of KP that covered two 3.52\,\AA\ wide intervals, centered on the k and h lines, with a step of 0.011\,\AA. The comparison was made at four wavelength ranges in the wings of the k and h lines, avoiding the emission part of the h and k lines. A wavelength shift of 0.83\,\AA\ was detected, consistent with the fact that the KP wavelengths are air wavelengths.  For the limb, KP give a range of $\mu$ from 0.249 to 0.184, centered around 0.23. This corresponds to a distance range of 939.1 to 953.1\arcsec\ from the disk center (for $R\sun=969.7$\arcsec) and we averaged our IRIS spectra accordingly. 

The results of the comparison are shown in Figure \ref{CalNUV}. We obtained similar conversion factors, of 1.92 and 2.13 $10^{12}$\,erg\,cm$^{-2}$\,s$^{-1}$\,sr$^{-1}$\,cm$^{-1}$ {\it per} IRIS counts/sec for $\mu=1.00$ and $\mu=0.23$ respectively and we used their average for our absolute calibration; this value is $\sim30$\% higher than that of the IRIS in-flight calibration provided by the {\tt iris\_get\_response} routine \citep{IRISguide}, in agreement with  \cite{2016SoPh..291...67V}. We add that the conversion factor for the k line was slightly higher (2\%) than that for the h line.

\begin{table}[h]
\caption{Absolute calibration factors (in~$10^{12}$\,erg\,cm$^{-2}$\,s$^{-1}$\,sr$^{-1}$\,cm$^{-1}$ {\it per} IRIS counts/sec)}
\label{Table02}
\begin{tabular}{lccccc}
\hline 
                   & NUV  &  C\,{\sc ii} 1335 & Cl\,{\sc i} 1352 & O\,{\sc i} 1356 & Si\,{\sc iv} 1400 \\
\hline 
SUMER         & --      &    0.64       &     0.45    & 0.38       &     0.34      \\  
HRTS QRL    & --      &    0.46       &     0.10    & 0.088      &     0.31      \\  
HRTS QR      & --      &    0.38       &     0.18    & 0.18       &     0.22      \\  
HRTS QRA    & --      &    1.11       &     0.39    & 0.34      &     0.41      \\  
IRIS in-flight & 1.53  &    0.75       &     0.52    & 0.47        &     0.27      \\  
Adopted values& 2.0 &    0.75       &     0.52    & 0.47        &     0.27      \\  
\hline 
\end{tabular}
\end{table}

For the intensity calibration of the FUV spectra we considered the {\it Solar Ultraviolet Measurements of Emitted Radiation} (SUMER) spectral atlas of \cite{2001A&A...375..591C}, the {\it High Resolution Telescope and Spectrograph} (HRTS) atlas of \cite{1993ApJS...87..443B} and the IRIS in-flight calibration. The SUMER atlas gives quiet-Sun spectra at the center of the disk, while the HRTS atlas provides spectra both near the center ($\mu=0.92$-0.98, labeled QRA and $\mu=0.85$-0.86, labeled QR) and near the limb ($\mu=0.17$-0.20, labeled QRL). We compared the SUMER and HRTS disk spectra with the averaged IRIS spectrum obtained at the disc center at 20:20 UT and the HRTS limb spectrum with the average spectrum of our run. In the IRIS FUV spectra the continuum was too weak to be reliably measured, therefore we used the integrated intensities of the strong lines of each spectral window, after subtraction of the continuum; in this way we derived four values of the calibration factor for each spectral window of IRIS. 

The calibration factors from SUMER, QR, QRA and QRL showed significant differences among them and with the IRIS in-flight calibration, with a dispersion of $\sim20$\% in FUV2 and $\sim45$\% in FUV1 ({\it cf.} the values in Table \ref{Table02}). In addition to the inherent difficulties of the calibration at these wavelengths, we note that there is an additional uncertainty due to the strong variation of the line intensity along the slit, which had an rms value of $\sim40$\%; moreover, there might be a variation with the solar cycle. In view of the above we decided to use the IRIS in-flight calibration values for the FUV; we note that these are not much different than those we computed from the SUMER disk center atlas. Given the uncertainties discussed above we do not consider the accuracy of this calibration to be better than a factor of two. 

\section{Results}
\subsection{Averaged Spectra and Bulk Parameters}\label{AvSpecProf}
\begin{figure}
\begin{center}
\parbox[b]{0.32\textwidth}{
\includegraphics[width=0.32\textwidth]{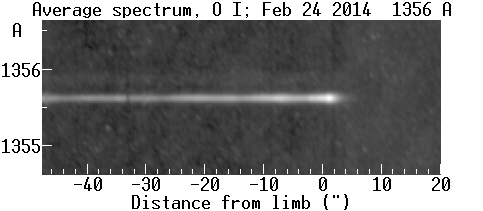}
\includegraphics[width=0.32\textwidth]{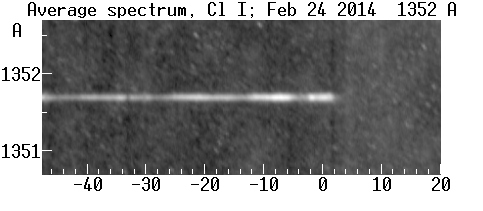}
\includegraphics[width=0.32\textwidth]{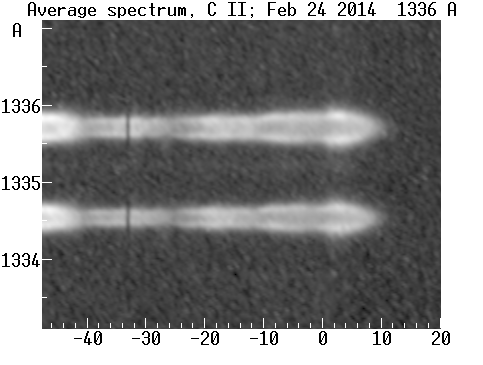}
\includegraphics[width=0.32\textwidth]{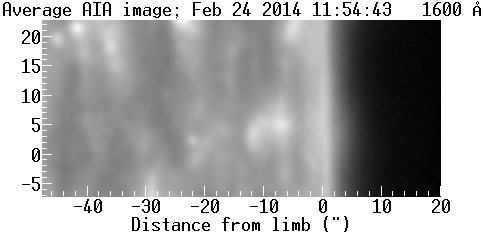}
}
\parbox[b]{0.32\textwidth}{
\includegraphics[width=0.32\textwidth]{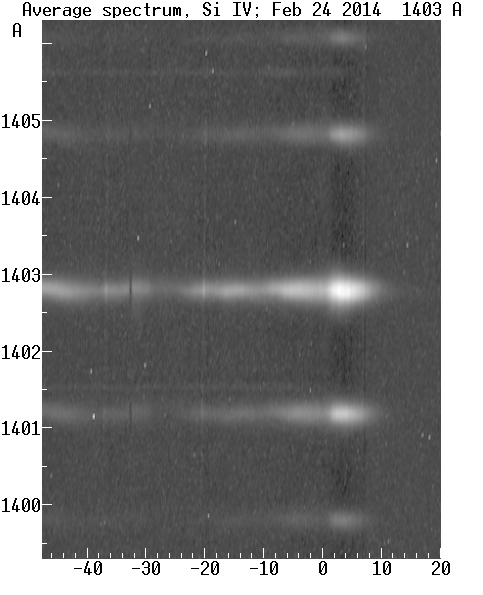}
\includegraphics[width=0.32\textwidth]{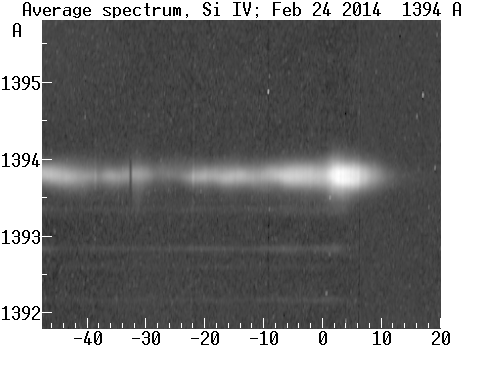}
\includegraphics[width=0.32\textwidth]{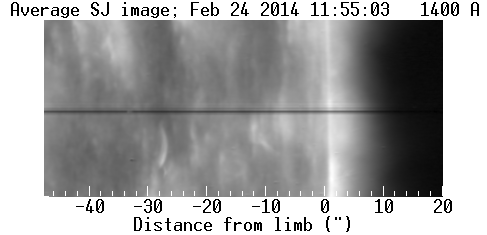}
}
\parbox[b]{0.32\textwidth}{
\includegraphics[width=0.32\textwidth]{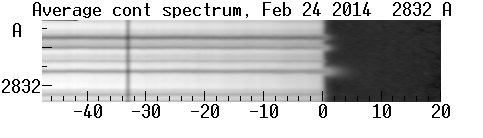}
\includegraphics[width=0.32\textwidth]{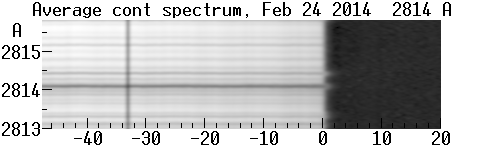}
\includegraphics[width=0.32\textwidth]{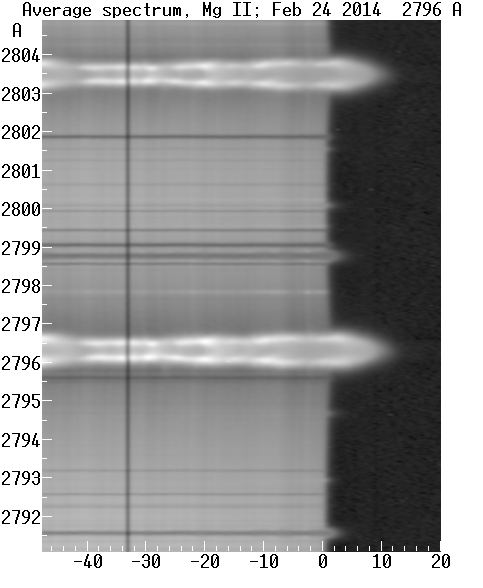}
\includegraphics[width=0.32\textwidth]{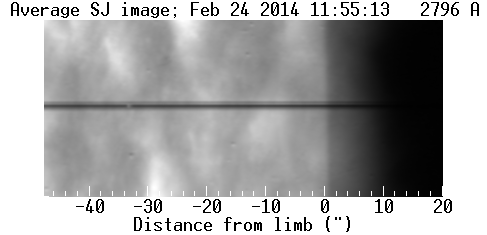}
}
\end{center}
\caption{Average IRIS spectra in all wavelength windows, shown with the dispersion in the vertical direction. The wavelengths are {\it in vacuo}. The fiducial mark is at $-33$\arcsec. The bottom row shows average images from the 1600 \AA\ AIA band as well as average IRIS slit-jaw images for comparison. Each frame is displayed with an appropriate value of the gamma correction, 
so that both bright and dark structures are visible.}
\label{AvSpec}
\end{figure}

For the computation of the average spectra we took particular care to avoid transient brightenings. For this purpose we made statistics of the intensity at each pixel and we dropped 5\% of the values from the low side of the distribution and another 5\% from the high side.  In addition to eliminating transients, this procedure removes most of the remaining hot pixels. The results are shown in Figure \ref{AvSpec}, where we also give portions of the average SJ images as well as the AIA 1600\,\AA\ band image; wavelengths are {\it in vacuo}. These images show that the IRIS slit went through mostly low intensity regions, crossing occasionally bright areas associated with the network.

We note that in addition to the main lines, several weak emission lines appear in the averaged spectra of the FUV band. The strongest of those are the O\,{\sc iv} lines at 1399.77 and 1401.16\,\AA\ and the Si\,{\sc iv} line at 1404.79\,\AA; for the identification of others we refer the reader to the SUMER atlas of \cite{2001A&A...375..591C}. 

\begin{figure}
\begin{center}
\includegraphics[width=\textwidth]{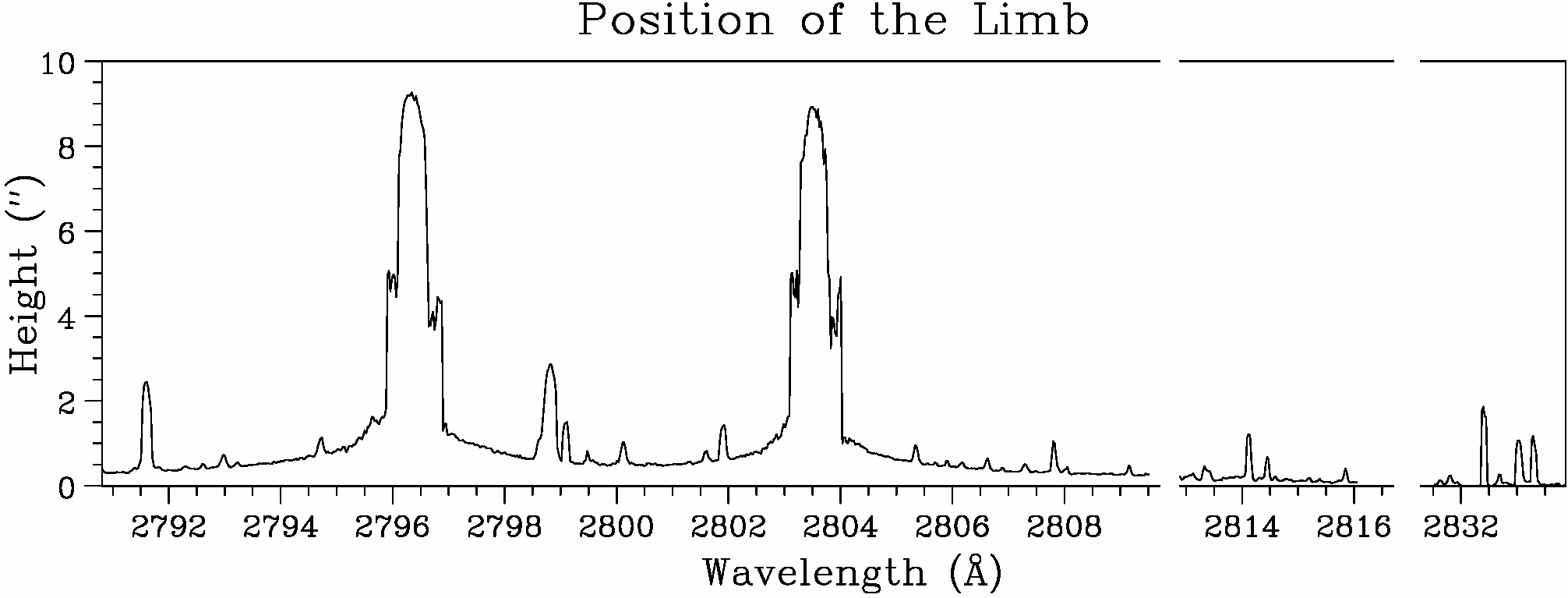}
\end{center}
\caption{The height of the limb as a function of wavelength for the three spectral windows in the NUV, computed from the maximum of the derivative of the intensity as a function of radial distance. The height is measured with respect to the limb position at 2832\,\AA.}
\label{LmbHt}
\end{figure}

In the NUV band there is emission beyond the limb in several weak lines, notably the Mg\,{\sc ii} triplet at 2791.60 and 2798.8\,\AA\ (the latter is a blend of the 2798.75 and 2798.82\,\AA\ lines, see \citealp{2015ApJ...806...14P}). There are other such lines in our spectra, {\it e.g.} the  Fe\,{\sc ii} line at 2832.40\,\AA, the Ti\,{\sc ii} line at 2832.99\,\AA\ and the  Fe\,{\sc i} line at 2833.27\,\AA\ in our  2832\,\AA\ spectral window and the 2814.42\,\AA\  Fe\,{\sc ii} line in our window around 2814\,\AA; identifications are from the National Institute of Standards and Technology (NIST) data base \citep{2015NIST}. The emission of these weak lines does not extend by more that 1-2\arcsec\ above the nearby continuum limb, as seen in  Figure \ref{LmbHt} which shows the height of the limb, measured from the position of the inflection point of the intensity variation, as a function of wavelength; the detection of emission beyond the limb from these weak lines became possible thanks to the high spatial resolution of IRIS.

For the emission lines we subtracted the background and for each position we computed numerically the integrated intensity, the peak intensity and its wavelength and the FWHM of the lines. For the single-peaked lines we also did a least square fit of the profiles to a Gaussian or Voigt function; this gave similar results to those of the numerical computation. For the h and k lines inside the disk, the background and integrated intensity are not meaningful and were not computed.

\subsection{Shape of the Line Profiles}\label{profiles}
\begin{figure}[h]
\begin{center}
\includegraphics[width=\textwidth]{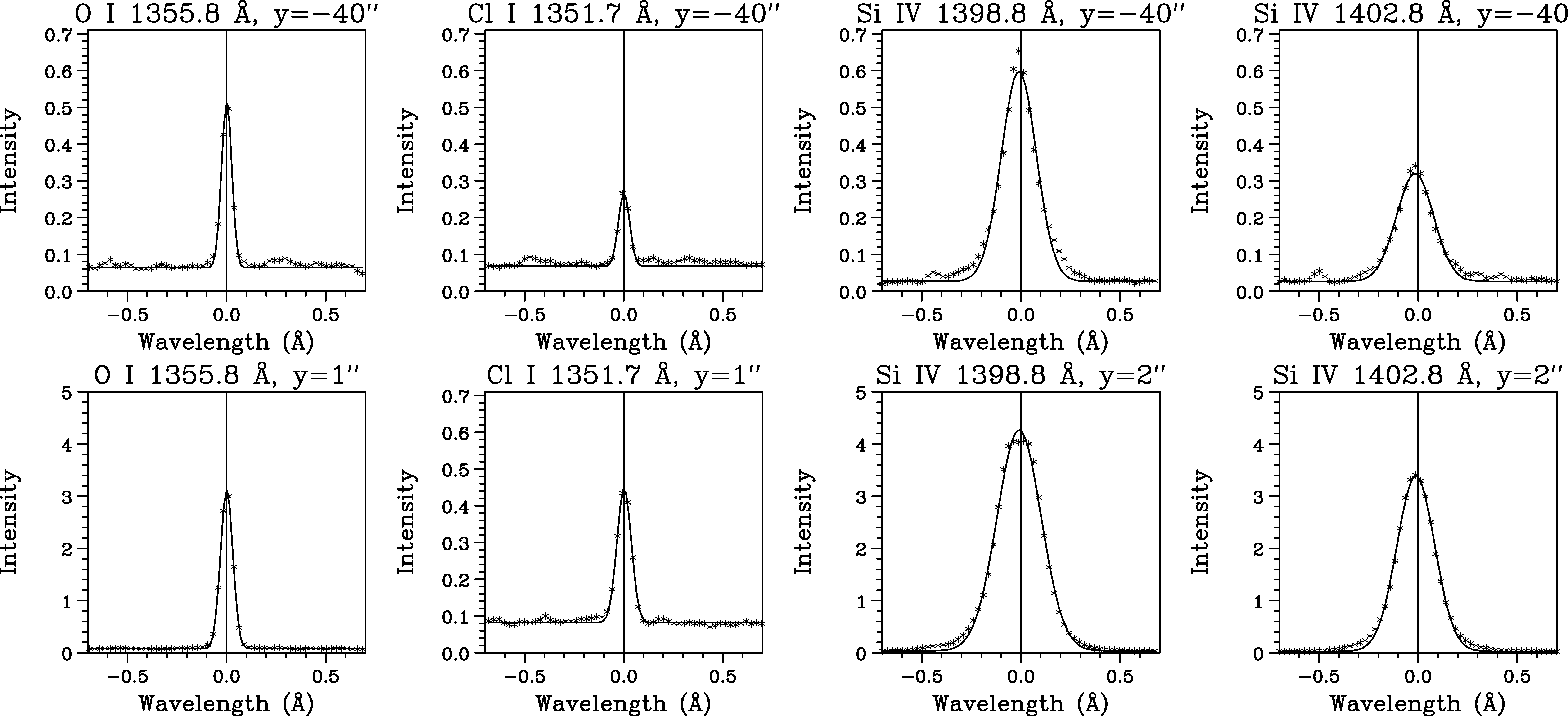}
\end{center}
\caption{Observed profiles (stars) and Gaussian fits (full lines) for the \OI, \Cl\ and the \Si\ lines on the disk (top row) and beyond the limb (bottom row), $y$ is the distance from the limb. The intensity is in units of $10^{12}$\,erg\,cm$^{-2}$\,s$^{-1}$\,sr$^{-1}$\,cm$^{-1}$.}
\label{profilesfig}
\end{figure}

As shown in Figure \ref{AvSpec}, \Mg\ and \Cb\ profiles are double-peaked on most of the disk, while the other emission lines have single-peaked profiles. Least square fits of single-peaked profiles with Gaussians  gave excellent results for the  \Cl\ and \OI\ lines, however, the Si\,{\sc iv} 1393.8 and 1402.8\,\AA\ line profiles on the disk had systematically a sharper peak and more extended wings than their Gaussian fit (Figure \ref{profilesfig}, top row). Fits with a Voigt function did not make any difference, due to the small value of the damping parameter. The good Gaussian fits of the \Cl\ and \OI\  profiles makes rather unlikely that the deviations are due to the spectrograph instrumental profile. Thus the observed shapes of the \Si\ profiles are probably due to deviations of the turbulent velocity distribution, which dominates the line profile, from the Gaussian shape (see also Section \ref{thin} below).

Just above the limb the profiles of the stronger \Si\ line (1393.8\,\AA) are flatter than their Gaussian fit, with a weak minimum at their core (Figure \ref{profilesfig}, bottom row). This is clearly an effect of opacity (see Section \ref{BulkDoublets} below), since the line profile is proportional to the profile of the absorption coefficient only in the optically thin case; moreover, the opacity beyond the limb is expected to increase, as material occulted by the photosphere becomes visible.
\begin{figure}[h]
\begin{center}
\includegraphics[height=12.3cm]{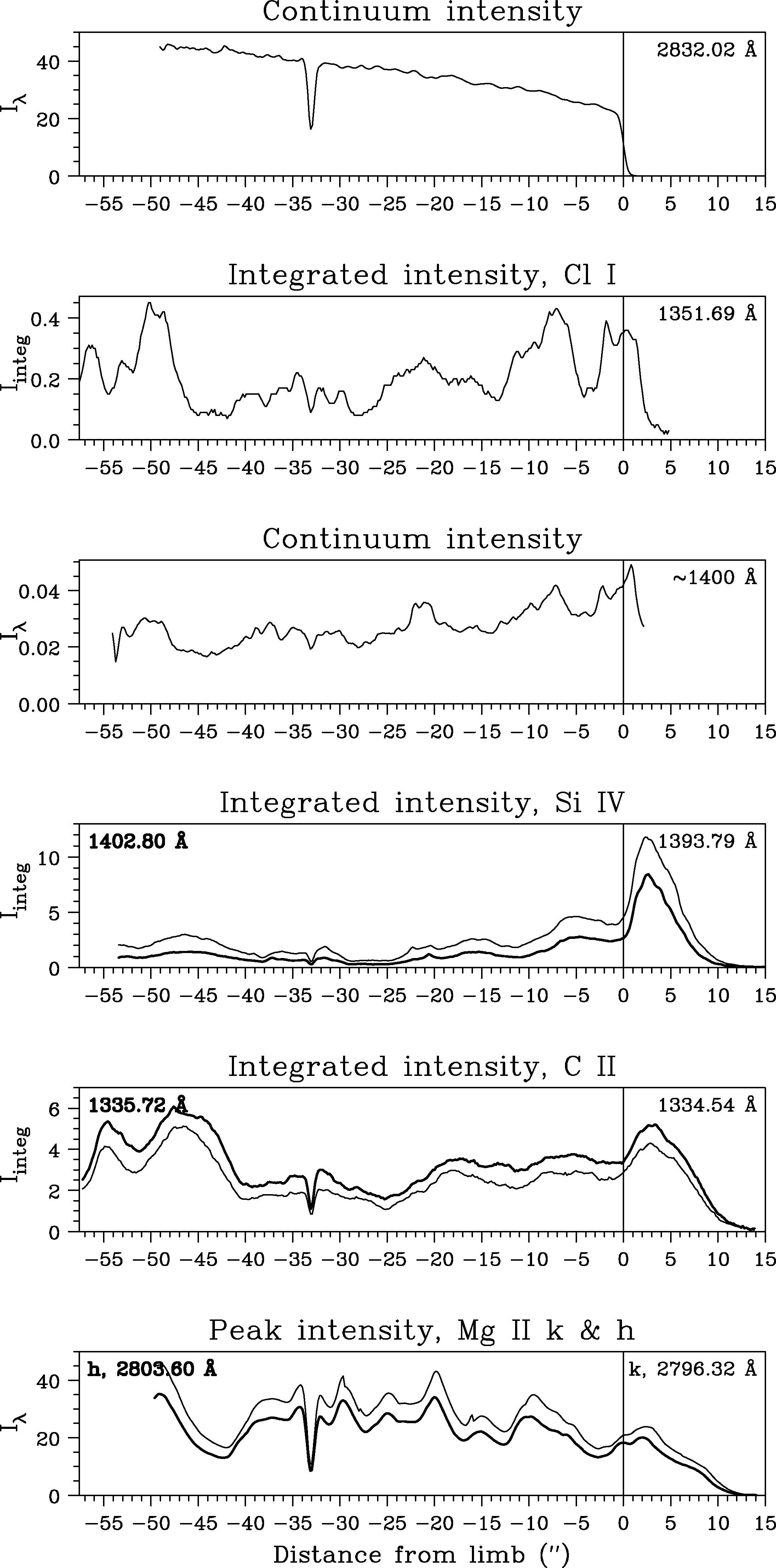}\hspace{.1cm}\includegraphics[height=12.3cm]{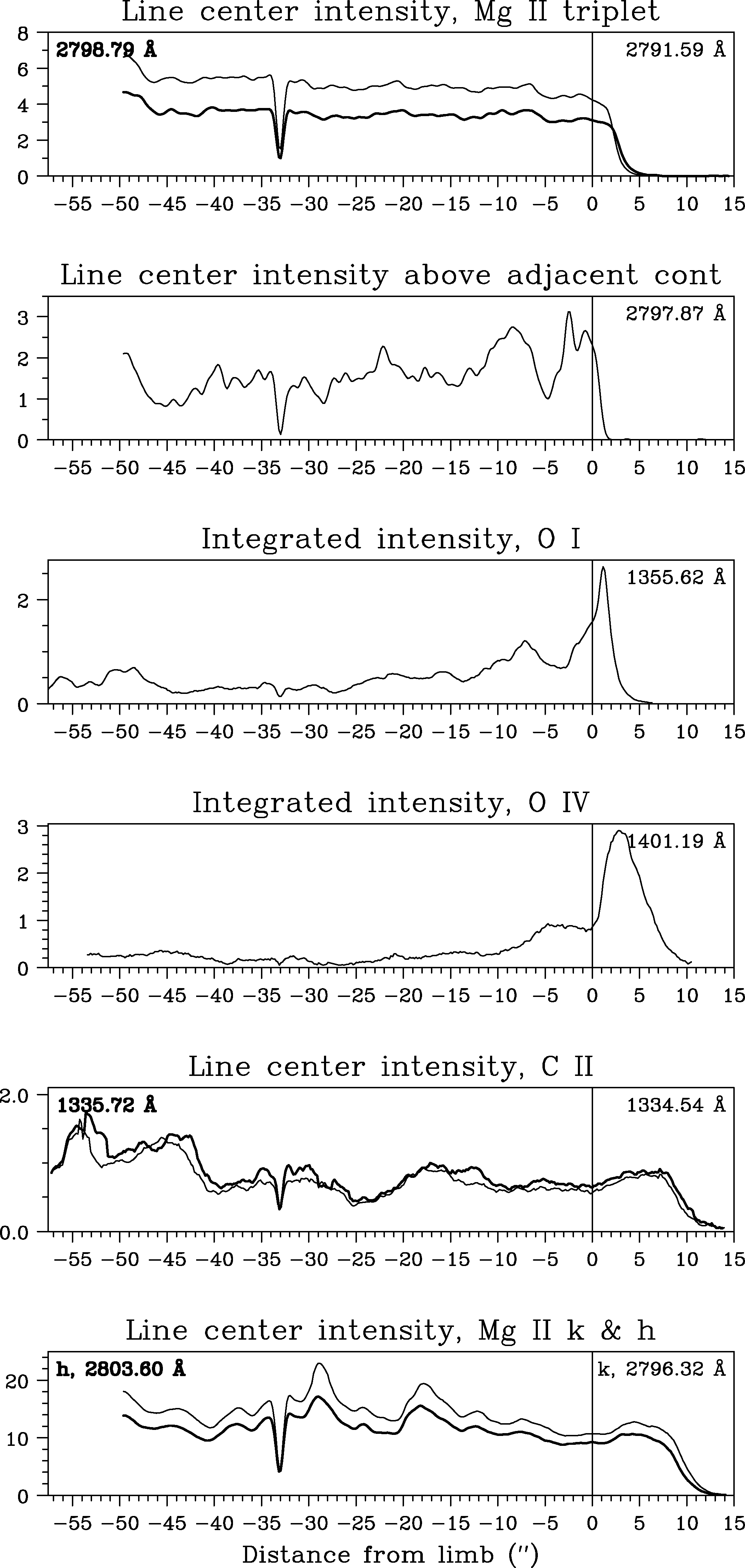}
\end{center}
\caption{The intensity as a function of radial distance from the limb. The specific intensity, $I_{\lambda}$, is in units of  $10^{12}$\,erg\,cm$^{-2}$\,s$^{-1}$\,sr$^{-1}$\,cm$^{-1}$, the integrated intensity, $I_{\rm integ}$, in $10^{3}$\,erg\,cm$^{-2}$\,s$^{-1}$\,sr$^{-1}$. Doublets are plotted in the same frame, with the thick curve corresponding to the line with longer wavelength. The dip at $-33$\arcsec\ is due to the fiducial mark.}
\label{IntLimb}
\end{figure}

\subsection{Radial Variation of the Intensity}\label{RadVarInt}
Figure \ref{IntLimb} gives the intensity as a function of radial distance for the strongest lines as well as for the Mg\,{\sc ii} triplet lines, which are in absorption on the disk but extend well above the photospheric limb as emission lines. We have included the line at 2797.87\,\AA, which is in emission on the disk and for which we could not find a reliable identification, the continuum near 1400\,\AA, and the O\,{\sc iv} line at 1401.19\,\AA\ ({\it cf.} Figure \ref{AvSpec}); the continuum intensity at 2832.02\,\AA\ is also given for reference. For the k and h lines we give the maximum intensity instead of the integrated intensity and for double-peaked lines we also give the intensity of their cores. 

On top of the fluctuations due to network structures, the O\,{\sc i}, Si\,{\sc iv}, and O\,{\sc iv} lines show a gradual enhancement towards the limb, close to the expected $1/\mu$ and a sharp rise after the photospheric limb, as expected for optically thin lines formed above the temperature minimum. The rise above the limb is less pronounced for the C{\sc\,ii} lines, while it is quite small in the core of the C{\sc\,ii} as well as the Mg{\sc\,ii} h and k lines which are optically thick. There is no clear jump for the Cl{\sc\,i} line; the structure is similar in the 2797.87\,\AA\ disk emission line and in the 1400\,\AA\ continuum, although these show increased intensity towards the limb. The Mg{\sc\,ii} triplet lines show little intensity variation on the disk and no peak near the limb.

The limb jump in optically thin FUV lines is due to the fact that material occulted by the photosphere becomes visible, as noted in the previous section. The peak intensity is at some distance above the photospheric limb, which apparently corresponds to the height with the maximum contribution to the line intensity, termed here as ``formation height''; then the position of the inflection points before and after the peak (termed here ``inner'' and ``outer'' limb) give the range of heights where the line is formed. In addition, the scale height of the emission at the outer limb, 
\be
H=I_\lambda\frac{dI_\lambda}{dr}^{-1},
\ee
gives information about the height structure of the atmosphere. 

These parameters can be computed numerically from the intensity profiles, however, for the outer limb and the scale height we preferred to fit the data to a hyperbolic tangent function in order to reduce noise in weak lines. For the optically thin lines we used the integrated intensity, for the continua the specific intensity, and for the optically thick lines, as well as the NUV lines, the specific intensity at the line core. We estimate the accuracy of this procedure to be better than one pixel. The finite spatial resolution of the instrument will not affect much the computed position of the limb, but will increase the computed scale height of the emission.

\begin{table}
\caption{Formation heights of selected lines and continua}
\label{Table03}
\begin{tabular}{lcccc}
\hline 
Line &  Inner limb& Peak position& Outer limb & Scale\\
       &        (\arcsec)            &   (\arcsec)      &        (\arcsec)        &(\arcsec)\\
\hline
2832.02 Cont               &  --  &  --  & 0.00 & 0.49 \\ 
2797.87  Line               &  --  &  --  & 0.73 & 0.66 \\ 
2800.09  Fe{\sc\,ii}               &  --  &  --  & 0.98 & 0.81 \\ 
2797.00 Cont               &  --  &  --  & 1.14 & 0.79 \\ 
1355.62 O{\sc\,i}          & 0.95 & 1.16 & 1.26 & 1.31 \\ 
1351.69 Cl{\sc\,i}         &  --  &  -- & 1.67 & 0.80 \\ 
FUV2 Fe{\sc\,ii} lines     &  $-0.3$~~~\,  & 0.95 & 1.8~\, & -- \\ 
2791.59 Mg{\sc\,ii} triplet&  --  &  --  & 2.35 & 0.81 \\ 
2798.79 Mg{\sc\,ii} triplet&  --  &  --  & 2.91 & 0.90 \\ 
1399.81 Si{\sc\,iv}        & 1.17 & 2.92 & 4.93 & 3.10 \\ 
1401.18 O{\sc\,iv}          & 1.24 & 2.83  & 4.99 & 3.15 \\ 
1402.80 Si{\sc\,iv}        & 1.16  & 2.59  & 5.26 & 2.96 \\ 
1393.79 Si{\sc\,iv}        & 1.51 & 2.37  & 5.81 & 2.65 \\ 
1334.54 C{\sc\,ii} core    &  --  &  --  & 8.85 & 1.79 \\ 
2803.58 Mg{\sc\,ii} h core &  --  &  --  & 9.02 & 1.82 \\ 
1335.72 C{\sc\,ii} core    &  --  &  --  & 9.34 & 1.72 \\ 
2796.32 Mg{\sc\,ii} k core &  --  &  --  & 9.53 & 1.88 \\ 
\hline 
\end{tabular}
\end{table}

Table \ref{Table03}, as well as Figure \ref{IntLimb}, show  that the outer limb for \Si\ is below that of \Mg\ and \Cb, although \Si\ is at a higher ionization state; this is, most probably, an optical depth effect, since both \Mg\ and \Cb\ are optically thick at low heights above the limb, hence their emission extends higher. Moreover, some authors (\citealp{2014ApJ...780L..12D}) claim that, because of departures from the Maxwellian distribution, the \Si\ lines could be formed at 10000\,K or below; others (\citealp{2014A&A...569L...7S}) found molecular absorption within \Si\ profiles, a possible indication of low temperatures.

The results for the lines shown in Figure \ref{IntLimb} and some weaker lines and continua are tabulated in Table \ref{Table03}, where the entries are ordered in increasing height of the outer limb. As mentioned above, we could find no identification for the 2797.87\,\AA\ line, which is in absorption on the disk and in emission beyond the limb. We have included three weak Fe{\sc\,ii} lines (1392.17, 1392.83, and 1405.63\,\AA), which we averaged due to their very low intensity and for which the scale height could not be reliably determined. The positions of the outer limb show that the Mg{\sc\,ii} k and h lines extend beyond 9.5\arcsec, slightly higher than the C{\sc\,ii} doublet. The Si{\sc\,iv} and the O{\sc\,iv} lines are formed at lower heights (outer limb from 4.93 to 5.81\arcsec, peak from 2.37 to 2.83\arcsec) and the Mg{\sc\,ii} triplet lines even lower (outer limb at around 2.6\arcsec). Weaker lines, including O{\sc\,i} and Cl{\sc\,i} still extend about 2\arcsec\ above the 2832\,\AA\ continuum ({\it c.f.} Figure \ref{LmbHt}). It is noteworthy that the behavior of these two lines that form at about the same height is quite different, with O{\sc\,i} showing strong limb emission and Cl{\sc\,i} not.

\begin{figure}[h]
\begin{center}
\includegraphics[width=\textwidth]{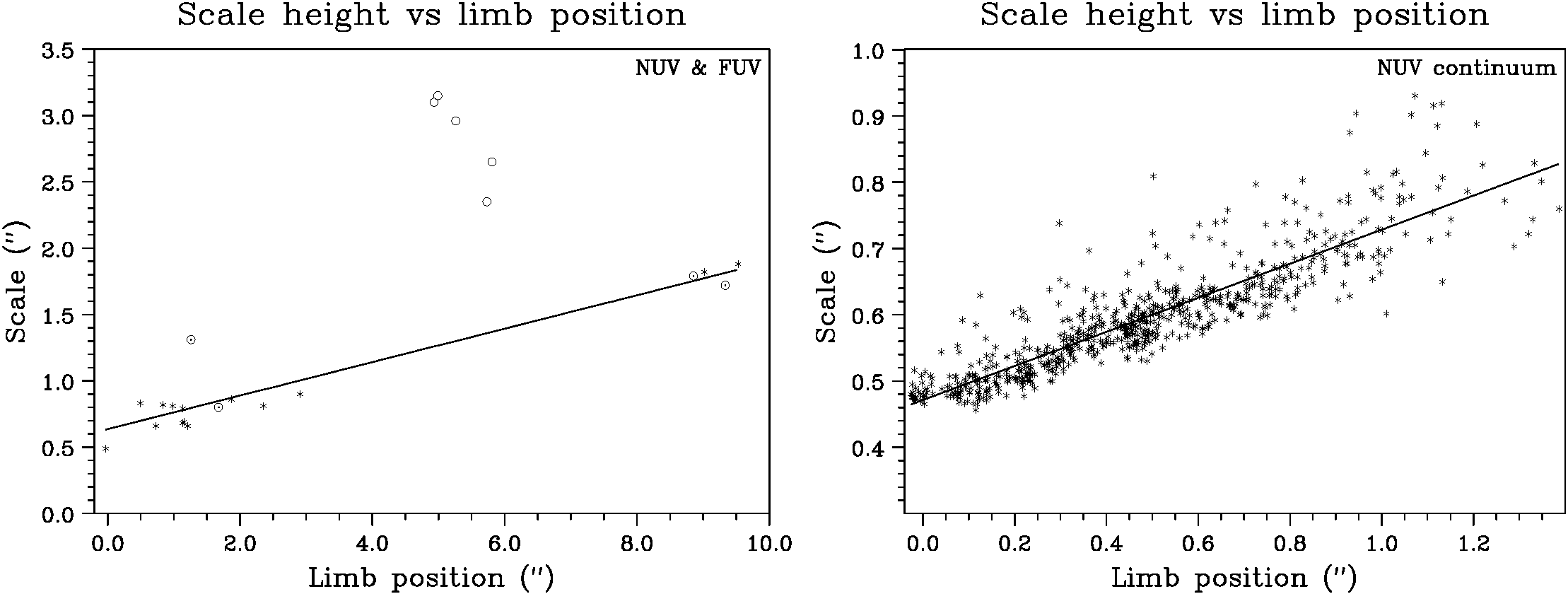}
\end{center}
\caption{The intensity scale height as a function of position of the outer limb for NUV and FUV lines (left). Asterisks mark the NUV and open circles the FUV lines respectively; asterisks and open circles with dots were used for the linear regression. The right panel shows the same relation for the NUV continuum; here the data points represent different wavelengths.}
\label{ScaleHeight}
\end{figure}

It is interesting to note that there is a tendency for the intensity scale height to increase with the height of the outer limb, as shown in the left plot of Figure \ref{ScaleHeight}. The notable exception are the Si{\sc\,iv} and the O{\sc\,iv} lines, probably due to their higher degree of ionization; the other point clearly above the regression line is from O{\sc\,i}. In the NUV continuum, which shows a clear rise close to the Mg{\sc\,ii} k and h lines (Figure \ref{LmbHt}), this association is better defined (Figure \ref{ScaleHeight}, right) but the slope of the regression line is two times higher (0.26 compared to 0.13). The scale-height {\it vs\/}. limb height association apparently reflects the increase of the density scale height with temperature in a complex manner, the analysis of which is beyond the scope of this article. The above analysis of the FUV lines formed in the TR aims at providing a range of values of their formation heights. It does not replace a full model of the TR, where its 2D structure should be taken into account, as proposed by \cite{1983SoPh...83...27D} and worked out by \cite{2009A&A...503..559B}. 

\begin{figure}[h]
\begin{center}
\includegraphics[width=.6\textwidth]{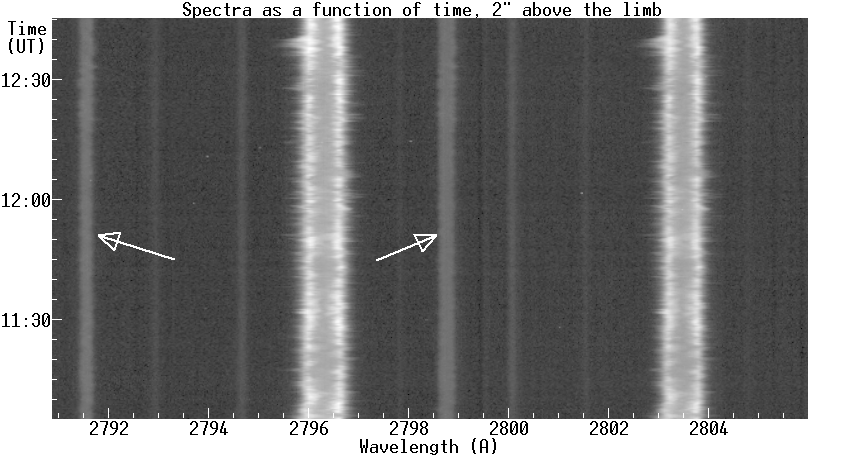}
\end{center}
\caption{The intensity as a function of wavelength and time in the region of k and h, at a distance of 2\arcsec\ above the limb, slit position 2. The width of the cuts is 0.5\arcsec. The arrows point to the \Mg\ triplet lines.}
\label{LimbCuts}
\end{figure}

Before closing this section we would like to point out that the \Mg\ triplet and other weak NUV lines that appear in emission beyond the limb have a very different temporal behavior than the k and h lines. This is demonstrated in Figure \ref{LimbCuts}, where we present spectral cuts around 2800\,\AA\ at a distance of 2\arcsec\ above the limb as a function of time. In contrast to the k and h lines, which show significant time variations due to spicular activity, the intensity in the triplet and other weak emission lines is flat, without any sign of spicules.

A similar remark on the spatial structure was made a long time ago by \cite{1965PASP...77..137P}, who noted that, in a flash spectrum, some rare-earth and other lines appeared structureless. Subsequently, \cite{1968SoPh....5..131P} found no time variations in a Er{\sc\,ii} line situated between Ca\,{\sc ii} H and H$_\varepsilon$; they concluded that rare-earth lines are not formed in spicules, suggesting the presence of an interspicular region in the chromosphere. Our results confirm this conclusion and extend it to emission lines with low height of formation. We will come back to this issue in Section \ref{AtmStruc}.

\subsection{Bulk Parameters of Doublets}\label{BulkDoublets}
\begin{figure}[ht]
\begin{center}
\includegraphics[height=6.6cm]{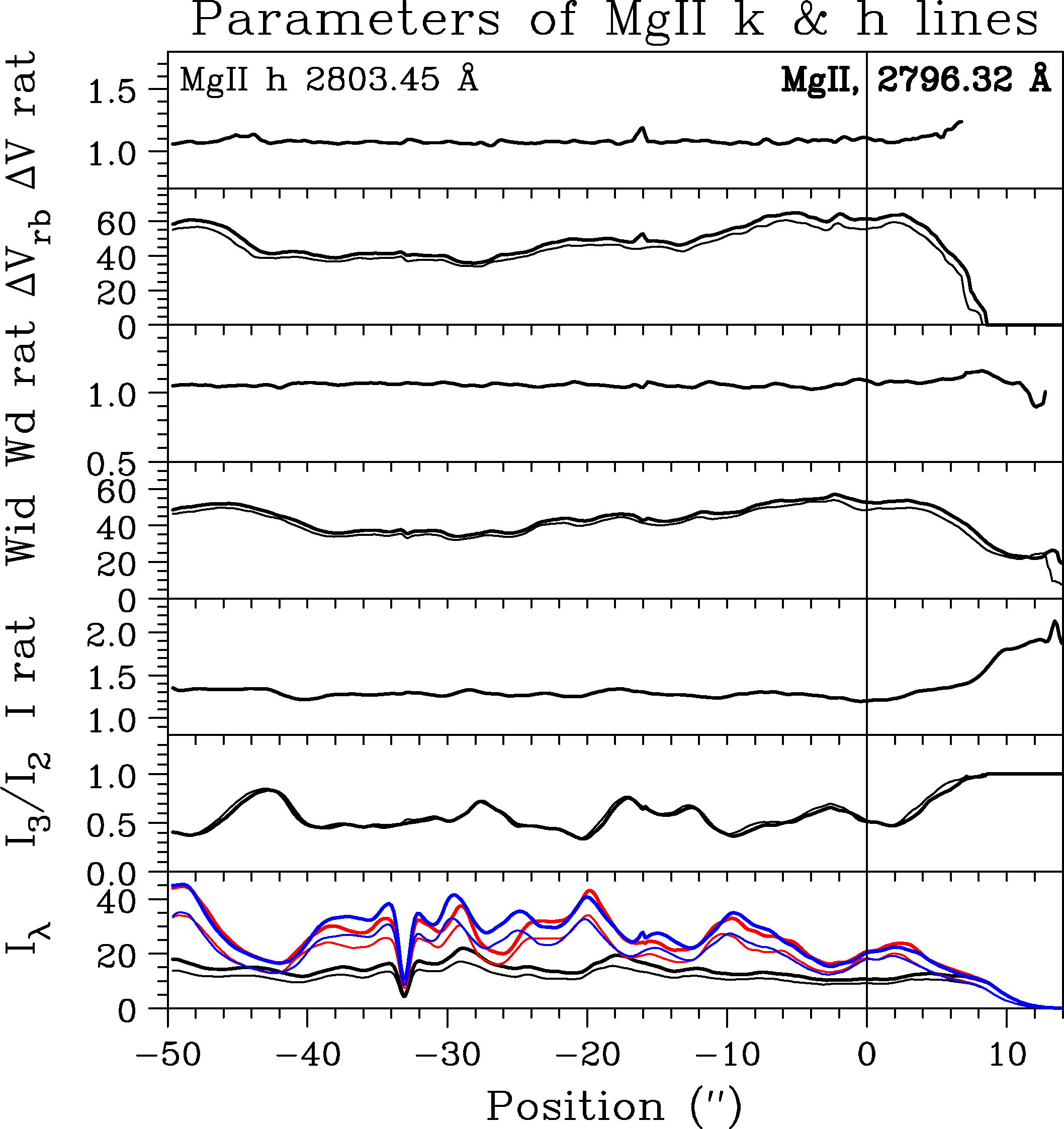}\hspace{.2cm}\includegraphics[height=6.6cm]{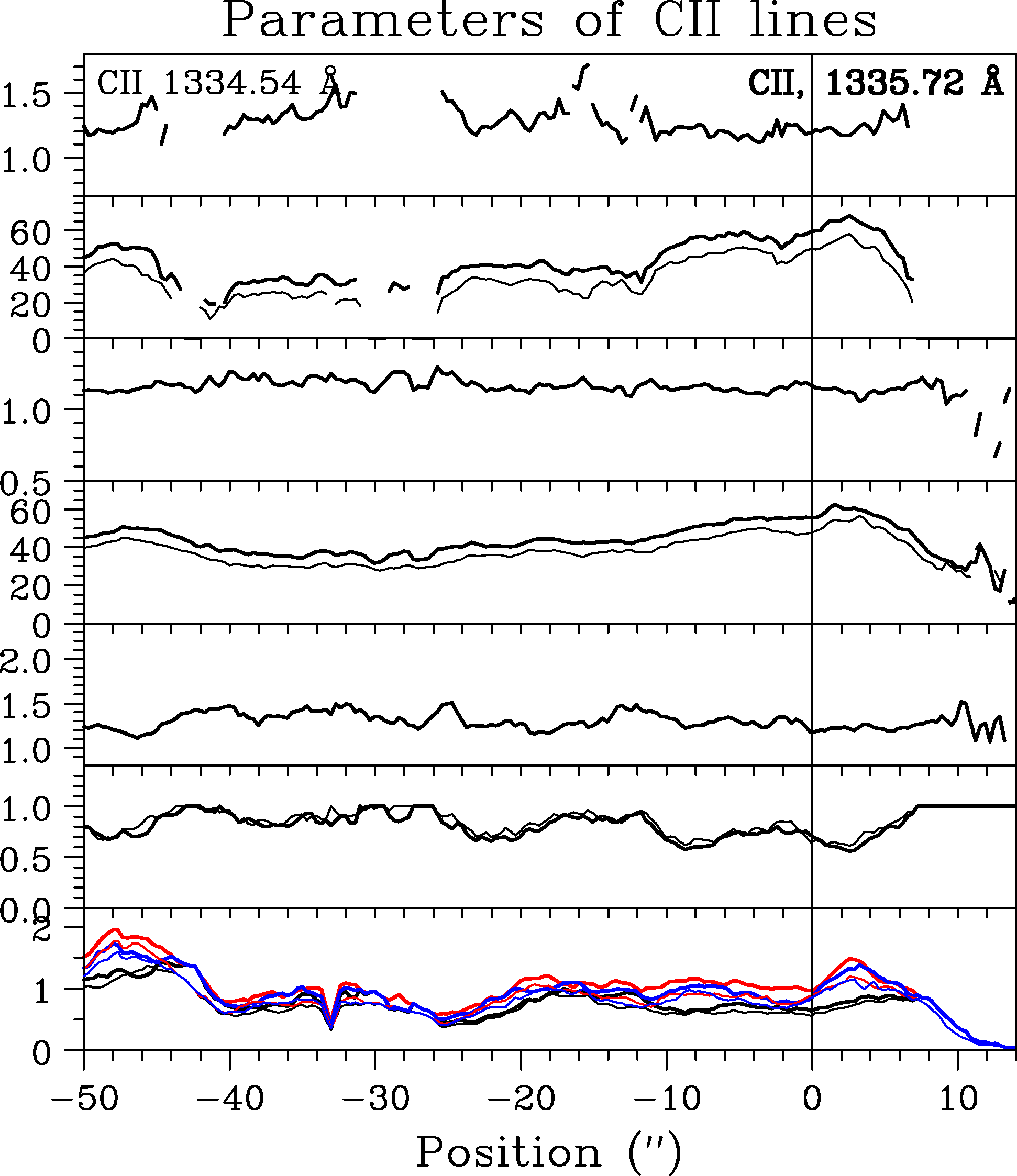}\\
\smallskip
\includegraphics[height=4.39cm]{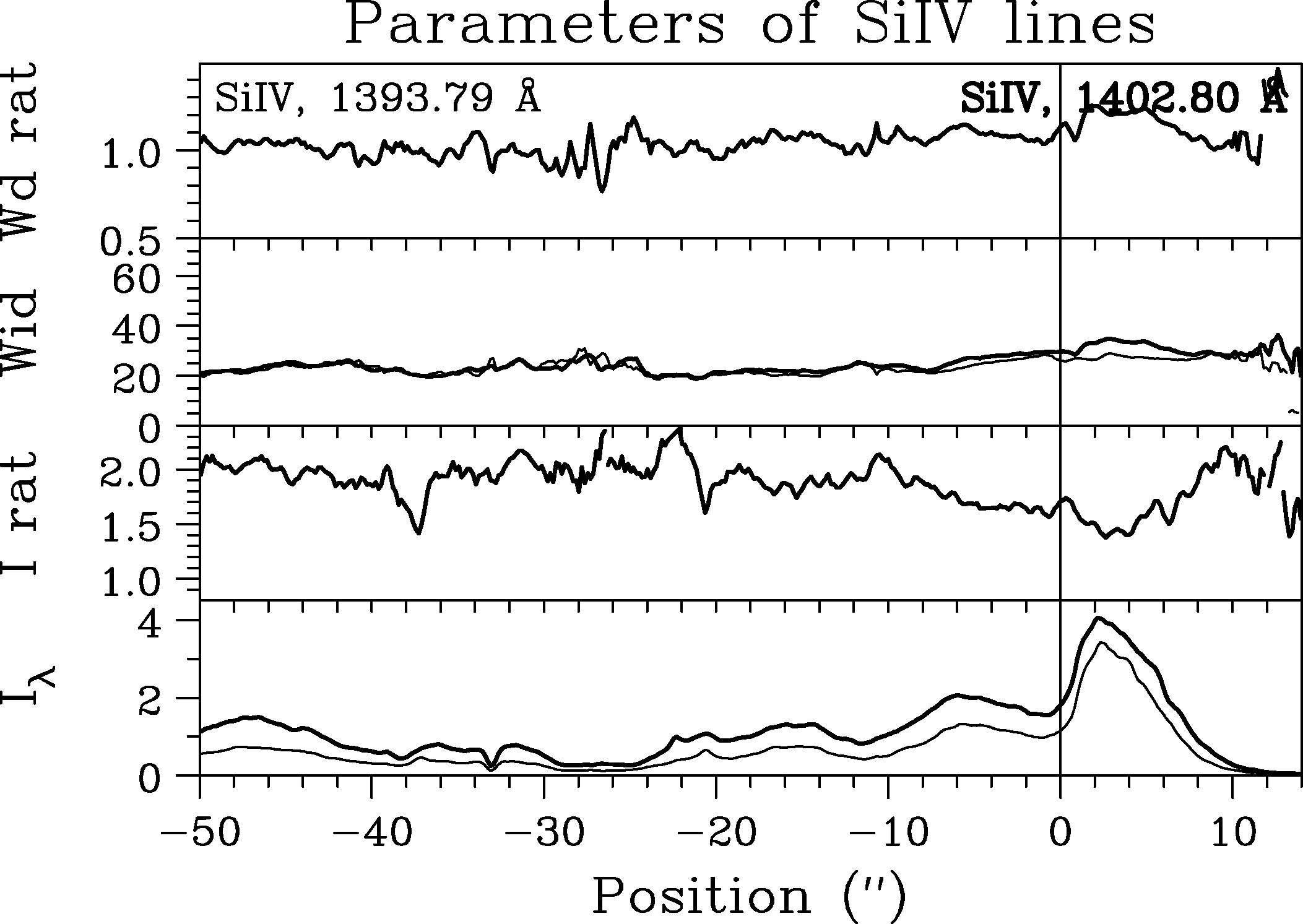}\hspace{.2cm}\includegraphics[height=4.39cm]{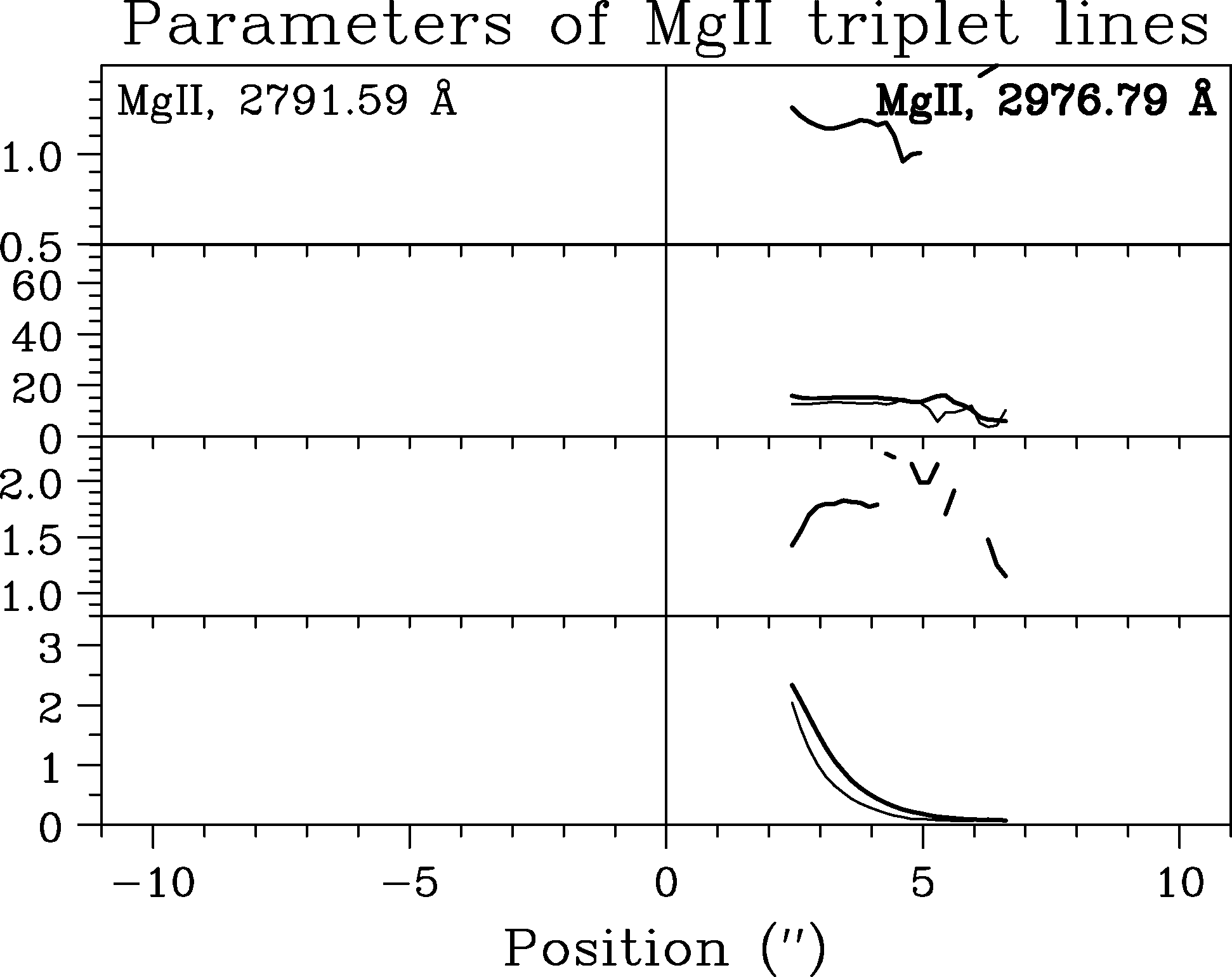}
\end{center}
\caption{Bulk parameters of doublets as a function of distance from the limb. Top row: Double-peaked lines. From bottom panel up: Intensity in the core, red peak (red curve) and blue peak (blue), in units of $10^{12}$\,erg\,cm$^{-2}$\,s$^{-1}$\,sr$^{-1}$\,cm$^{-1}$; core-to-peak intensity ratio; line intensity ratio; line width in \kms; line width ratio; LOS velocity difference of the peaks in \kms; ratio of LOS velocity difference. Bottom row: The pertinent parameters for single-peaked lines. Thick curves correspond to the strongest line of the doublet. Note that the position scale for the Mg{\sc\,ii} triplet is expanded.}
\label{doublets}
\end{figure}

The comparison of doublet parameters gives important information, for example the intensity ratio is a good diagnostic of the optical depth. In Figure \ref{doublets} we give graphs of bulk parameters as a function of distance from the limb for the 
Mg{\sc\,ii}, C{\sc\,ii}, and Si{\sc\,iv} doublets as well as for the Mg{\sc\,ii} triplet lines. For double-peaked lines (top row), in addition to the line intensities, their ratio, the line widths and their ratio, we provide the contrast between the line core intensity, $I_3$, and that of the peaks, $I_2$, the distance between the peaks, and the ratio of this distance between the doublet lines. The line width, corrected for instrumental broadening, is expressed in terms of velocity, which facilitates the comparison between NUV and FUV lines; it is equal to FWHM/$\sqrt{\ln 16}$, so that for an optically thin Gaussian line it coincides with the Doppler width.

As seen in the figure, the \Mg\ lines are double peaked until 8\arcsec\ above the limb ({\it cf.} Figure \ref{AvSpec}); up to about that point, the ratio of the k/h core intensity is flat, without any clear correlation with the intensity variations due to local structures and has an average value of $1.22\pm0.04$. In the optically thin regime above the limb the ratio increases and reaches the value of 1.96, very close to the opacity ratio of the two lines which is equal to 2.

As expected, the k line is broader  than the h line, with an average ratio of $1.06\pm0.01$ on the disk, decreasing near unity above the limb. Starting near $-30$\arcsec, the line widths increase towards the limb. Our results show two peaks on either side of the limb, unlike those of \cite{2014ApJ...792L..15P}, which show a single peak beyond the limb. We measured the FWHM from the plot in  \cite{2014ApJ...792L..15P} and compared it with our measurements; our values are the same as those of the above authors beyond the limb, but slightly higher just inside. This difference could be due to their spectrum being a better representation of the average quiet Sun than ours ({\it c.f.} discussion in Section \ref{intro}), although  the region near the slit appears fairly quiet in the average SJ image in Figure \ref{AvSpec}.

The separation of the emission peaks, $\Delta v_{b}$, also increases as the limb is approached; the corresponding k/h ratio is also flat on the disk and low above the limb, with an average value of $1.08\pm0.02$. As for the line of sight (LOS) velocity, this shows small fluctuations around the average value in all lines. Finally, the core-to-peak ratio, $I_3/I_2$ is practically identical in the two lines and shows fluctuations that are anticorrelated with the peak intensity, going below 50\% in bright regions. We add that, in spite of the averaging, there remains some red/blue asymmetry, $I_r/I_b$, in the individual profiles (not shown in Figure \ref{doublets}). This ranges from 0.67 to 1.16, while the average of all our disk profiles gives $I_r/I_b=0.95$ (see also Figure \ref{CalNUV}).

The behavior of the \Cb\ lines is similar to that of \Mg, but the data are more noisy. For this and the other doublets we give the ratio of integrated intensities rather than of core intensities in Figure \ref{doublets}. Compared to \Mg, the line ratio is higher on the disk ($1.31\pm0.09$) and near 1.5 in the optically thin region above the limb. We note that the oscillator strength ratio for these lines is 1.12.   
The width of the stronger \Cb\ line is almost the same as that of \Mg\ k, but the width ratio has a higher value ($1.16\pm0.04$). The peak separation is  smaller, and the corresponding ratio higher ($1.29\pm0.13$). On the average, the red peak is stronger here, with $I_r/I_b=1.06$ for the 1334.54\,\AA\ line and 1.13 for the 1335.72\,\AA\ line. 

The \Si\ doublet is a good example of optically thin lines, as evidenced from the fact that, up to a distance of $\sim10$\arcsec\ from the limb, the line width ratio is practically unity ($1.04\pm0.08$) and the line ratio is near the opacity ratio of 2. Closer to and beyond the limb the intensity ratio decreases and the width ratio increases; this is an opacity effect, due to the increase of the column density along the line of sight. Assuming that the source function is constant and the same for the two lines of the doublet, the line ratio gives a maximum optical depth of 1.5 for the weaker line at a height of 2\,--\,4\arcsec\ above the limb. As noted in Section  \ref{profiles}, at that location we have slightly flat-topped profiles for the stronger line. We add that the line width is about one half of that of \Mg\ and \Cb. 

For the \Mg\ triplet we could only measure the bulk parameters in a small region from 2.5 to 6\arcsec, where the lines are in emission and unobstructed by the nearby continuum. Here the line intensity ratio is around 1.8, the width slightly smaller than that of \Si\ ($\sim15$\kms) and the width ratio around 1.16.

\begin{figure}[h]
\begin{center}
\includegraphics[height=6.3cm]{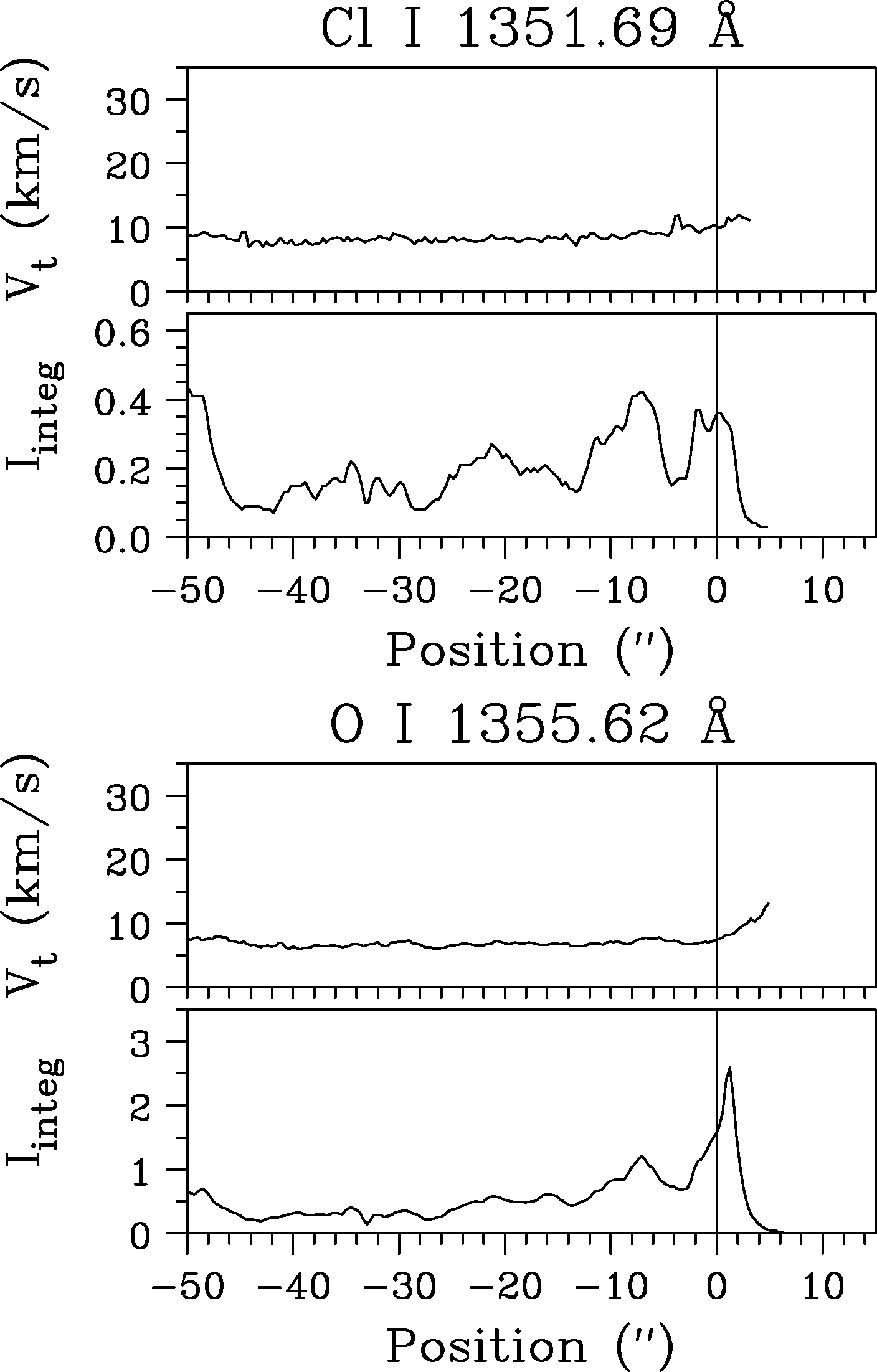}\hspace{0.2cm}\includegraphics[height=6.3cm]{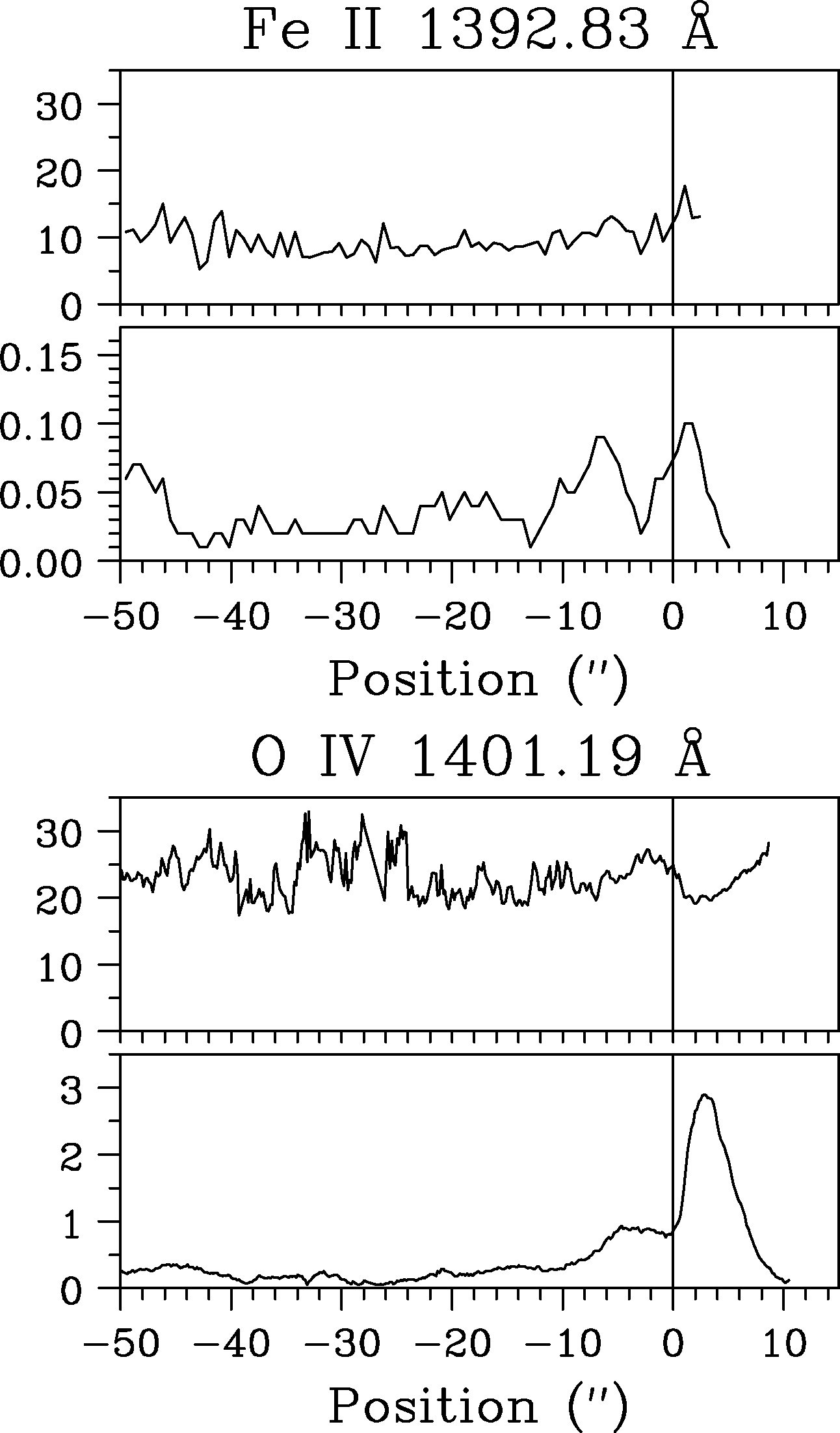}\hspace{0.2cm}\includegraphics[height=6.3cm]{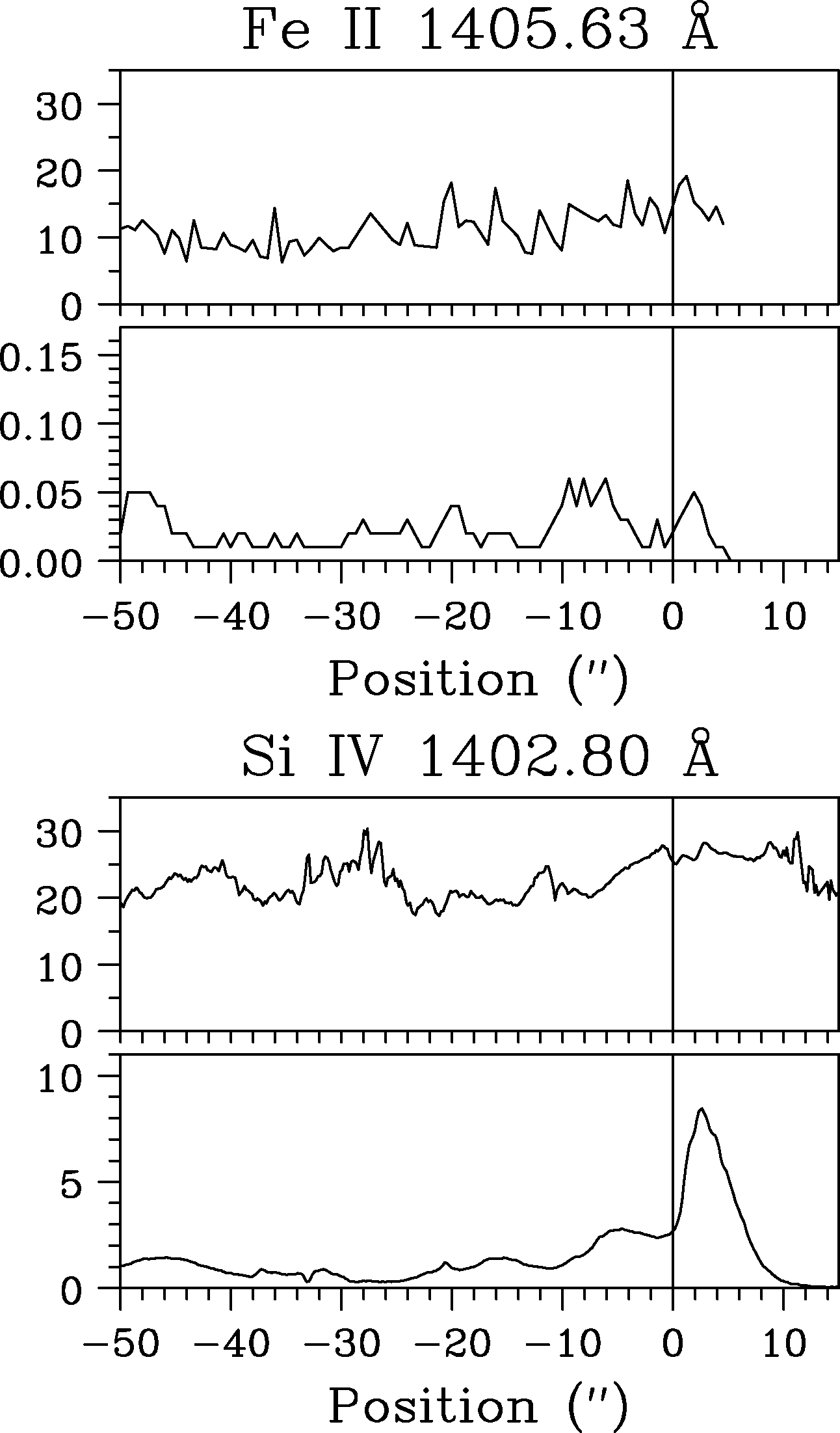}
\end{center}
\caption{Turbulent velocity as a function of distance from the limb for a number of FUV emission lines. The integrated intensity in $10^{3}$\,erg\,cm$^{-2}$\,s$^{-1}$\,sr$^{-1}$ is also shown for reference.}
\label{Vturb}
\end{figure}

\subsection{Optically Thin Lines: Turbulent Velocity and \OIV\ Ratios}\label{thin}
In optically thin lines, which is probably the case of all weak emission lines on the disk in the FUV, we can measure the Doppler width and from that, together with the temperature of line formation, compute the turbulent velocity, $V_t$. We give the results in Figure \ref{Vturb}, where, in addition to the \Cl\ and \OI\ lines, the \Si\ 1402.8\,\AA\ line as well as weak Fe\,{\sc ii} and O\,{\sc iv} lines in FUV2 are included.  For the computation of the turbulent velocity we adopted temperature values of $10^4$, $1.6\times10^4$, and $6.4\times10^4$\,K for \Cl--Fe\,{\sc ii}, \OI\ and \Si--O\,{\sc iv} respectively (\citealp{1993SoPh..144..217D, 1998ApJ...505..957C, 2014SoPh..289.2733D}). In any case, the thermal velocity is small compared to the Doppler width velocity, ranging from 2.2 for \Cl\ to 8.2\kms\ for O\,{\sc iv}.

For the low ionization lines the turbulent velocity on the disk is very flat, unaffected by the intensity variations, with average values from 6.9 for \Cl\ to 10\kms\ for Fe\,{\sc ii}. There is a slight increase as we approach the limb, which could be an opacity effect. In \OI\ $V_t$ is reliably measured up to 5\arcsec\ beyond the limb, where it rises to 13\kms. Within the margins of the dispersion, $V_t$ is identical in the higher ionization lines, $\sim22$\kms.

\begin{table}[h]
\caption{Values of the turbulent velocity}
\label{Tab:Vturb}
\begin{tabular}{lccccccc}
\hline
      &   &\multicolumn{2}{c}{This work} & \multicolumn{2}{c}{DM}&\multicolumn{2}{c}{CSL}\\		
\hline 
Line  & $\lambda$& T        &   $V_t$   &T        &   $V_t$  &T        &$V_t$\\
      &  \AA\    & $10^4$\,K&   \kms\   &$10^4$\,K&   \kms\  &$10^4$\,K&\kms\ \\ 
\hline
\OI\        & 1355.62 & 1.5 & $ 6.9\pm 0.4$ & 1.6 & $ 4\pm 3$& 1.62 &  7.0\\
\Cl\        & 1351.69 & 1.0 & $ 8.4\pm 0.8$ &  --   &   --       &   --   &  --   \\
Fe\,{\sc ii}& 1392.83 & 1.0 & $ 9.5\pm 2.0$ &  --   &   --       &   --   &  --   \\
Fe\,{\sc ii}& 1405.63 & 1.0 & $10.9\pm 2.9$ &  --   &     --     &   --   &  --   \\
\Si\        & 1402.80 & 6.4 & $22.1\pm 2.6$ & 6.6 & $22\pm 7$& 7.08 & 23.2\\
O\,{\sc iv} & 1404.19 & 6.4 & $23.3\pm 3.0$ &16.8 & $16\pm 7$& --     & --  \\
\hline
\end{tabular}
\end{table}

Our results are summarized in Table \ref{Tab:Vturb}, together with those of \citealp{1993SoPh..144..217D} (DM) from HRTS and \citealp{1998ApJ...505..957C} (CSL) from SUMER. For the lines in common they are consistent, except for the O\,{\sc i} and  O\,{\sc iv} results of \cite{1993SoPh..144..217D}; the discrepancy of the latter is partly due to the different temperatures assumed. Our \Si\ results are close to those of \cite{2015ApJ...799L..12D} from IRIS, who give $V_t\simeq20$\kms, assuming $T=8\times10^4$\,K.

In a recent article, \cite{2016A&A...594A..64P} discussed the importance of the O\,{\sc iv} and \Si\ line ratios near 1400\,\AA\ as electron density diagnostics, largely independent on temperature. They considered four ratios,  $R_1=\frac{\rm{\OIV}\-\,1399.78}{\rm{\OIV}\,1401.16}$, $R_2=\frac{\rm{\OIV}\,1401.16}{\rm{\OIV}\,1404.81}$ $R_3=\frac{\rm{\Si}\,1405.85}{\rm{\Si}\,1406.06}$ and $R_4=\frac{\rm{\Si}\, 1402.77}{\rm{\OIV}\,1401.16}$, pointing out that $R_4$ gives much higher densities than the others. In our case, although the 1399.78\,\AA\ line is quite weak, we could get meaningful values for $R_1$ near the limb by averaging four spectral rows. We obtained a value of $R_1=0.20$ at 1\arcsec\ above the limb which, according to Figure 2 of \cite{2016A&A...594A..64P} computed for $\log T_{\rm \OIV}=5.15$, gives 
$N_e=9.3\times10^9$\,cm$^{-3}$. The ratio drops with height, indicating a density decrease, reaching a value of 0.18 at 5.5\arcsec\ that corresponds to $N_e=2.4\times10^9$\,cm$^{-3}$. These values should be considered as indicative only, because our measurements are near the low density sensitivity limit of $R_1$.

\section{Spatial Structure and Temporal Variations Near the Limb}\label{AtmStruc}
Although our spectra do not provide two-dimensional information, they can give images of the intensity as a function of position along the slit and time, for selected wavelengths. As the wavelength can be selected at will, we have the flexibility to probe the entire region in which the NUV and FUV radiation forms. We will first examine the lower and subsequently the higher atmospheric layers 

\subsection{Lines and Continua Formed in the Low Atmosphere}
\begin{figure}[h]
\begin{center}
\includegraphics[width=\textwidth]{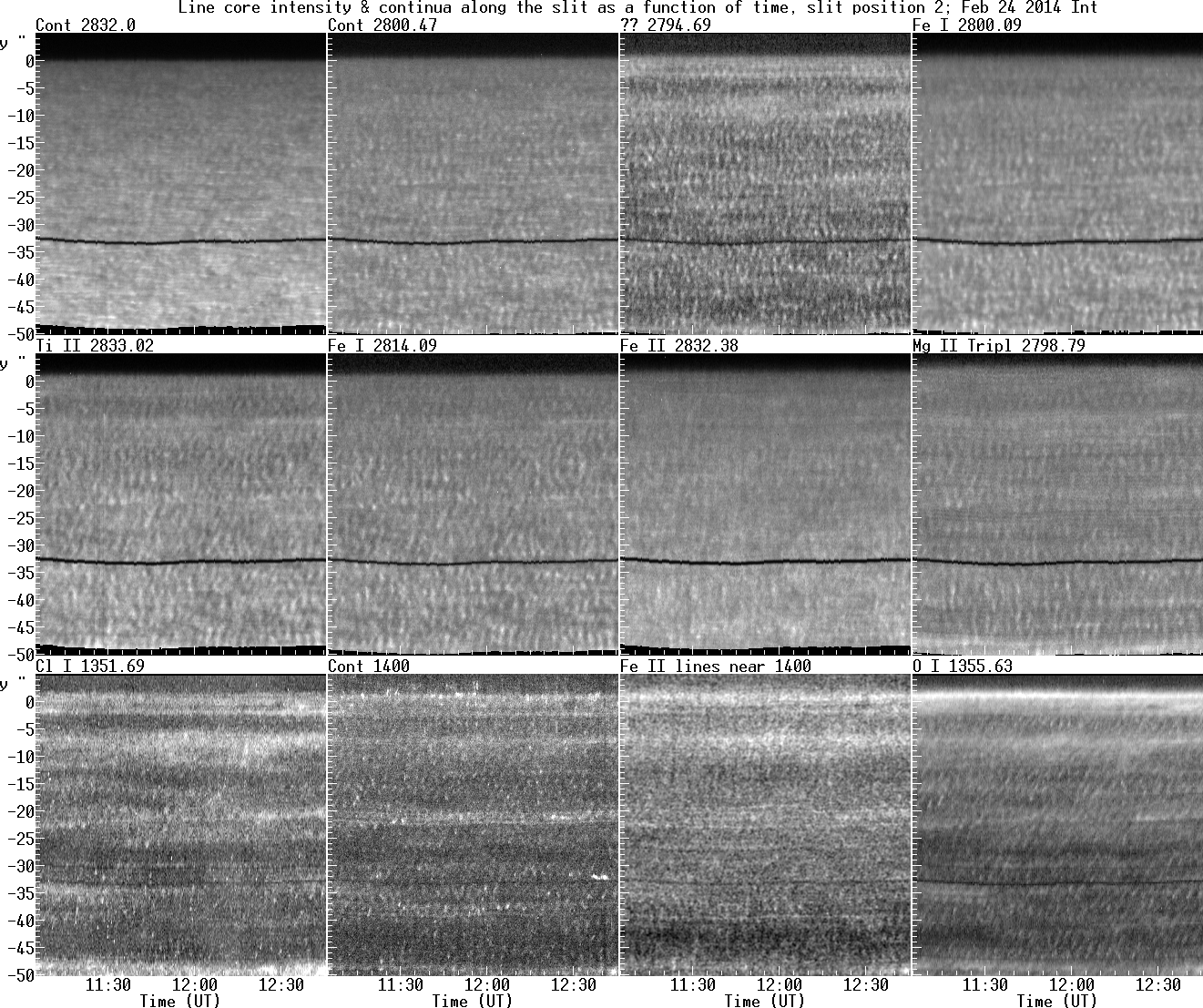}
\end{center}
\caption{Intensity as a function of position and time for lines and continua formed low in the atmosphere. South is up.}
\label{CutsLow}
\end{figure}

In this section we will examine NUV continua and lines in emission above the limb as well as low ionization FUV lines and the continuum around 1400\,\AA. Images of the intensity as a function of position and time are given in Figure \ref{CutsLow}, ordered in increasing height of the limb, first in the NUV and then in the FUV. The core intensity is displayed for the NUV lines,  the integrated intensity for the FUV; the image of the Fe\,{\sc ii} in the FUV is the sum of the integrated intensities of the three weak lines at 1392.17, 1392.83 and 1405.63\,\AA\ ({\it cf.} Section \ref{RadVarInt}), while for the 1400\,\AA\ continuum we integrated all FUV2 spectral regions devoid of lines (a similar operation for FUV1 gave too noisy results). For some of these lines the average center-to-limb variation is given in Figure \ref{IntLimb}.

A first remark is that none of these lines show time variations associated with spicules beyond the limb, confirming the result presented in Section \ref{RadVarInt} and Figure \ref{LimbCuts} and extending them to the NUV lines; some activity is present in the 1400\,\AA\ continuum, which is probably due small impulsive events on the disk rather than spicules (see next section). 

In the 2832\,\AA\ continuum we see slight variations of the intensity with time, which are probably associated with the granulation. As we go higher, in the quasi-continuum between the k and h lines, we see temporal variations reminiscent of the p-mode oscillations, similar to those seen with the {\it Transition Region and Coronal Explorer} (TRACE) (see,{\it  e.g.\/}, Figure 7 of \citealp{2001A&A...379.1052K}). The same type of structure is present in all NUV lines, with some variations from one line to another, being more prominent and sharper in the unidentified 2794.69\,\AA\ line. A one-dimensional power spectrum analysis shows increased power in the frequency range of 2.5 to 8\,mHz, while the average autocorrelation function peaks in the range of 180 (2797.85\,\AA\ line) to 235\,s (Fe\,{\sc ii} 2832.40\,\AA). In the FUV, only the \OI\ line has adequate signal-to-noise ratio to clearly show the oscillations. We note that image cuts in the 1600 and 1700\,\AA\ AIA bands have similar appearance, with the autocorrelation peaking at a time lag of 225\,s.

In addition to p-mode oscillations, the \OI\ position-time image shows short inclined features, suggestive of small bright structures moving towards the limb. They last for 2\,--\,8 min and their average apparent velocity, measured through 2D autocorrelation, is 1.8 \kms. Similar features are seen in the more noisy 1400\,\AA\ continuum position-time image, but are absent from all others. They are apparently low-lying structures, since we see nothing close and above the limb. In the lack of 2D imaging, we cannot be more specific on their nature.

\subsection{Lines and Continua Formed in the High Atmosphere}
\begin{figure}[h]
\begin{center}
~~\includegraphics[width=.98\textwidth]{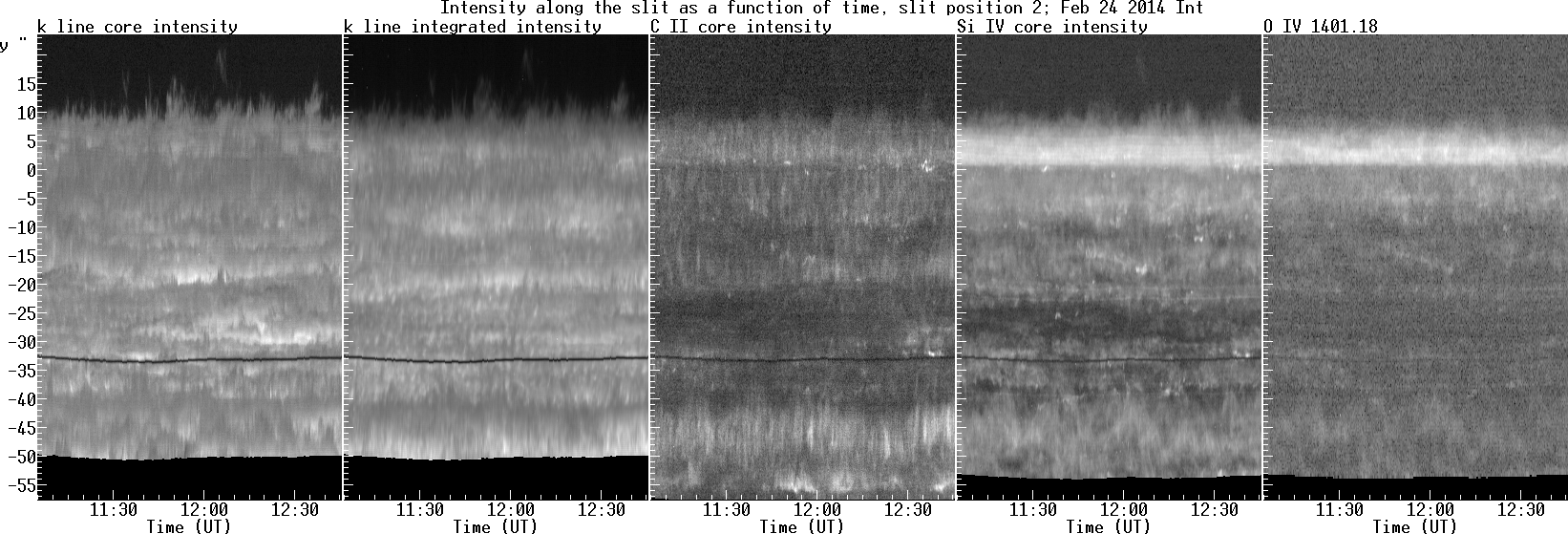}\\
\smallskip
\includegraphics[width=\textwidth]{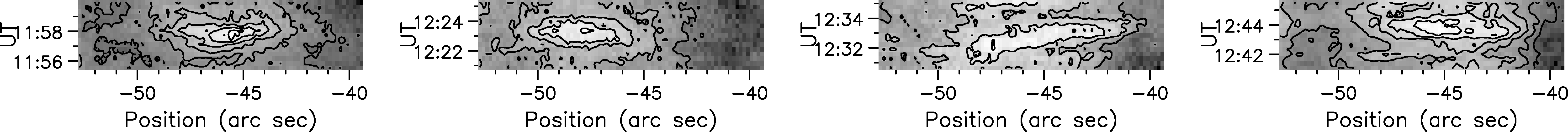}
\end{center}
\caption{Intensity as a function of position and time for lines formed high in the atmosphere. South is up. The bottom row shows contour/gray plots of bright features seen in the \Cb\ image between $y=-51$ and $-39$\arcsec.
}
\label{CutsHigh}
\end{figure}

The \Mg\ k and h, as well as the \Cb, \OIV, and \Si\ lines fall in this group. The corresponding position-time images for slit position 2 are shown in Figure \ref{CutsHigh}, where we have averaged the two \Cb\ lines and the two \Si\ lines to reduce noise; apart from the intensity values, the images for the components of each doublet are practically identical. The integrated intensity images of these lines as well as \OIV\ are very similar to the core intensity images and are not shown in the figure. For the k line the integrated intensity covers an 1.1\,\AA\ interval centered at the core and includes both emission peaks. The images for the h line are very similar to those for k and are not shown here.

All lines show clear spicular structure, extending at least 8\arcsec\ above the limb in \Mg\ and \Cb\ and somewhat lower in \Si\ and \OIV\ ({\it cf.} Table \ref{Table03}). Individual structures go up to 15\arcsec\ and one structure that crossed the slit from the side at 12:05 UT (best seen in the k line core) went up to 21\arcsec. On the disk and up to 2\arcsec\ from the limb we have occasionally impulsive events, best visible in the \Cb\ and \Si\ lines, also seen in the 1400\,\AA\ continuum image (Figure \ref{CutsLow}); these are smaller than 1\arcsec\ and shorter than 4 min and appear similar to the one reported by \cite{2015SoPh..290.2871T}. There is no obvious association of these events with spicules and, in any case, they are much less numerous.

We will now focus our attention to the bright, almost vertical structures that appear on the disk in the \Cb, the most prominent of which are located between $-51$ and $-39$\arcsec; they are also visible in the k and h integrated intensity images, but they are hard to see in \Si. We note that their footpoints are near a persistent network structure, which appears as a bright band at $y\simeq-50$\arcsec\ in the \Cl\ image of Figure \ref{CutsLow}. Other, weaker, network structures ({\it e.g.\/}, at $y\simeq-22$\arcsec) are also associated with (weaker) \Cb\ features. They appear at regular intervals of $\sim250$\,s, as derived from an autocorrelation analysis. 

The contour plots shown in the bottom row of Figure \ref{CutsHigh} suggest that a region 5 to 10\arcsec\ along the slit brightens almost simultaneously within 1 to 2 min. In a few cases there are measurable motions: the feature shown in the third plot has an apparent ascending speed of $\sim85$\kms, the ones in the second and fourth plot are descending with $\sim60$ and $\sim115$\kms\ respectively, while the first is probably a blend of a descending and an ascending feature. These values, as well as the overall similarity with features above the limb, are suggestive of fast (type II) spicules seen in emission on the disk. 

\section{Comparison of k and h Parameters with Radiative Transfer Models}
The high-quality IRIS spectral data provide an opportunity for comparison with radiative transfer models. In this section we describe the NTLE model that we used, the selection of IRIS and Balmer line data and we present the results of the comparison.
  
\subsection{The Model}
We compared observed \Mg\ line profiles and bulk parameters with the results of the theoretical model PROM57Mg, developed at the {\it Institut d' Astrophysique Spatiale}. This is a one-dimensional, static, isothermal and isobaric NLTE model with a vertical slab geometry, originally developed for prominences.

We note that NLTE computations of resonance lines of H and Ca\,{\sc ii} have been performed in a cylindrical geometry (fitting rather well the geometrical structure of spicules) through Monte Carlo methods which have the advantage of treating any complex geometry (\citealp{1968ApJ...152..493A, 1969JQSRT...9..519A, 1969JQSRT...9.1579H}). A spicule model based upon Ca\,{\sc ii} K has been built by \cite{1969SoPh...10...88A}, where the spicule is approximated by a cylinder within which the thermodynamic parameters (temperature and density) vary with altitude. A peculiar feature of these computations consists in ``isolating'' one structure but taking into account the incident radiation from the chromosphere. 

However, major drawbacks of these methods are that they consider only a two-level atom and also that they do not treat consistently the hydrogen atom and particularly its ionization. Actually, the relevant issue concerns the need of a multidimensional treatment of radiative transfer as discussed in \cite{1973ApJ...185..167J}. As \cite{1980ApJS...42..221J} put it, when assessing the spicule (cylinder) computations of \cite{1969SoPh...10...88A}: ``the density (hence, roughly, the opacity) scale height is approximately 5\,Mm while the cylinder diameter was 0.8\,Mm. Thus escape through the cylinder side limits the scale for axial exchange effects to a distance small compared to the scale for density and temperature variations.'' Their scaling law approach was confirmed by the 2D NLTE computations of \cite{1978ApJ...220.1001M} with a two-level atom, and later applied to prominences for the case of the H, Ca\,{\sc ii} and Mg\,{\sc ii} lines by \cite{1982ApJ...253..330V}. In summary, we estimate that the geometrical approach of a spicule as a vertical 1D slab is sufficient for building a thermodynamic model and consequently can benefit from the extensive 1D modeling of prominences (see, {\it e.g.\/}, \citealp{2015ASSL..415..103H, 2015ASSL..415..131L}), where the essential ingredient is the proper and detailed treatment of the incident radiation. 

The PROM57Mg program that we used is a variant of PROM7\footnote{for details see \tt{https://idoc.ias.u-psud.fr/MEDOC/Radiative\%20transfer\%20codes/PROM7}}, in which the H atom is represented by 20 levels plus the continuum and Mg{\sc\,ii} with 2 bound levels plus 3 continua. The input parameters are the temperature, the gas pressure, the geometrical thickness of the slab, the turbulent velocity, $\xi$, and the height above the photosphere. The direction of the line of sight is also a parameter, but in our case this is fixed since at the limb the line of sight is perpendicular to the vertical slab; the average 30\degr\ inclination of spicules with respect to the vertical \citep{1968SoPh....3..367B} was not taken into account. The output of the model is line profiles of the Lyman, Balmer, and Paschen lines as well as of the Mg{\sc\,ii} h and k lines.

Being resonant lines, h and k are strongly affected by the incident radiation. Model computations with the incident radiation set to zero show that the intensity drops by 70\,--\,75\%,  for input parameters near our best model at the height of 5400\,km (see below). This indicates that most of the emission is due to scattering, the selection of the incident profile is therefore crucial. A natural choice is the calibrated IRIS profile from the center of the disk (top panel of Figure \ref{CalNUV}) which is representative of the quiet Sun, while the averaging along the slit assures that both network and cell interior regions are included.

\subsection{IRIS Observations Selected for Modeling}
\begin{figure}[h]
\begin{center}
\includegraphics[width=\textwidth]{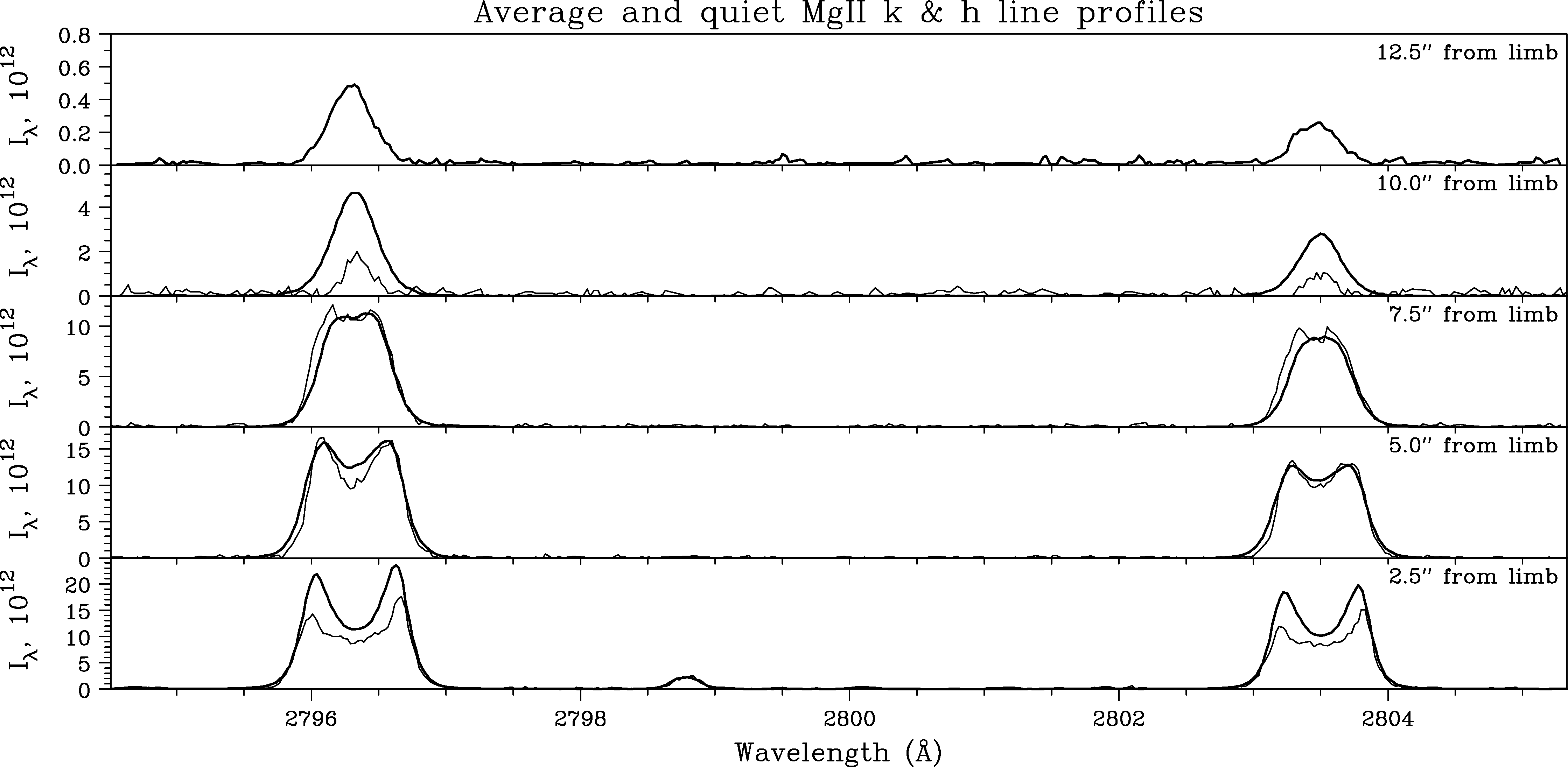}
\end{center}
\caption{Average (thick line) and quiet k and h profiles selected for modeling.}
\label{SelProf}
\end{figure}

We compared the model computations with average profiles (Section \ref{AvSpecProf}) of the h and k lines at 2.5, 5.0, 7.5, 10.0, 12.5\arcsec\ above the limb (corresponding to heights of 1800, 3600, 5400, 7200, and 9000\,km). In order to check how typical these profiles are, we also selected profiles from the spectrum at 11:20:24 UT, slit position 2 ({\it c.f.} Section \ref{obs}), which was typically quiet, devoid of any transient events; for this spectrum we could get meaningful profiles at the four lower heights only. The selected profiles are shown in Figure \ref{SelProf}; the ``quiet" profiles are close to the average ones at 5 and 7.5\arcsec\ above the limb, but notably weaker at other heights. 

\begin{table}[!h]
\caption{Bulk parameters of selected average and quiet profiles}
\label{BulkSel}
{\renewcommand{\tabcolsep}{0.05cm}
\begin{tabular}{l@{~~~}cccccc@{~~~}cccccc}
\hline 
&\multicolumn{6}{c}{Average Profiles}&\multicolumn{6}{c}{Quiet Profiles}\\
Height &$I_{2b}$&$I_{3}$&$I_{2r}$&$I_{int}$&FWHM&$V_r\!-\! V_b$&$I_{2b}$&$I_{3}$&$I_{2r}$&$I_{int}$&FWHM&$V_r\!-\!V_b$\\
~~km & $10^{12}$ & $10^{12}$ & $10^{12}$ &   $10^{4}$  &   \AA\    &   km/s& $10^{12}$ & $10^{12}$ & $10^{12}$ &   $10^{4}$  &   \AA\    &   km/s \\
\hline                                                                                                          
1800,  k   & 22.13 & 11.47 & 23.83 &  14.53  &  0.83  &  63.8  & 14.93 &   8.72 & 17.78 & 10.57   &  0.89  & 70.6 \\
1800,  h   & 18.26 & 10.03 & 19.52 &  11.65  &  0.77  &  58.7  & 12.17 &   7.60 & 15.56 &  8.59   &  0.81  & 64.6 \\
ratios     &  1.21 &  1.14 &  1.21 &   1.25  &  1.08  &   1.09 &  1.23 &   1.15 &  1.14 &  1.23   &  1.10  & 1.09 \\
\hline                                                                                                          
3600,  k   & 16.06 & 12.53 & 16.31 &  11.57  &  0.77  &  50.1  & 17.57 &   9.04 & 16.43 & 10.75   &  0.74  & 56.0 \\
3600,  h   & 12.57 & 10.54 & 12.67 &   8.63  &  0.71  &  44.3  & 14.10 &   9.40 & 13.20 &  8.47   &  0.68  & 39.7 \\
ratios     &  1.28 &  1.14 &  1.18 &   1.32  &  1.09  &   1.13 &  1.26 &   0.96 &  1.25 &  1.27   &  1.09  & 1.41 \\
\hline                                                                                                          
5400,  k   & 11.09 & 10.91 & 11.40 &   6.65  &  0.57  &  16.3  & 12.98 &  10.31 & 11.89 &  7.46   &  0.62  & 30.3 \\
5400,  h   &  8.76 &  8.63 &  8.88 &   4.59  &  0.50  &   8.2  & 10.15 &   8.10 & 10.03 &  5.50   &  0.56  & 23.3 \\
ratios     &  1.26 &  1.26 &  1.29 &   1.45  &  1.15  &   1.99 &  1.28 &   1.27 &  1.19 &  1.36   &  1.10  & 1.31 \\
\hline
7200,  k   &  4.64 &  4.64 &  4.64 &   1.96  &  0.38  &  0.0   &  2.53 &   2.53 &  2.53 &  0.59   &  0.19  &  0.0 \\
7200,  h   &  2.72 &  2.72 &  2.72 &   1.09  &  0.36  &  0.0   &  1.25 &   1.25 &  1.25 &  0.29   &  0.25  &  0.0 \\
ratios     &  1.70 &  1.70 &  1.70 &   1.81  &  1.07  &   --   &  2.03 &   2.03 &  2.03 &  2.03   &  0.77  &  --  \\
\hline 
9000,  k   &  0.48 &  0.48 &  0.48 &   0.20  &  0.35  &  0.0   &  --   &   --   &  --   &   --    &   --   &  --   \\
9000,  h   &  0.25 &  0.25 &  0.25 &   0.10  &  0.39  &  0.0   &  --   &   --   &  --   &   --    &   --   &  --   \\
ratios     &  1.95 &  1.95 &  1.95 &   2.00  &  0.92  &   --   &  --   &   --   &       &   --    &   --   &  --   \\
\hline 
\end{tabular}
}
\end{table}

In order to facilitate the comparison with the model, we computed bulk parameters which are listed in Table \ref{BulkSel}. These include the intensity at the blue and red peaks as well as at the line core ($I_{2b}, I_{2r}, I_3$), the integrated intensity ($I_{int}$), the line width (FWHM), and the separation of the red and blue peaks of the profiles, $V_r-V_b$.

\subsection{Balmer Line Data}
Since the program provides information on the Balmer lines, we took into consideration reported values of their bulk parameters. Calibrated observations are both scarce and inhomogeneous; we give in Table \ref{Table06} results from
tables or figures of \citealp{1971SoPh...17..355K} (KK71), \citealp{1972ARA&A..10...73B} (B72), \citealp{1973SoPh...32..345A} (A73), \citealp{1976SoPh...46...93K} (KBB76) and \citealp{1988ChA&A..12..136D} (DN88).  The value of B72 is a compilation of previous results, A73 and KBB76 used spectra from the \'echelle spectrograph of the {\it Vacuum Tower Telescope} (now the {\it Dunn Solar Telescope}) of Sacramento Peak, KK71 from the solar telescope of Pulkovo and DN88 from the coronograph of the Norikuta Solar Observatory. The results from A73 are averaged along the slit (hence closer to what we measure with IRIS), while the others are averages over individual spicules. Our adopted values (last column in Table \ref{Table06}) are from a second degree fit of the data in the table, excluding the KK71 values which are considerably lower than the others; the average $\sigma$ of this fit was $0.8\times10^4$\,\,erg\,cm$^{-2}$\,s$^{-1}$\,sr$^{-1}$. The only reliable result we could find for \hb\ was from A73 at 5400\,km, with a value of $2.24\times10^4$\,\,erg\,cm$^{-2}$\,s$^{-1}$\,sr$^{-1}$.

\begin{table}[!h]
\caption{\ha\ integrated intensity ($10^{4}$\,erg\,cm$^{-2}$\,s$^{-1}$\,sr$^{-1}$)}
\label{Table06}
\begin{tabular}{ccccccc}
\hline 
Height& KK71  &B72 &  A73  &KBB76 & DN88  &Adopted\\
   km                             &               \\ 
\hline 
3800  &  --   & -- &  --   & --   &  --   & 29.22 \\
4000  &  --   & -- &  --   & --   & 26.7  &  --   \\
5000  &  --   & -- &  --   & 18   & 17.5  &  --   \\
5400  &  --   & -- & 16.25 & --   &  --   & 16.25 \\
6000  & 3.41  & 21 &  --   & --   &  6.03 &  --   \\
7000  & 2.93  & -- &  --   &  5.4 &  5.86 &  --   \\
7200  &  --   & -- &  --   & --   &   --  &  5.17 \\
8000  & 1.59  & -- &  --   & --   &  2.46 &  --   \\
9000  &  --   & -- &  --   &  1.8 &  3.00 &  2.02 \\
\hline 
\end{tabular}
\end{table}

\subsection{Methodology and Results}
\begin{figure}[!h]
\begin{center}
\includegraphics[width=\textwidth]{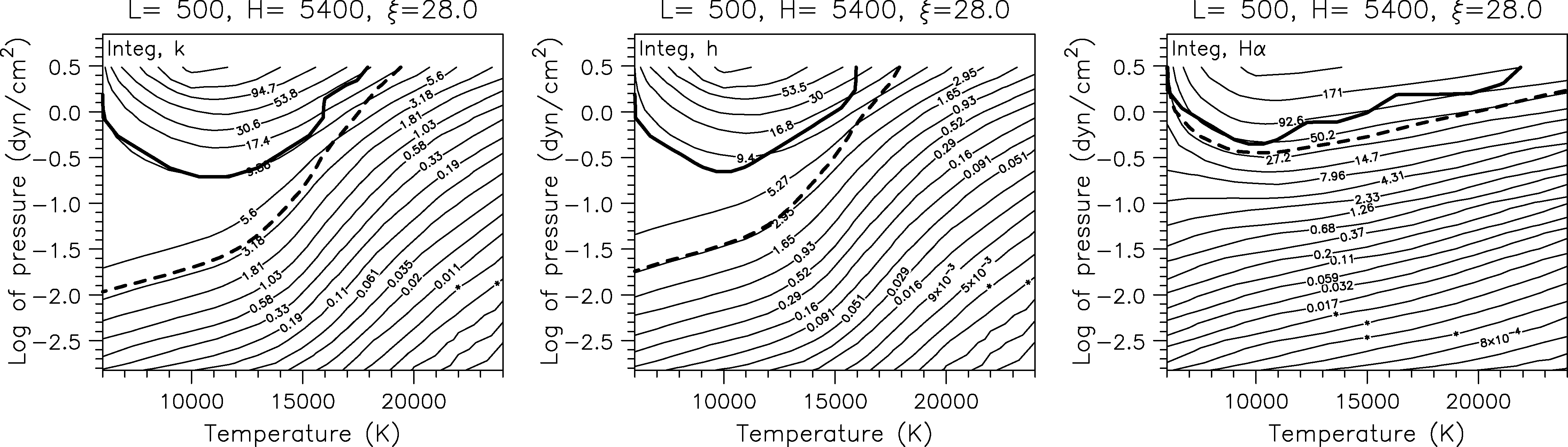}
\end{center}
\caption{Contours of integrated intensity as a function of temperature and pressure for the Mg{\sc\,ii} k and h lines and H$_\alpha$, computed with the PROM57Mg model. $L$ is the thickness of the slab in km, $H$ the height in km, and $\xi$ the turbulent velocity in \kms.  Contours are logarithmically spaced, in $10^{4}$\,erg\,cm$^{-2}$\,s$^{-1}$\,sr$^{-1}$. Profiles above the thick line are double-peaked. The dashed line separates optically thick ($\tau_0>1$, above) from optically thin ($\tau_0<1$, below) profiles.}
\label{grid1}
\end{figure}

Out of the five input parameters of the models, the height above the photosphere is fixed from the observations; moreover, the height affects only the dilution factor of the incident radiation, which varies little in our case. We adopted a geometrical thickness of 500\,km, typical for spicules. This leaves the temperature, the gas pressure, and the turbulent velocity as free parameters. Following Beckers (1968, 1972), we computed profiles and bulk parameters for grids of models as a function of temperature and gas pressure, for fixed values of the turbulent velocity; we did this for the k and h lines, as well as for H$_\alpha$ and H$_\beta$. An example is given in Figure \ref{grid1}, which shows contours of the computed integrated intensity as a function of temperature and pressure (TP plots) for the  Mg{\sc\,ii} k and h lines and H$_\alpha$ for a turbulent velocity  $\xi=28$\,km\,s$^{-1}$ and for the intermediate height $H=5400$\,km which will be our starting point.

It is obvious that a particular observed value of a bulk parameter defines a line in the temperature-pressure (TP) plot and that lines for different bulk parameters should cross, providing a unique pair of temperature and density. However, as suggested by the plots of Figure \ref{grid1}, these lines are almost parallel to one another for Mg{\sc\,ii} and a unique crossing point cannot be defined. In order to remove the ambiguity we need additional information, such as the intensity of the Balmer lines which, due to differences in their process of formation, have different slopes than the Mg{\sc\,ii} doublet (Figure \ref{grid1}) and can thus intersect with the latter. 

\begin{figure}[!h]
\begin{center}
\includegraphics[height=3.5cm]{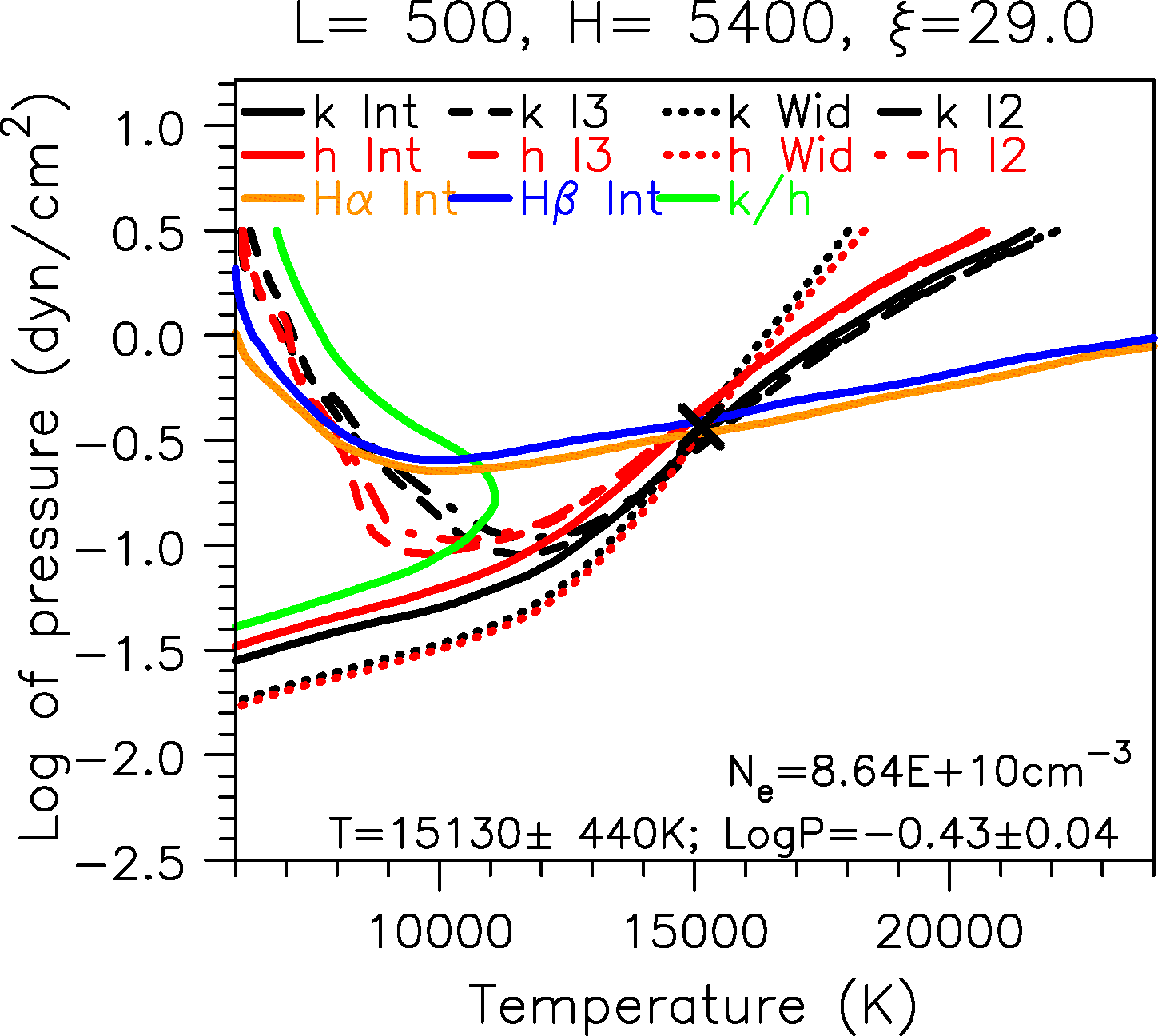}~
\includegraphics[height=3.5cm]{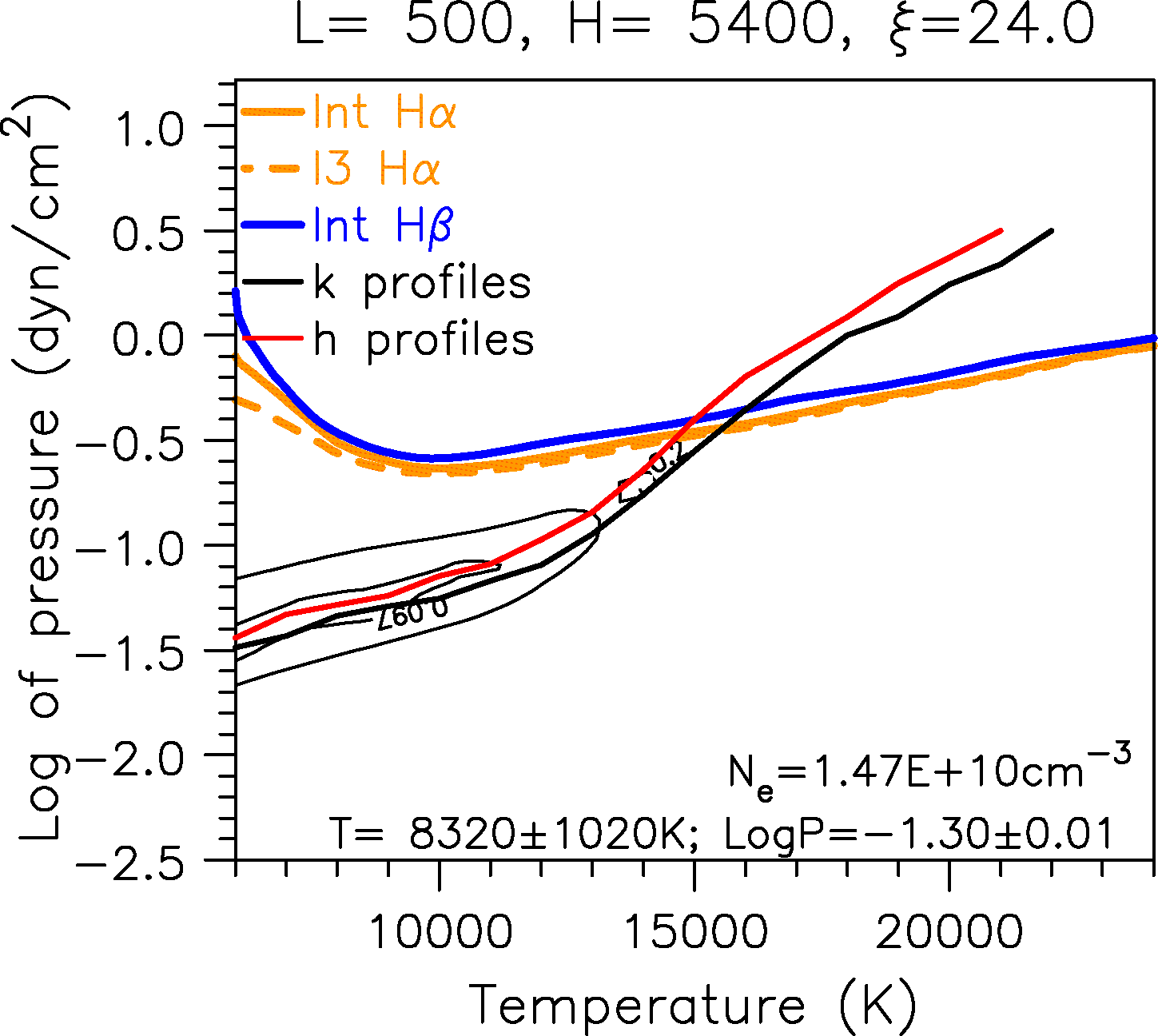}~
\includegraphics[height=3.5cm]{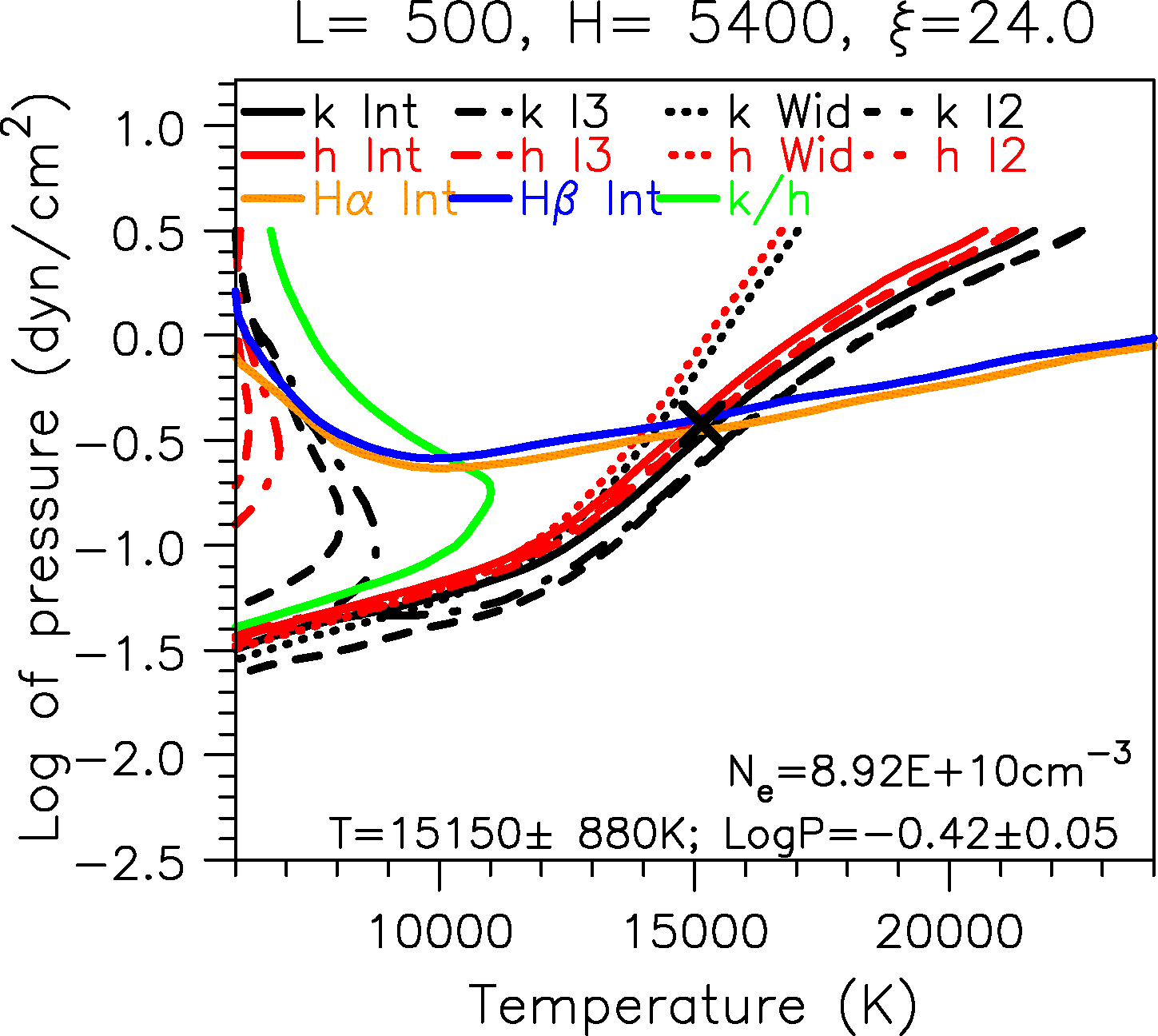}
\end{center}
\caption{Left: Temperature-pressure plot for the Mg{\sc\,ii} k and h lines (integrated intensity, core intensity (I3), peak intensity (I2), k/h ratio, and line width), as well as the integrated intensity of H$_\alpha$ and H$_\beta$ from \cite{1973SoPh...32..345A}. X marks the average position of the intersection of the Balmer line curves with the Mg{\sc\,ii} curves; the corresponding $T$, $P_g$, and $N_e$ values are given at the bottom of the plot. Center: TP plot of the minimum deviation of observed and model  Mg{\sc\,ii} profiles, together with the \ha\ and \hb\ integrated intensity curves. Right: Same as the left plot for $\xi=24$\kms.}
\label{Balmer}
\end{figure}

The left panel of Figure \ref{Balmer} shows TP plots for $H=5400$\,km of the bulk parameters of the k and h lines, as well as the \ha\ and \hb\ integrated intensity. We note that the the \ha\ and \hb\ curves are almost identical, indicating a good agreement between the model predictions and the observations for these lines. The value $\xi$  was chosen so that the dispersions minimize. The Mg{\sc\,ii} curves cross at practically the same point with the Balmer line curves; the $\times$ symbol marks the average of the intersections ($T\simeq15100$\,K, $P_g\simeq0.37$\dcm\ and $N_e\simeq8.6\times10^{10}$\,cm$^{-3}$) and the errors correspond to the dispersions of the crossing positions. For this height \cite{1973SoPh...32..345A} deduced electron temperatures between 12000 and 15000\,K, densities from 6 to $12\times10^{10}$\,cm$^{-3}$ and an average value of $\xi=18$\kms\ from 36 spicules observed simultaneously in \ha, \hb\ and Ca{\sc\,ii} K.

An alternative approach is to compute the average deviation between the observed and model k and h line profiles and seek its minimum in the $T$, $P_g$, and $\xi$ space. For each model profile we compute the deviations, $\chi^2_k$ and $\chi^2_h$, for the \Mg\ lines, normalized by the observed core intensity and combined, to obtain:
\be
\sigma_{hk}=\sqrt{\chi^2_k/I_{3k}^2+\chi^2_h/I_{3h}^2}
\ee
For a particular value of $\xi$,  the region of minimum  $\sigma_{kh}$ is located along a trough, extending diagonally on a TP plot. In the central panel of Figure \ref{Balmer} we present the location of the minimum deviation for k and h, as well as two contours around the combined $\sigma_{kh}$ minimum; the curves for \ha\ and \hb\ are also plotted and we selected the value of $\xi$ which minimized the minimum $\sigma_{kh}$. Although the curves for k and h intersect with those of the Balmer lines at about the same $P_g$ and $T$ as in the bulk parameter TP plot of the left panel of the figure, the actual \begin{figure}[!h]
\begin{center}
\includegraphics[height=3.5cm]{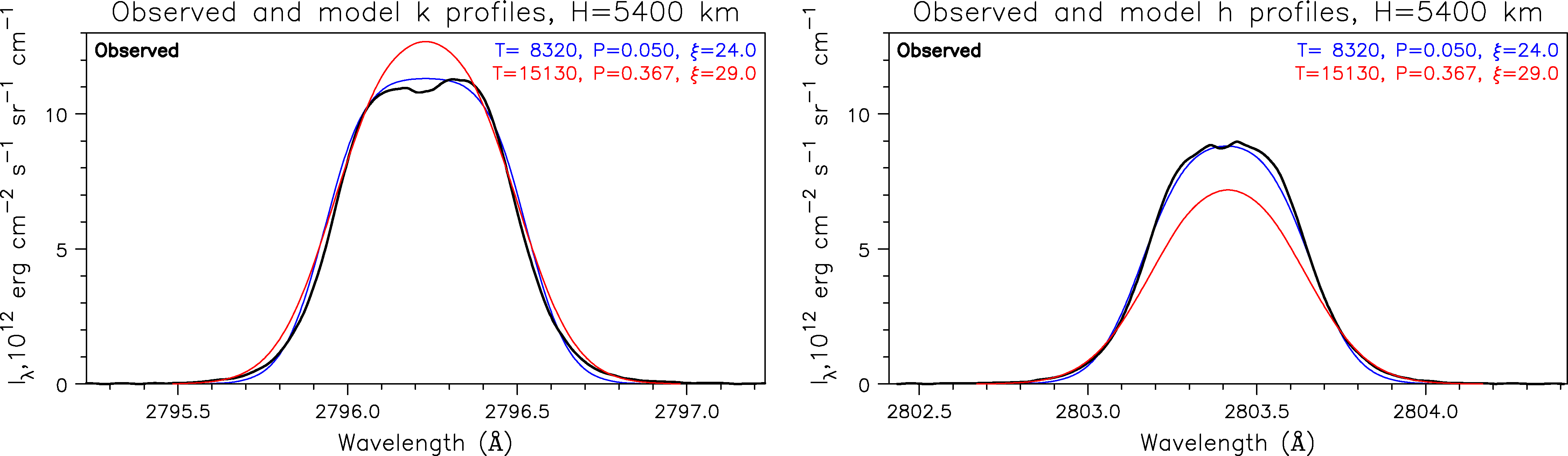}
\end{center}
\caption{Observed (thick lines) and model profiles of the k and h lines. Blue lines correspond to the profiles deduced from the minimization of the residuals, red lines from the \Mg\ and Balmer line bulk parameters. The corresponding values of $T$ (K), $P_g$ (\dcm) and $\xi$ (\kms) are marked on the plots.}
\label{pro5400}
\end{figure}
minimum of $\sigma_{kh}$ is at a very different location: $T\simeq8300$\,K and $P_g\simeq0.050$\dcm. For this combination of physical parameters the model predicts integrated intensities of 1.13 and 0.12\intun\ for \ha\ and \hb\ respectively, more than an order of magnitude smaller than the observed; it is hence incompatible with the Balmer observed line intensity. On the other hand, the model deduced from the bulk parameter TP plot gives profiles that do not reproduce well the peak intensity (Figure \ref{pro5400}), although they reproduce the line width.

It is appropriate at this point to discuss how much a change in the Balmer line intensities affects the derived values of the physical parameters, considering the fact that the Balmer lines were not observed simultaneously with the Mg{\sc\,ii} lines, as well as the scatter of the reported values. We found that a 10\% increase in the Balmer intensity produced a 7\% increase in $P_g$ and only a 1\% increase in $T$. It is also interesting to examine the effect of the geometrical thickness; changing $L$ from 500 to 1000\,km, we had a temperature increase of 2\% and a pressure decrease of 30\%. 

\begin{figure}[!h]
\begin{center}
\includegraphics[height=2.61cm]{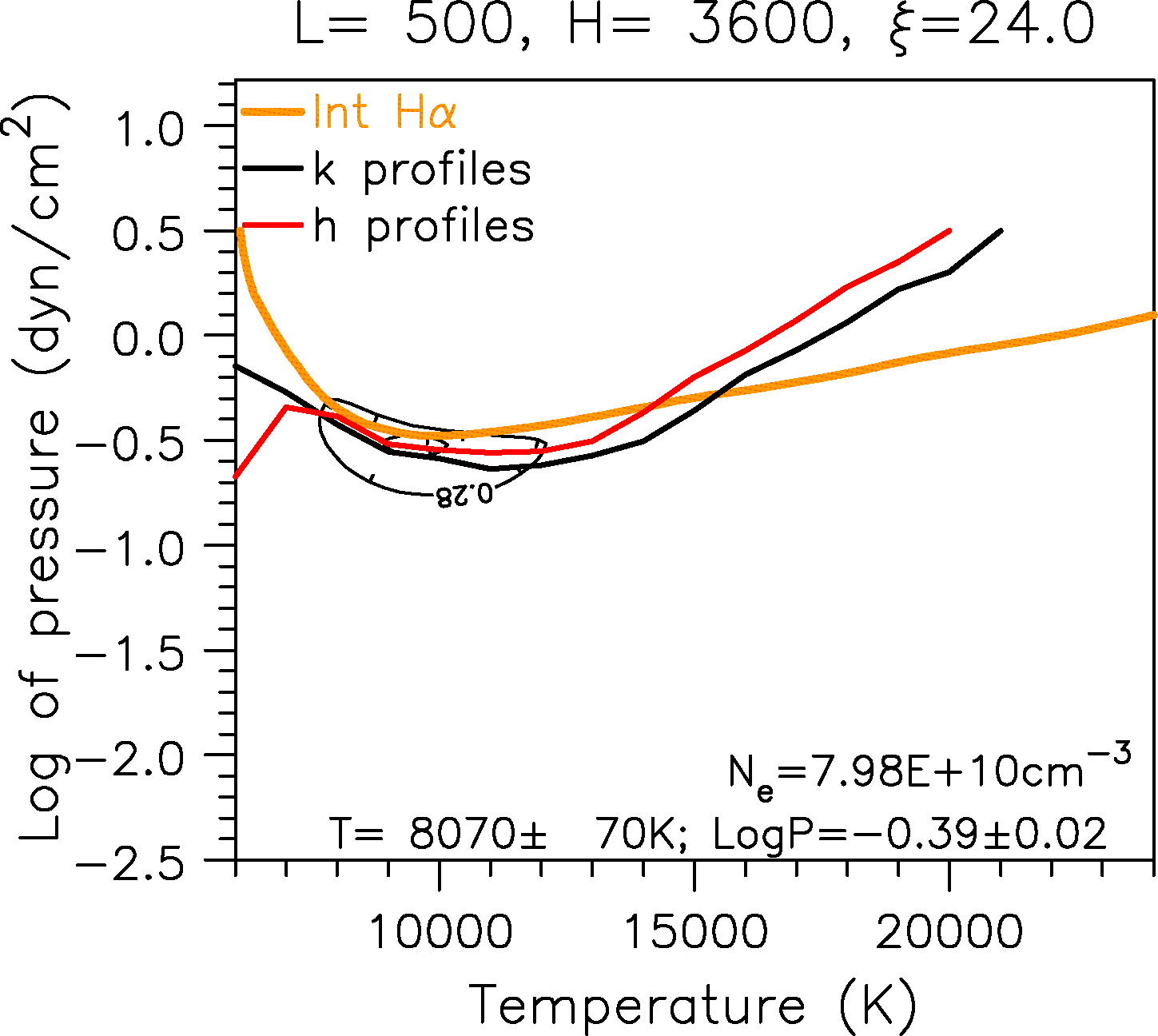}\,
\includegraphics[height=2.61cm]{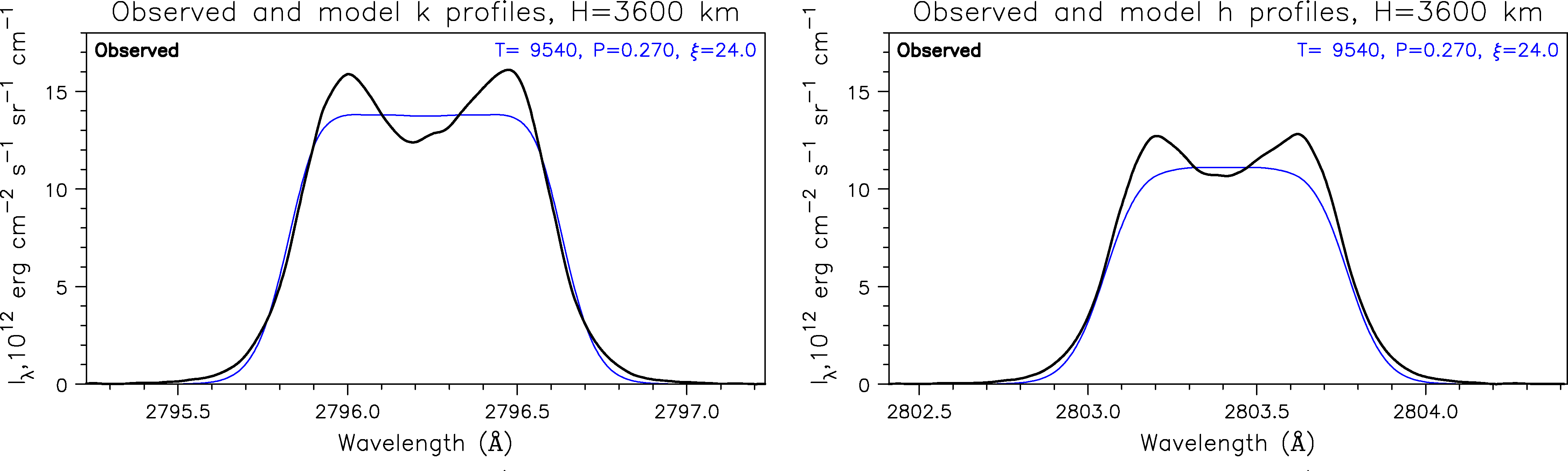}
\end{center}
\caption{TP plot for residuals (left) and \Mg\ line profiles (center and right) at a height of 3600\,km.}
\label{3600}
\end{figure}

\begin{figure}[!h]
\begin{center}
\includegraphics[height=2.61cm]{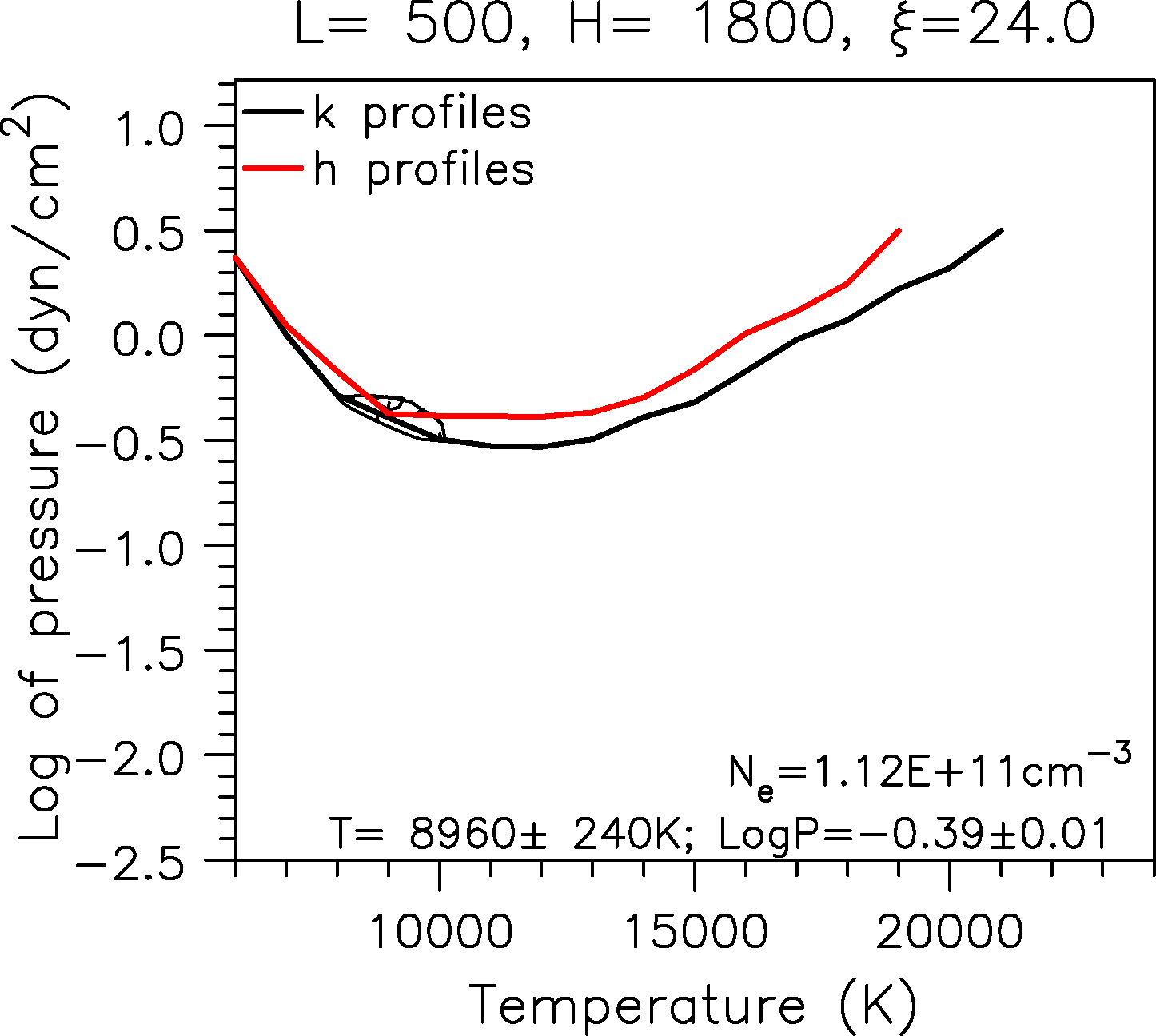}\,
\includegraphics[height=2.61cm]{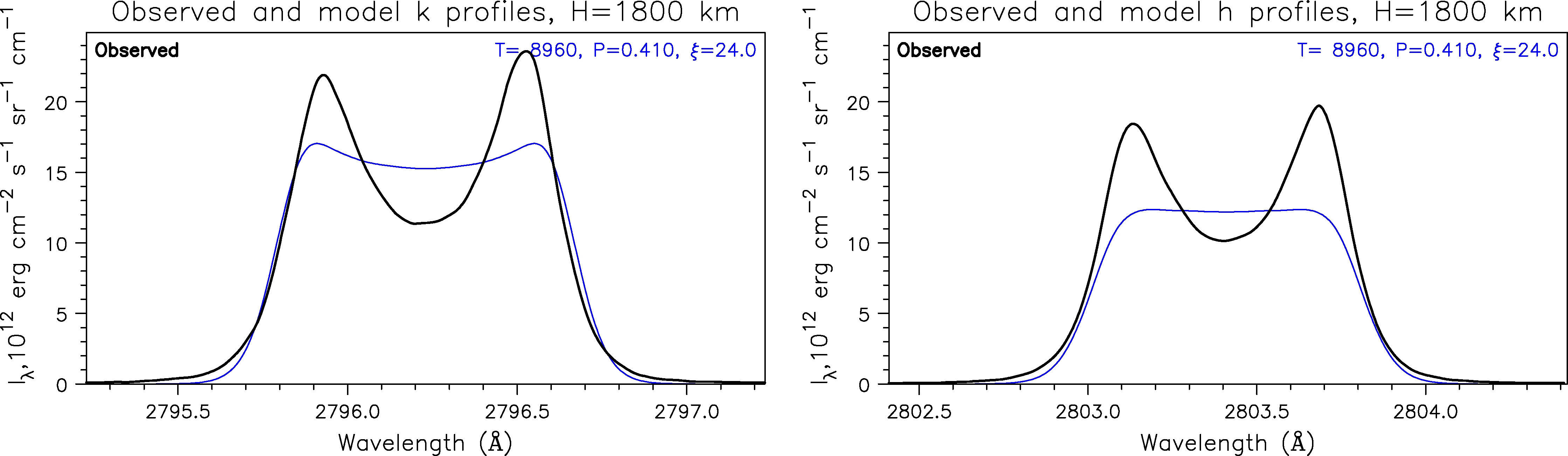}
\end{center}
\caption{TP plot for residuals (left) and \Mg\ line profiles (center and right) for the height of 1800\,km.}
\label{1800}
\end{figure}

Going to lower heights, Figure \ref{3600} shows our results for $H=3600$\,km. It is interesting that here the \ha\ curve passes near the minimum $\sigma_{kh}$, so that the result is compatible with the Balmer line observations. The corresponding model \Mg\ profiles reproduce well the observed line width, but are flat-topped rather than double-peaked. We note that there is a second crossing point around $T=14700$\,K and $P_g=0.49$\dcm, but this predicts profiles too narrow compared to the observed ones. The situation is similar for $H=1800$\,km, where we have no observed \ha\ values (Figure \ref{1800}). 

The absence of central reversals in the computed profiles at low heights could be due to the fact that our model assumes a geometrical thickness of 500\,km, which is supposed to be representative of individual spi\-cu\-les. As shown in Figure \ref{grid1}, model profiles are single-peaked over most of the parameter space, for reasons that have to do with low optical thickness and a rather constant source function inside the isothermal structure. The observed reversals at low heights may be due to overlapping spicules (\citealp{2015ApJ...806..170S}), which increase the opacity, or to varying thermodynamic conditions inside the structure that the isothermal model cannot take into account.

\begin{figure}[!h]
\begin{center}
\includegraphics[height=2.61cm]{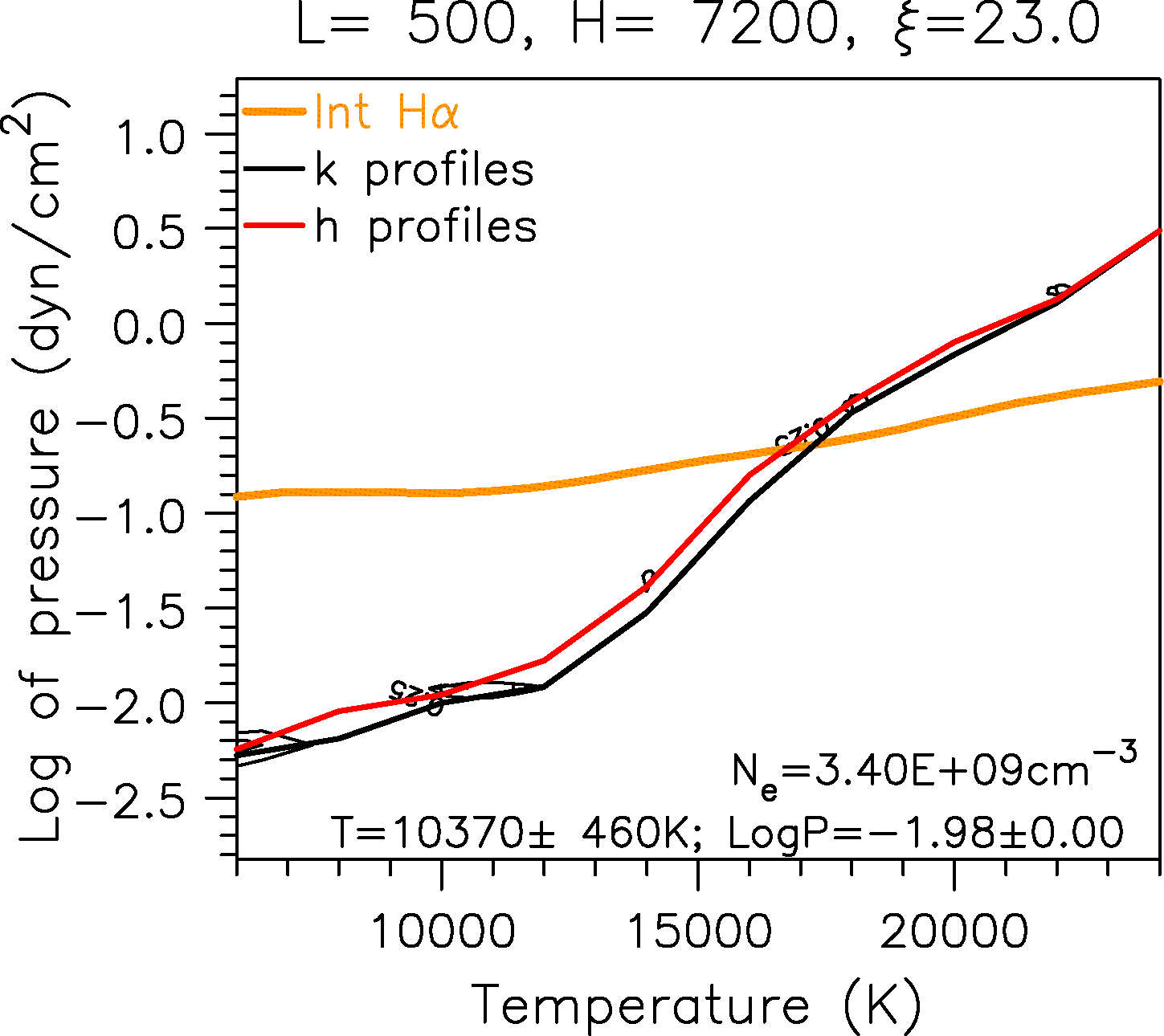}\,
\includegraphics[height=2.61cm]{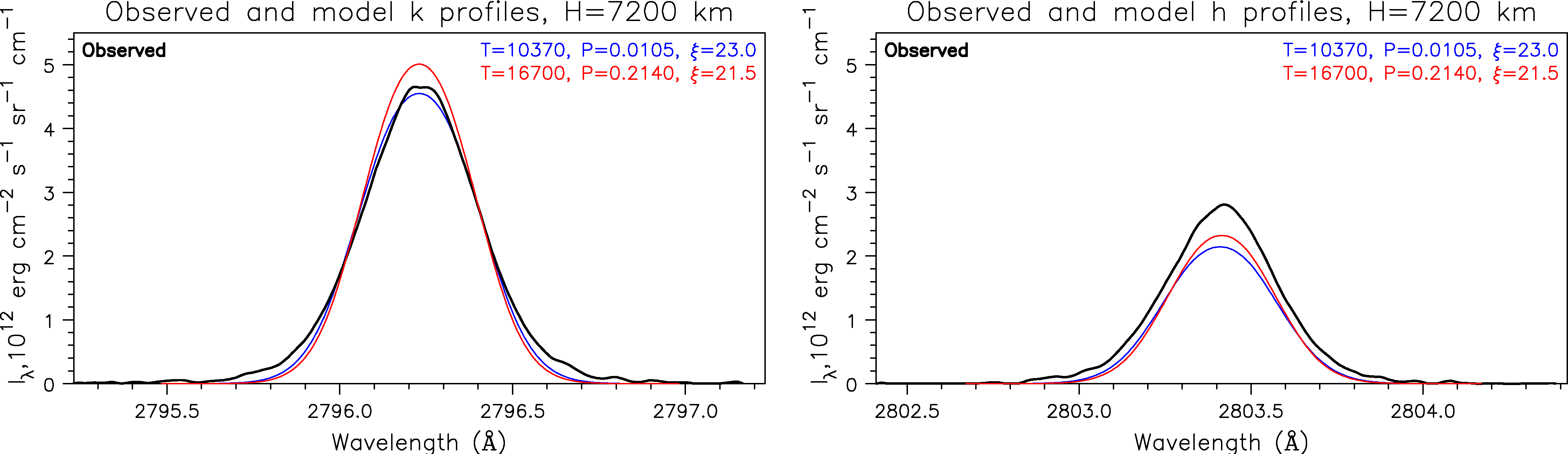}
\end{center}
\caption{TP plot for residuals (left) and \Mg\ line profiles (center and right) for the height of 7200\,km.}
\label{7200}
\end{figure}

Above 5400\,km the region of low $\sigma_{kh}$ is more extended with several minima and the choice of the most pertinent is difficult. At 7200\,km (left panel of Figure \ref{7200}) we have a low temperature minimum around 10400\,K which, as at 5400\,km, is far from the \ha\ curve; at this height the TP plot using both \Mg\ and \ha\ bulk parameters gives a higher temperature, 15700\,K. The corresponding k and h profiles, shown in the middle and right panels of  Figure \ref{7200}, are very similar.

\begin{figure}[!h]
\begin{center}
\includegraphics[height=2.61cm]{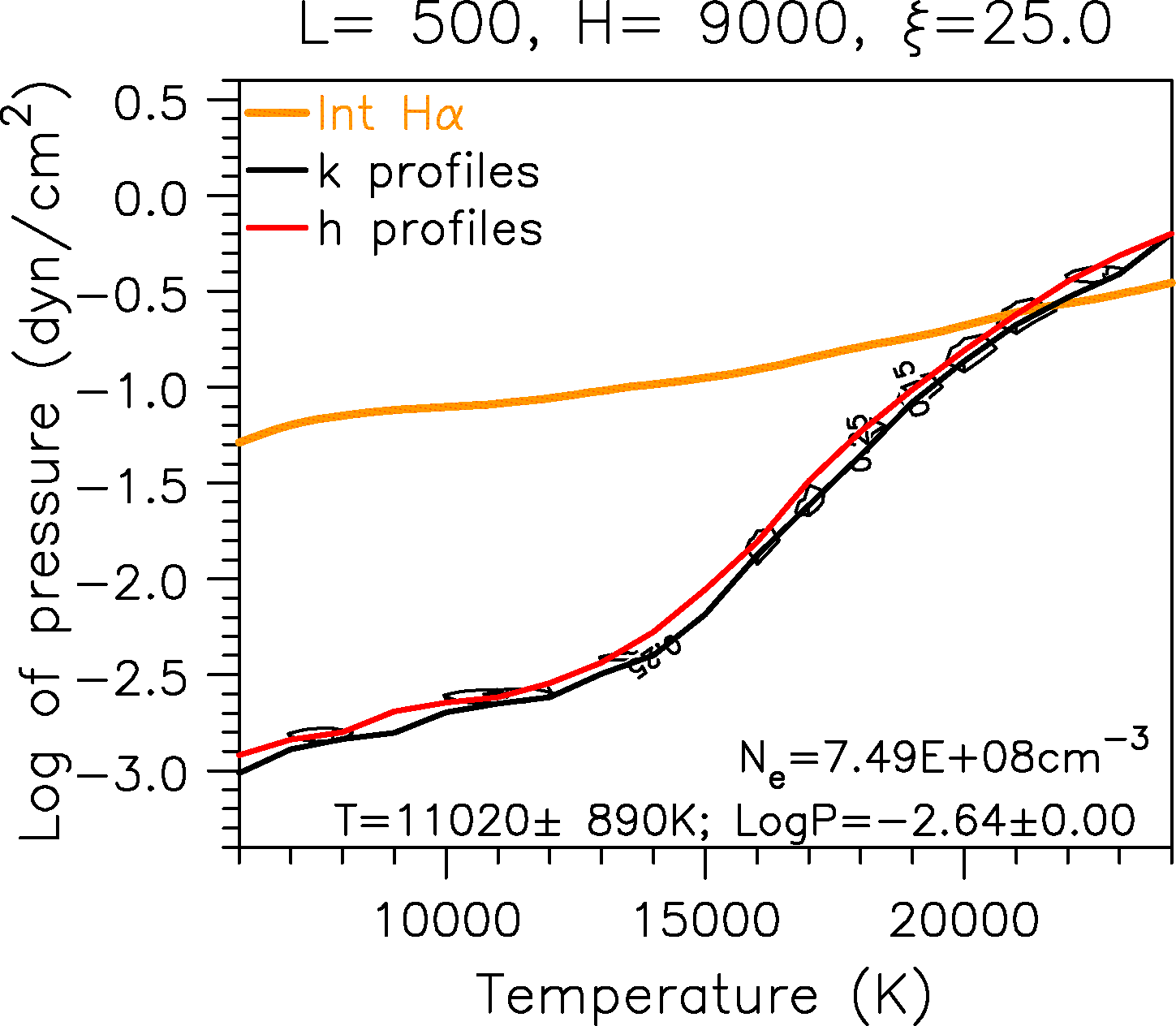}\,
\includegraphics[height=2.61cm]{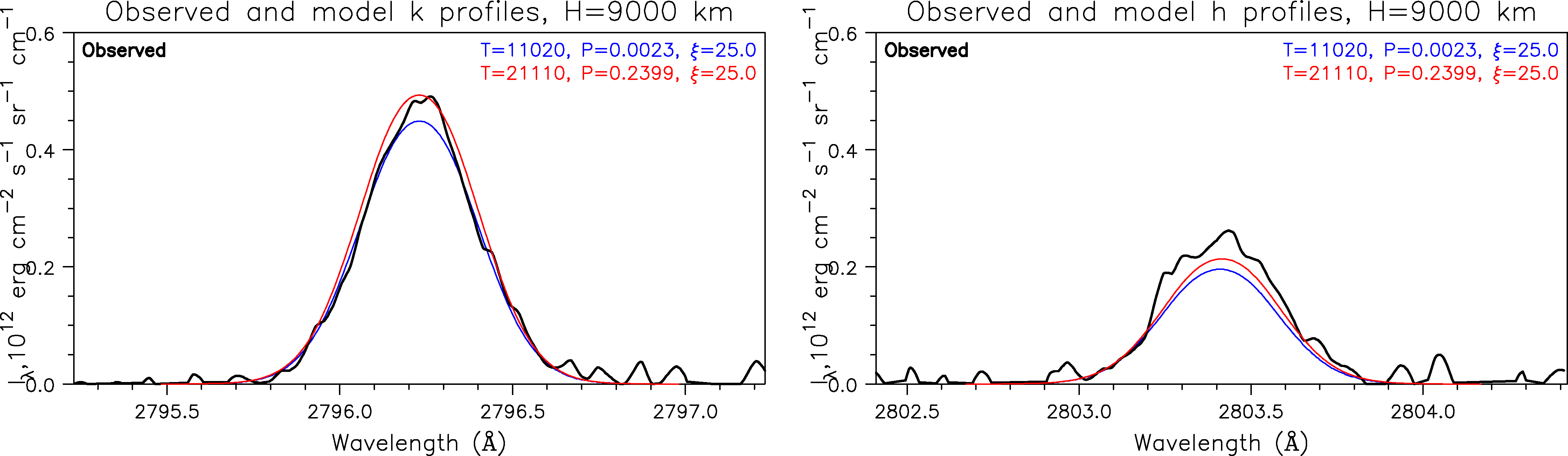}
\end{center}
\caption{TP plot for residuals (left) and \Mg\ line profiles (center and right) for the height of 9000\,km. The pressure scale in the left panel is shifted with respect to the previous figures.}
\label{9000}
\end{figure}

Similarly, at 9000\,km we have a low temperature solution around 11000\,K, which is not consistent with the \ha\ data and a high temperature solution around 21100\,K, when both \Mg\ and \ha\ are considered (Figure \ref{9000}). Again the differences between the respective line profiles are minimal.

Our results are summarized in Table \ref{derived}, together with the estimates of \cite{1972ARA&A..10...73B},  \cite{1976SoPh...46...93K}, and \cite{1988SoPh..117...21M}; the last authors derived $T_e$ and $\xi$ by comparing the widths of high order Balmer lines with those of Mg{\sc\,i} and Ti{\sc\,ii} in eclipse spectra. We did not include in the table our low temperature solutions for $H=7200$ and 9000\,km which predict too low densities without improving the agreement between observed and model \Mg\ profiles, but we retained the one for 5400\,km which gives a better fit to the k and h profiles.

\begin{table}[!ht]
\caption{Derived physical parameters}
\label{derived}
{\renewcommand{\tabcolsep}{0.07cm}
\begin{tabular}{cccccccccccccc}
\hline
 \multicolumn{4}{c}{This work}%
&\multicolumn{3}{c}{Beckers (1972)}%
&\multicolumn{4}{c}{Krall etal (1976)}
&\multicolumn{3}{c}{MH88}\\
Height&$\xi$& $T_e$ & $N_e$   &Height& $T_e$ & $N_e$   &Height&$\xi$&$T_e$ & $N_e$   &Height&$\xi$&$T_e$ \\
  km  & km/s&    K  &$10^{10}$&  km  &   K   &$10^{10}$&  km  &km/s &   K  &$10^{10}$&  km  &km/s &   K  \\
\hline
1800  & 24.0& ~9000 & 11.2 & ~2000  &~9000 & 16~ &   --  & -- &   --  &  -- & 2160 &  8.9& 9040\\
3600  & 24.0& ~8100 & 7.98 & ~4000  &17000 & 15~ &   --  & -- &   --  &  -- & 3730 & 13.3& 5440\\
5400  & 24.0& ~8300 & 1.47 &   --   &  --  & --  &   --  & -- &   --  &  -- & 4680 & 10.5& 7120\\
5400  & 29.0& 15130 & 8.64 & ~6000  &14000 & 8.9 & ~6000 & 14 & 13000 &  7  &  --  & --  &   --\\
7200  & 21.5& 16700 & 4.61 & ~8000  &16000 & 4.3 & ~8000 & 15 & 14000 &  4  &  --  & --  &   --\\
9000  & 25.0& 21100 & 4.11 & 10000  & --   & 3.4 & 10000 & 18 & 16000 &  2  &  --  & --  &   --\\
\hline
\end{tabular}
}
\end{table}

Our values for $T_e$ and $N_e$ are close to those of \cite{1972ARA&A..10...73B}, \cite{1973SoPh...32..345A}, and \cite{1976SoPh...46...93K}, except for $H=3600$; for $H=1800$ we are also close to \cite{1988SoPh..117...21M}. Our density values are higher, but comparable to those derived from the O\,{\sc iv} line ratios in Section \ref{thin}. The estimates for $\xi$ are systematically higher that previously reported values; they are also higher than the values we derived in this work for low ionization lines (Table \ref{Tab:Vturb}) and close to those for O\,{\sc iv}. Still, they are compatible with the observed width of the k and h lines at heights where they are optically thin: a FWHM of $\sim0.37$\,\AA\ (Table \ref{BulkSel}) gives $\xi\simeq23.5$\kms\ for $T_e\sim20000$\,K.

Before closing this section we will discuss a number of theoretical and observational issues that could influence our results:
\begin{enumerate}
\item Vertical motions will shift the incident profile; its FWHM of $\sim0.6$\,\AA\ corresponds to $\sim65$\kms, hence this effect will be important for fast (type II) spicules.
\item Spicules are not isolated structures, therefore incident radiation from the side could be important.
\item Integration over many structures along the line of sight can be important at low heights.
\item Spicules, intruding into the corona, should have their own transition region; hence the isothermal assumption may not be accurate enough. Relaxing the isothermal assumption may allow the \Mg\ and the Balmer line radiation to form in different regions, as suggested a long time ago (see \citealp{1968SoPh....3..367B,1972ARA&A..10...73B} for references) and may bring closer the low and high temperature solutions we found for $H=5400$\,km. 
\item \Mg\ and Balmer line observations are not simultaneous and may be affected by factors such as the location on the Sun and the solar cycle.
\item The geometrical location of the limb, which defines the origin of the height scale, may not be the same for \Mg\ and Balmer line observations. 
\end{enumerate}  

\section{Summary and Conclusions}
The high spatial and spectral resolution IRIS spectra and the associated slit-jaw images gave us the opportunity to conduct a detailed study of the solar chromosphere and low TR near and beyond the limb, to perform accurate measurements of the intensity and compute bulk spectral parameters. We took particular care to correct for instrumental effects, such as pointing, jitter, and diffuse light beyond the limb. Absolute intensity calibration was performed using the spectra of \cite{1976ApJ...205..599K} in the NUV and the IRIS in-flight calibration in the FUV spectral range.

In this article we focused on the average spectral properties, retaining the dynamics for future work. Although our emphasis is on the strong \Mg, \Cb, and \Si\ lines, we found that weaker lines (such as the \Mg\ triplet) and the NUV continuum contain valuable information on the variation of the average physical parameters as a function of height in the region of their formation. 

The profiles of the \Mg\ k and h lines as well as of the \Cb\ doublet are double peaked up to a height of about 8\arcsec\ above the limb, while the \Si\ and weak lines, including NUV lines in emission beyond the limb, are optically thin. Departures of the \Si\ profiles on the disk from the Gaussian shape indicate that the distribution of turbulent velocity is non-Gaussian.

We computed the variation of intensity with position near the limb and from that we estimated the effective height of formation (from the peak of the curve) and the range of formation heights (from the inflection points). We found that for the k  line the  outer inflection point (outer limb) is at a height of 9.5\arcsec, for the 1334.54\,\AA\ \Cb\ line at 8.9\arcsec, and for the 1393.79\,\AA\ \Si\ line at 5.8\arcsec. Here the height is measured with respect to the inflection point of the intensity curve at 2832\,\AA, which is the brightest region of the spectrum in the IRIS NUV range. We also computed the scale height of the line intensity and found that it increased with the height of the outer limb. A similar behavior was found for the scale height of the NUV continuum intensity, with a slope of the regression line two times higher. The increase of the intensity scale height is apparently due to the increase of the temperature with height, thus our results provide potentially useful information to be exploited in future works on atmospheric modeling. 

Although both \Si\ lines are optically thin on the disk, opacity effects are noticeable in the stronger of the two lines close to the limb; beyond the limb, we estimate a maximum optical depth of 1.5 for the weaker line at a height of 2\,--\,4\arcsec. For \Si\ and other optically thin lines we estimated the value of turbulent velocity, which we found to be between $\sim7$\kms\ for O\,{\sc i} and 23\kms for O\,{\sc iv}. From the ratio of the 1399.78 and 1401.16\,\AA\ O\,{\sc iv} lines we made a rough estimate of the electron density of $9.3\times10^9$\,cm$^{-3}$ at 1\arcsec\ and $2.4\times10^9$\,cm$^{-3}$ at 5.5\arcsec\ above the limb.

Although the spectrograph slit was placed at two fixed positions, spatio-temporal information could be obtained by creating 
images of the intensity as a function of position and time. Such images for lines and continua forming low in the atmosphere show clear evidence of p-mode oscillations on the disk. It is important to note that no spicular structure is visible beyond the limb in lines such as the \Mg\ triplet, O\,{\sc i} and others (Figures \ref{CutsLow} and \ref{LimbCuts}), in agreement with the finding of \cite{1968SoPh....5..131P} for rare-earth lines.  As pointed out by these authors, this suggests the presence of an interspicular region in the chromosphere. In addition to that, this observation may indicate that spicules are formed above the height of 2\arcsec, the chromosphere being more homogeneous up to that height. Slow-moving features are visible in the O\,{\sc i} position-time images, the origin of which we could not identify. Lines formed higher in the atmosphere (\Mg, \Cb, and \Si) show spicular structure extending more than 8\arcsec\ above the limb with individual features up to 15\,--\,20\arcsec. Bright, fast-moving structures with their footpoints near network points on the disk are prominent in \Cb, apparently associated with fast moving spicules.

We selected \Mg\ k and h profiles at specific heights from the average spectra and from a quiet spectrum for comparison with model computations. We used the PROM57Mg model developed at IAS, which is an isothermal and isobaric NLTE model of a vertical slab that treats properly and in detail the incident radiation and computes profiles of the \Mg\ as well as of the Balmer lines. We found that 70\,--\,75\% of the k and h emission is due to scattering. The comparison of the model computations with the observations and the determination of the physical parameters was done by two methods: (a) plotting contours for each bulk parameter of k and h as well as for the Balmer lines from the literature on a $P_g(T)$ diagrams (TP plots) and determining their crossing point, and (b) computing the deviation,  $\sigma_{kh}$, between the model and observed \Mg\ profiles and locating its minimum, again in TP plots.

In general we obtained reasonable agreement between model and observations. At the height of 5400\,km we found that the best model derived from the combined \Mg\ and Balmer line bulk parameters had a greater deviation between observed and model k and h profiles than the best model derived from the minimum $\sigma_{kh}$ alone which, however, predicted Balmer line intensities more than a factor of ten smaller than those reported in the literature. Still the two methods gave compatible results both lower and higher in the atmosphere. At low heights the model gave flat-topped rather than the observed double peaked profiles. Our derived values of the electron temperature range from $\sim8000$\,K at small heights to $\sim20000$\,K at large heights and for the electron density from $1.1\times10^{11}$ to $4\times10^{10}$\,cm$^{-3}$; they are broadly consistent with previously reported values for spicules. Our estimate for the turbulent velocity is $\sim24$\kms, while values in the literature are smaller than 20\kms.

This work opens a number of issues that could be further exploited in the future. For example, the location of the limb in lines and continua and the intensity scale height that can provide input for improved modeling of the low chromosphere; the presence of a homogeneous layer, devoid of spicules in the low chromosphere; the origin of the slow-moving features in O\,{\sc i} and fast moving features in \Cb. Last but not least, we mention the need for improved NLTE modeling, taking into account vertical motions and relaxing the isothermal assumption.

\begin{acks}
The authors gratefully acknowledge use of data from the IRIS and SDO (AIA and HMI) databases. IRIS is a NASA small explorer mission developed and operated by LMSAL with mission operations executed at NASA Ames Research center and major contributions to downlink communications funded by ESA and the Norwegian Space Centre. C.E.A and A.K. wish to thank the Institut d' Astrophysique Spatiale for their warm hospitality during their stay in Orsay.
\end{acks}

\medskip\noindent{\footnotesize {\bf Disclosure of Potential Conflicts of Interest} The authors declare that they have no conflicts of interest.}

\end{article} 

\begin{thebibliography}{}

\bibitem[\protect\citeauthoryear{Alissandrakis}{1973}]
{1973SoPh...32..345A}
Alissandrakis, C.E.: 1973, {\it Solar Phys.} {\bf 32}, 345, DOI: 10.1007/BF00154947. 

\bibitem[\protect\citeauthoryear{Avery and House}{1968}]
{1968ApJ...152..493A}
Avery, L.W.~and House, L.L.: 1968, {\it Astrophys. J.} {\bf 152}, 493, DOI: 10.1086/149566. 

\bibitem[\protect\citeauthoryear{Avery and House}{1969}]
{1969SoPh...10...88A}
Avery, L.W.~and House, L.L.: 1969, {\it Solar Phys.} {\bf 10}, 88,  DOI: 10.1007/BF00146158. 

\bibitem[\protect\citeauthoryear{Avery, House, and Skumanich}{1969}]
{1969JQSRT...9..519A}
Avery, L.W., House, L.L., Skumanich, A.: 1969, {\it J. Quant. Spectrosc. Rad. Trans.} {\bf 9}, 519, DOI: 10.1016/0022-4073(69)90004-1. 

\bibitem[\protect\citeauthoryear{Beckers}{1968}]
{1968SoPh....3..367B}
Beckers, J.M.: 1968, {\it Solar Phys.} {\bf 3}, 367, DOI: 10.1007/BF00171614.

\bibitem[\protect\citeauthoryear{Beckers}{1972}]
{1972ARA&A..10...73B}
Beckers, J.M.: 1972, {\it Annu. Rev. Astron. Astrophys.} {\bf 10}, 73, DOI: 10.1146/annurev.aa.10.090172.000445 . 

\bibitem[\protect\citeauthoryear{Brekke}{1993}]
{1993ApJS...87..443B}
Brekke, P.: 1993, {\it The Astrophys. Journal Supplement Series} {\bf 87}, 443, DOI: 10.1086/191810. 

\bibitem[\protect\citeauthoryear{Buchlin and Vial}{2009}]
{2009A&A...503..559B}
Buchlin, E.~and Vial, J.-C.: 2009, {\it Astron. Astrophys.} {\bf 503}, 559, DOI: 10.1051/0004-6361/200811588. 

\bibitem[\protect\citeauthoryear{Chae, Sch{\"u}hle, and Lemaire}{1998}]
{1998ApJ...505..957C}
Chae, J., Sch{\"u}hle, U., Lemaire, P.: 1998, {\it Astrophys. J.} {\bf 505}, 957, DOI: 10.1086/306179. 

\bibitem[\protect\citeauthoryear{Curdt \emph{et al.}}{2001}]
{2001A&A...375..591C}
Curdt, W., Brekke, P., Feldman, U., Wilhelm, K., Dwivedi, B.N., Sch{\"u}hle, U., Lemaire, P.: 2001, {\it Astron. Astrophys.} {\bf 375}, 591, DOI: 10.1051/0004-6361:20010364. 

\bibitem[\protect\citeauthoryear{De Pontieu \emph{et al.}}{2007}]
{2007PASJ...59S.655D}
De Pontieu, B., McIntosh, S., Hansteen, V.H., Carlsson, M., Schrijver, C.J., Tarbell, T.D., Title, A.M., Shine, R.A., Suematsu, Y., Tsuneta, S., Katsukawa, Y., Ichimoto, K., Shimizu, T., Nagata, S.: 2007, {\it Pub. Astron. Soc. Japan} {\bf 59}, S655, DOI:  10.1093/pasj/59.sp3.S655. 

\bibitem[\protect\citeauthoryear{De Pontieu \emph{et al.}}{2014}]
{2014SoPh..289.2733D}
De Pontieu, B., Title, A.M., Lemen, J.R., Kushner, G.D., Akin, D.J., Allard, B., Berger, T., Boerner, P., Cheung, M., Chou, C., Drake, J.F., Duncan, D.W., Freeland, S., Heyman, G.F., Hoffman, C., Hurlburt, N.E., Lindgren, R.W., Mathur, D., Rehse, R., Sabolish, D., Seguin, R., Schrijver, C.J., Tarbell, T.D., W{\"u}lser, J.-P., Wolfson, C.J., Yanari, C., Mudge, J., Nguyen-Phuc, N., Timmons, R., van Bezooijen, R., Weingrod, I., Brookner, R., Butcher, G., Dougherty, B., Eder, J., Knagenhjelm, V., Larsen, S., Mansir, D., Phan, L., Boyle, P., Cheimets, P.N., DeLuca, E.E., Golub, L., Gates, R., Hertz, E., McKillop, S., Park, S., Perry, T., Podgorski, W.A., Reeves, K., Saar, S., Testa, P., Tian, H., Weber, M., Dunn, C., Eccles, S., Jaeggli, S.A., Kankelborg, C.C., Mashburn, K., Pust, N., Springer, L., Carvalho, R., Kleint, L., Marmie, J., Mazmanian, E., Pereira, T.M.D., Sawyer, S., Strong, J., Worden, S.P., Carlsson, M., Hansteen, V.H., Leenaarts, J., Wiesmann, M., Aloise, J., Chu, K.-C., Bush, R.I., Scherrer, P.H., Brekke, P., Martinez-Sykora, J., Lites, B.W., McIntosh, S.W., Uitenbroek, H., Okamoto, T.J., Gummin, M.A., Auker, G., Jerram, P., Pool, P., Waltham, N.: 2014, {\it Solar Phys.} {\bf 289}, 2733, DOI: 10.1007/s11207-014-0485-y. 

\bibitem[\protect\citeauthoryear{De Pontieu \emph{et al.}}{2015}]
{2015ApJ...799L..12D} 
De Pontieu, B., McIntosh, S., Martinez-Sykora, J., Peter, H., Pereira, T.~M.~D.: 2015, \apjl, 799, L12, DOI: 10.1088/2041-8205/799/1/L12. 

\bibitem[\protect\citeauthoryear{Dere and Mason}{1993}]
{1993SoPh..144..217D}
Dere, K.P.~and Mason, H.E.: 1993, {\it Solar Phys.} {\bf 144}, 217, DOI: 10.1007/BF00627590. 

\bibitem[\protect\citeauthoryear{Du and Nakagomi}{1988}]
{1988ChA&A..12..136D}
Du, J.-S.~and Nakagomi, Y.: 1988, {\it Chinese Astron. Astrophys.} {\bf 12}, 136, DOI: 10.1016/0275-1062(88)90008-2. 

\bibitem[\protect\citeauthoryear{Dumont \emph{et al.}}{1983}]
{1983SoPh...83...27D}
Dumont, S., Pecker, J.-C., Mouradian, Z., Vial, J.-C., Chipman, E.: 1983, {\it Solar Phys.} {\bf 83}, 27, DOI: 10.1007/BF00148241. 

\bibitem[\protect\citeauthoryear{Dud{\'{\i}}k \emph{et al.}}{2014}]
{2014ApJ...780L..12D}
Dud{\'{\i}}k, J., Del Zanna, G., Dzif{\v c}{\'a}kov{\'a}, E., Mason, H.E., Golub, L.: 2014, {\it Astrophys. J.} {\bf 780}, L12, DOI:  10.1088/2041-8205/780/1/L12. 

\bibitem[\protect\citeauthoryear{Heinzel}{2015}]
{2015ASSL..415..103H}
Heinzel, P.: 2015, In: Vial, J.-C., Engvold, O. (eds.), {\it Solar Prominences, Astrophys. Space Scien. Lib.} {\bf 415}, 103, DOI: 10.1007/978-3-319-10416-4\_5. 

\bibitem[\protect\citeauthoryear{House and Avery}{1969}]
{1969JQSRT...9.1579H}
House, L.L.~and Avery, L.W.: 1969, {\it J. Quant. Spectrosc. Rad. Trans.} {\bf 9}, 1579, DOI: 10.1016/0022-4073(69)90096-X. 

\bibitem[\protect\citeauthoryear{Jones and Skumanich}{1980}]
{1980ApJS...42..221J}
Jones, H.P.~and Skumanich, A.: 1980, {\it Astrophys. J. Supp. Ser.} {\bf 42}, 221, DOI: 10.1086/190649. 

\bibitem[\protect\citeauthoryear{Jones and Skumanich}{1973}]
{1973ApJ...185..167J}
Jones, H.P.~and Skumanich, A.: 1973, {\it Astrophys. J.} {\bf 185}, 167,DOI: 10.1086/152406. 

\bibitem[\protect\citeauthoryear{Kramida \emph{et al.}}{2015}]
{2015NIST}
Kramida, A., Ralchenko, Yu., Reader, J., NIST ASD Team: 2015,
NIST Atomic Spectra Database (ver. 5.3). Available: {\tt{http://physics.nist.gov/asd}} National Institute of Standards and Technology, Gaithersburg, MD.

\bibitem[\protect\citeauthoryear{Kohl and Parkinson}{1976}]
{1976ApJ...205..599K}
Kohl, J.L.~and Parkinson, W.H.: 1976, {\it Astrophys. J.} {\bf 205}, 599, DOI: 10.1086/154317. 

\bibitem[\protect\citeauthoryear{Krall, Bessey, and Beckers}{1976}]
{1976SoPh...46...93K}
Krall, K.R., Bessey, R.J., Beckers, J.M.: 1976, {\it Solar Phys.} {\bf 46}, 93, DOI: 10.1007/BF00157556. 

\bibitem[\protect\citeauthoryear{Krat and Krat}{1971}]
{1971SoPh...17..355K}
Krat, V.A.~and Krat, T.V.: 1971, {\it Solar Phys.} {\bf 17}, 355, DOI: 10.1007/BF00150038. 

\bibitem[\protect\citeauthoryear{Krijger \emph{et al.}}{2001}]
{2001A&A...379.1052K}
Krijger, J.M., Rutten, R.J., Lites, B.W., Straus, T., Shine, R.A., Tarbell, T.D.: 2001, {\it Astron. Astrophys.} {\bf 379}, 1052, DOI: 10.1051/0004-6361:20011320. 

\bibitem[\protect\citeauthoryear{Labrosse}{2015}]
{2015ASSL..415..131L}
Labrosse, N.: 2015, In: Vial, J.-C., Engvold, O. (eds.), {\it Solar Prominences, Astrophys. Space Scien. Lib.} {\bf} 415, 131, DOI: 10.1007/978-3-319-10416-4\_6. 

\bibitem[\protect\citeauthoryear{Matsuno and Hirayama}{1988}]
{1988SoPh..117...21M}
Matsuno, K.~and Hirayama, T.: 1988, {\it Solar Phys.} {\bf 117}, 21, DOI: 10.1007/BF00148569. 

\bibitem[\protect\citeauthoryear{Mihalas, Auer, and Mihalas}{1978}]
{1978ApJ...220.1001M}
Mihalas, D., Auer, L.H., Mihalas, B.R.: 1978, {\it Astrophys. J.} {\bf 220}, 1001, DOI: 10.1086/155988. 

\bibitem[\protect\citeauthoryear{Pasachoff, Noyes, and Beckers}{1968}]
{1968SoPh....5..131P}
Pasachoff, J.M., Noyes, R.W., Beckers, J.M.: 1968, {\it Solar Phys.} {\bf 5}, 131, DOI: 10.1007/BF00147962. 

\bibitem[\protect\citeauthoryear{Pereira, De Pontieu, and Carlsson}{2012}]
{2012ApJ...759...18P}
Pereira, T.M.D., De Pontieu, B., Carlsson, M.: 2012, {\it Astrophys. J.} {\bf 759}, 18, DOI: 10.1088/0004-637X/759/1/18. 

\bibitem[\protect\citeauthoryear{Pereira \emph{et al.}}{2014}]
{2014ApJ...792L..15P}
Pereira, T.M.D., De Pontieu, B., Carlsson, M., Hansteen, V., Tarbell, T.D., Lemen, J., Title, A., Boerner, P., Hurlburt, N., W{\"u}lser, J.P., Mart{\'{\i}}nez-Sykora, J., Kleint, L., Golub, L., McKillop, S., Reeves, K.K., Saar, S., Testa, P., Tian, H., Jaeggli, S., Kankelborg, C.: 2014, {\it Astrophys. J.} {\bf 792}, DOI: 10.1088/2041-8205/792/1/L15.

\bibitem[\protect\citeauthoryear{Pereira \emph{et al.}}{2015a}]
{2015ApJ...806...14P}
Pereira, T.M.D., Carlsson, M., De Pontieu, B., Hansteen, V.: 2015, {\it Astrophys. J.} {\bf 806}, 14, DOI: 10.1088/0004-637X/806/1/14. 

\bibitem[\protect\citeauthoryear{Pereira \emph{et al.}}{2015b}]
{IRISguide}
Pereira, T.M.D, McIntosh, S.W., De Pontieu, B., Hansteen, V., Carlsson, M. and Boerner, P: 2015, 
A User¢s Guide to IRIS Data Retrieval, Reduction \& Analysis, Release 1.0.  Available: {\tt http://iris.lmsal.com/itn26/itn26.pdf}

\bibitem[\protect\citeauthoryear{Pierce}{1965}]
{1965PASP...77..137P}
Pierce, A.K.: 1965, {\it Pub. Astron. Soc. Pacific} {\bf 77}, 137, DOI: 10.1086/128179. 

\bibitem[\protect\citeauthoryear{Polito \emph{et al.}}{2016}]
{2016A&A...594A..64P}
Polito, V., Del Zanna, G., Dud{\'{\i}}k, J., Mason, H.E., Giunta, A., Reeves, K.K.: 2016, {\it Astron. Astrophys.} {\bf 594}, A64, DOI: 10.1051/0004-6361/201628965. 

\bibitem[\protect\citeauthoryear{Secchi}{1875}]
{1875leso.book.....S}
Secchi, A.: 1875, Le Soleil. Gauthier-Villars, Paris DOI: 10.3931/e-rara-14748.

\bibitem[\protect\citeauthoryear{Shibasaki, Alissandrakis, and Pohjolainen}{2011}]
{2011SoPh..273..309S}
Shibasaki, K., Alissandrakis, C.E., Pohjolainen, S.: 2011, {\it Solar Phys.} {\bf 273}, 309, DOI: 10.1007/s11207-011-9788-4. 

\bibitem[\protect\citeauthoryear{Skogsrud \emph{et al.}}{2015}]
{2015ApJ...806..170S}
Skogsrud, H., Rouppe van der Voort, L., De Pontieu, B., Pereira, T.M.D.: 2015, {\it Astrophys. J.} {\bf 806}, 170, DOI: 10.1088/0004-637X/806/2/170. 

\bibitem[\protect\citeauthoryear{Schmit \emph{et al.}}{2014}]
{2014A&A...569L...7S}
Schmit, D.J., Innes, D., Ayres, T., Peter, H., Curdt, W., Jaeggli, S.: 2014, {\it Astron. Astrophys.} {\bf 569}, L7, DOI: 10.1051/0004-6361/201424432. 

\bibitem[\protect\citeauthoryear{Sterling}{2000}]
{2000SoPh..196...79S}
Sterling, A.C.: 2000, {\it Solar Phys.} {\bf 196}, 79, DOI: 10.1023/A:1005213923962. 

\bibitem[\protect\citeauthoryear{Tavabi, Koutchmy, and Golub}{2015}]
{2015SoPh..290.2871T}
Tavabi, E., Koutchmy, S., Golub, L.: 2015, {\it Solar Phys.} {\bf 290}, 2871, DOI: 10.1007/s11207-015-0771-3. 

\bibitem[\protect\citeauthoryear{Tsiropoula \emph{et al.}}{2012}]
{2012SSRv..169..181T}
Tsiropoula, G., Tziotziou, K., Kontogiannis, I., Madjarska, M.S., Doyle, J.G., Suematsu, Y.: 2012, 
{\it Space Sci. Rev.} {\bf 169}, 181, DOI: 10.1007/s11214-012-9920-2. 

\bibitem[\protect\citeauthoryear{Tsiropoula and Tziotziou}{2004}]
{2004A&A...424..279T}
Tsiropoula, G.~and Tziotziou, K.: 2004, {\it Astron. Astrophys.} {\bf 424}, 279. DOI: 10.1051/0004-6361:20035794. 

\bibitem[\protect\citeauthoryear{Vernazza, Avrett, and Loeser}{1981}]
{1981ApJS...45..635V}
Vernazza, J.E., Avrett, E.H., Loeser, R.: 1981, {\it Astrophys. J. Supp. Ser} {\bf 45}, 635, DOI: 10.1086/190731. 

\bibitem[\protect\citeauthoryear{Vial}{1982}]
{1982ApJ...253..330V}
Vial, J.C.: 1982, {\it Astrophys. J.} {\bf 253}, 330, DOI: 10.1086/159639. 

\bibitem[\protect\citeauthoryear{Vial \emph{et al.}}{2016}]
{2016SoPh..291...67V}
Vial, J.-C., Pelouze, G., Heinzel, P., Kleint, L., Anzer, U.: 2016, {\it Solar Phys.} {\bf 291}, 67, DOI: 10.1007/s11207-015-0820-y. 

\bibitem[\protect\citeauthoryear{Zaqarashvili and Erd{\'e}lyi}{2009}]
{2009SSRv..149..355Z}
Zaqarashvili, T.V.~and Erd{\'e}lyi, R.: 2009, {\it Space Sci. Rev.} {\bf 149}, 355, DOI: 10.1007/s11214-009-9549-y. 

\bibitem[\protect\citeauthoryear{Zhang \emph{et al.}}{2012}]
{2012ApJ...750...16Z}
Zhang, Y.Z., Shibata, K., Wang, J.X., Mao, X.J., Matsumoto, T., Liu, Y., Su, J.T.: 2012, {\it Astrophys. J.} {\bf 750}, 16, DOI: 10.1088/0004-637X/750/1/16.

\end{thebibliography}
\end{document}